\documentclass[a4paper]{article}

\usepackage{jcappub}
\usepackage{amsmath}
\usepackage{bm}
\usepackage{graphicx}
\usepackage{enumitem}
\usepackage{geometry}
\usepackage{xspace}
\usepackage{xcolor}
\usepackage{hyperref}
\usepackage{verbatim}
\usepackage{soul}

\allowdisplaybreaks 

\makeatletter
\gdef\@fpheader{}
\g@addto@macro\bfseries{\boldmath}
\makeatother

\newcommand{\deflen}[2]{%
    \expandafter\newlength\csname #1\endcsname
    \expandafter\setlength\csname #1\endcsname{#2}%
}

\usepackage{caption}
\usepackage{subcaption}
\usepackage{mathrsfs}
\usepackage{graphicx}

\newcommand{\exd}{{ \mathrm{d}} }
\def\be{\begin{equation}}
\def\ee{\end{equation}}
\def\bea{\begin{eqnarray}}
\def\eea{\end{eqnarray}}

\def\cC{{\mathcal{C}}}
\def\cO{{\mathcal{O}}}

\def\cE{{\mathcal{E}}}

\def\cI{{\cal I}}
\def\cJ{{\mathcal{J}}}

\def\cL{{\cal L}}

\def\cR{{\mathcal{R}}}
\def\cS{{\mathcal{S}}}

\def\cW{{\cal W}}
\def\cX{{\cal X}}
\def\cY{{\cal Y}}
\def\cZ{{\cal Z}}

\def\ssA{{\scriptscriptstyle A}}
\def\ssB{{\scriptscriptstyle B}}

\def\ssI{{\scriptscriptstyle I}}

\def\ssR{{\scriptscriptstyle R}}

\def\mfa{{\mathfrak a}}
\def\mfb{{\mathfrak b}}
\def\mfc{{\mathfrak c}}
\def\mfh{{\mathfrak h}}

\def\mfp{{\mathfrak p}}

\def\mfz{{\mathfrak{z}}}

\def\bmk{{\bm{k}}}

\def\bmp{{\bm{p}}}
\def\bmq{{\bm{q}}}
\def\bmx{{\bm{x}}}

\def\scrE{\mathscr{E}}
\def\scrH{\mathscr{H}}

\def\kUV{{k_{\UV}}}

\def\Tr{\mathrm{Tr}}

\def\Trenv{\underset{(e)}{\mathrm{Tr}}}

\def\pref#1{(\ref{#1})}
\def\Mp{M_p}

\def\slrl{\epsilon_1}
\def\IR{{\scriptscriptstyle IR}}
\def\UV{{\scriptscriptstyle UV}}
\def\nn{\nonumber}
\newcommand{\roughly}[1]{\mathrel{\raise.3ex\hbox{$#1$\kern-0.75em
\lower1ex\hbox{$\sim$}}}}
\newcommand{\lsim}{\roughly<}

\def\w{\tilde v}
\def\p{\tilde p}
\def\sys{{(s)}}
\def\env{{(e)}}
\def\in{\mathrm{in}}

\subheader{}

\title{Tensor Catalyzed Decoherence of Primordial Scalar Fluctuations}

\author[a,b,c]{C.P.~Burgess,}
\author[a,b]{D.~Dineen,}
\author[d]{R.~Holman,}
\author[e,f]{Greg Kaplanek}

\affiliation[a]{Department of Physics \& Astronomy,  McMaster University, Hamilton, ON, Canada
}

\affiliation[b]{Perimeter Institute for Theoretical Physics, Waterloo, ON, Canada
}
 
\affiliation[c]{School of Theoretical Physics, Dublin Institute for Advanced Studies, 
Dublin, Ireland}

\affiliation[d]{Minerva University, 14 Mint Plaza, San Francisco, CA, 
USA}

\affiliation[e]{Department of Electrical Engineering and Computer Science, Syracuse University, NY 13210, USA}

\affiliation[f]{Institute for Quantum \& Information Sciences, Syracuse University, NY 13210, USA}

\emailAdd{cburgess@pitp.ca}
\emailAdd{ddineen@pitp.ca}
\emailAdd{rfcholman@gmail.com}
\emailAdd{gkaplane@syr.edu}

\date{today}

\begin{document}

\sloppy

\abstract{We provide the first complete calculation of the leading contribution to the decoherence of primordial fluctuations in the scalar part of the metric fluctuations within simplest, single-clock, inflationary models, assuming decoherence comes from gravitational interactions with an environment made up of the other, unmeasured, short-wavelength scalar and tensor  modes. We include the contributions from {\it all} of the contributing interactions at leading order in powers of $(H/\Mp)^2$ and of the slow-roll parameter $\slrl$. Unlike all extant calculations in the literature we find a result that is {\it not} slow-roll suppressed (and so is the same order of magnitude as tensor-mode decoherence). The difference arises because the dominant decoherence comes from interactions that involve {\it both} the tensor and scalar environments simultaneously (and so are missed when they are examined separately). We verify that the leading contributions to decoherence are UV finite (as they must be). We confirm that the interactions driving decoherence become time-local and Gaussian in the deep super-Hubble inflationary regime and show how this can be used to reliably compute the evolution of the purity in the late-time regime relevant for observations (where naive perturbation theory is known to break down). The same derivation shows explicitly why the effects that undermine perturbation theory for decoherence do not also undermine the basic inflationary prediction for the amplitude of primordial fluctuations (in agreement with general arguments).}

\maketitle

\section{Introduction}
\label{sec:intro}

It is a remarkable part of the modern cosmological story that quantum fluctuations in the very early universe provide the best current explanation for the pattern of correlations we see in primordial density fluctuations in the later universe (for textbook reviews see {\it e.g.}~\cite{Brandenberger:1994ce, Mukhanov:2005sc, Weinberg:2008zzc, Baumann:2022mni}). These are the same density fluctuations whose gravitational self-interactions ultimately seed (and explain the properties of) the riot of structure we find in the sky around us. Quantum gravitating effects in this most conservative of pictures are not only observable: {\it they have already been observed!}

This observation has invited many proposals to try to find observational tests of the quantum nature of primordial fluctuations \cite{Martin:2012pea, Maldacena:2015bha, Martin:2015qta, Martin:2017zxs, Choudhury:2016cso, Martin:2018zbe, Green:2020whw,Espinosa-Portales:2022yok, brahma2022universal,Martin:2021znx,Martin:2022kph,Piotrak:2025zhy,Micheli:2025yux,Haque:2025pav,Liu:2026mzz,Ireland:2026txt,Cielo:2026njg}, and (starting in very early days \cite{Sakagami:1987mp,Brandenberger:1990bx,Matacz:1992mk,Lombardo:1995fg,Barvinsky:1995va,Polarski:1995jg,Calzetta:1995ys,Kubotani:1997xa,Kiefer:1998jk,Barvinsky:1998cq,Kiefer:1998qe,Bellini:2001jm,Lombardo:2005iz,Burgess:2006jn,Prokopec:2006fc,Sharman:2007gi,Weenink:2011dd,Kiefer:2008ku,Kiefer:2010pb,Franco:2011fg,Burgess:2014eoa,Boyanovsky:2015tba,Nelson:2016kjm,Hollowood:2017bil,Shandera:2017qkg,boddy2017decoherence,Bao:2019ghe,Brahma:2020zpk,Banerjee:2021lqu,Brahma:2022yxu,Colas:2022hlq,Burgess:2022nwu,DaddiHammou:2022itk,Colas:2022kfu,Sou:2022nsd,Boutivas:2023mfg,Tinwala:2024wod,Colas:2024xjy,Colas:2024ysu,deKruijf:2024ufs,Burgess:2024eng,Brahma:2024yor,Burgess:2025dwm,Cespedes:2025zqp,deKruijf:2025jya,Sano:2025ird,Lopez:2025arw,Cielo:2025ibc,Christie:2025knc,Choudhury:2025ssc,Haque:2026bby,Christie:2026dwx} has also stimulated many estimates of the rate with which these fluctuations decohere due to the presence of the cosmic environment in which they are found. Decoherence in this context is usually formulated in terms of the rate of change of the system's purity, defined by 
\be \label{PurityDef}
   \gamma := \hbox{Tr } \varrho^2 \,,
\ee
where $\varrho$ denotes the system's reduced density matrix (after the environment has been traced out). $\gamma$ as defined by \pref{PurityDef} lies between zero and unity and $\gamma = 1$ $\leftrightarrow$ $\varrho$ describes a pure state.

What emerges from these studies is this: the rate of decoherence found depends on the interaction strength between the observed modes and their environment, with stronger interactions generating faster decoherence. Since gravitational interactions are the weakest interactions of all they are often regarded as providing an estimate of the slowest possible decoherence rate. A very natural choice for the environment is the enormous number of short wavelength modes for each of the fields present since these are usually not measured in late-time cosmology. The decoherence process is necessarily subtle though, because unobserved short-wavelength modes always exist (even now) and do not observably decohere the world around us outside of cosmology. As a result significant decoherence only starts once modes become stretched to Hubble size \cite{DaddiHammou:2022itk, Burgess:2024heo, Burgess:2024eng}. 

Explicit calculations of the decoherence rate due to gravitational interactions with shorter-wavelength modes within minimal inflationary cosmologies usually start with the cubic interactions among scalar and tensor perturbations of the metric and inflaton fields, such as are listed in \cite{Maldacena:2002vr}. The Lagrangian density for these has the generic form
\be \label{LintFactors}
   \cL_{\rm int}(x) = \sum_A \cO_\ssA(x)  \otimes \cO_\ssB(x)
\ee
where the local operator $\cO_\ssA$ acts on the `system' degrees of freedom of interest (those whose decoherence is to be computed) and $\cO_\ssB$ similarly acts on the environment. 

For calculations of the observed primordial density fluctuations the system degrees of freedom are the space of scalar fluctuations of the metric whose wavelength is long enough to be cosmologically relevant. Long-wavelength tensor modes can also be included to the extent that these are observable. The environment then consists of all of the short-wavelength scalar and tensor modes to which we have no access in cosmology. Each field in the problem can therefore be split into system and environmental contributions by separating long and short wavelength modes: $\zeta(x) = \zeta_{\sys}(x) + \zeta_{\env}(x)$.

General power-counting arguments show that the rate of gravitationally driven decoherence must be proportional to $(H/\Mp)^2$ -- where $H$ is the local Hubble parameter -- which is controllably small (though not negligible) during inflation. General arguments \cite{Burgess:2022nwu} then imply that (for simple inflationary models) the dominant interactions through which short-wavelength environments can decohere long-wavelength system evolution must involve only cubic interactions, such as
\be \label{LintSample}
   \cL_{\rm int} \ni c_1 \, \zeta \, \partial^\mu \zeta \, \partial_\mu \zeta \,, \qquad \hbox{or} \qquad
   \cL_{\rm int} \ni c_2 \, \gamma^{\mu\nu} \, \partial_\mu \zeta \, \partial_\nu \zeta \,,
\ee
or any of the other cubic interactions listed in \cite{Maldacena:2002vr}. Here $c_i$ are calculable coefficients and $\zeta$ is the scalar perturbation and $\gamma_{\mu\nu}$ the tensor perturbation that are defined in more detail in \S\ref{sec:OpenEFT} below. 

Early calculations start by working through the implications of a specific interaction for decoherence. For instance refs.~\cite{Nelson:2016kjm, Burgess:2022nwu, Sou:2022nsd} explore the decoherence of long-wavelength scalar perturbations, $\zeta_{\sys}$, interacting with the short-wavelength scalar environment, $\zeta_{\env}$, through the $\zeta (\partial \zeta)^2$ interaction of \pref{LintSample}. On the other hand, refs.~\cite{ye2018quantum, Burgess:2022nwu} explore the implication of a tensor environment on scalar-mode decoherence (and the implications of a scalar environment for tensor mode decoherence) generated by the coupling $\gamma (\partial \zeta)^2$ appearing in \pref{LintSample}. Some open-system studies involving bilinear $(\partial \gamma)^2$ \cite{deKruijf:2024ufs} and trilinear $\gamma (\partial \gamma)^2$ \cite{brahma2022universal} tensor interactions have also been done. 

For tensor perturbations these discussions agree that the leading contribution to decoherence arise with $\partial_t \gamma \sim H^2/\Mp^2$ while for scalar perturbations they instead give $\partial_t\gamma \sim \slrl H^2/\Mp^2$ and so are slow-roll suppressed. So far all calculations remain partial inasmuch as none of them includes the contributions of {\it all} interactions relevant at the order in $H/\Mp$ and $\slrl$ being computed. More recently \cite{Burgess:2025dwm} has studied the implications of {\it all} trilinear scalar interactions for the decoherence of scalar fluctuations, but this also remains incomplete because scalar-tensor interactions can also contribute at this order (such as those computed in \cite{Burgess:2022nwu}). Although partial calculations of decoherence can identify how the evolution of the purity depends on variables like $H/\Mp$, $\slrl$ and time, a full discussion of the overall numerical coefficient or of divergence cancellation requires knowing how {\it all} of the interactions contribute at a given order in $H/\Mp$ and $\slrl$. 

We here report on a calculation with a somewhat surprising result: we find a contribution to scalar decoherence that gives $\partial_t \gamma \sim H^2/\Mp^2$ and so -- unlike extant calculations in the literature -- is {\it not} slow-roll supressed (and so is the same order of magnitude as what is expected for the decoherence of tensor modes). Furthermore, we identify {\it all} of the interactions that can contribute at this order and so provide for the first time a complete calculation of the dominant decoherence rate of scalar perturbations in minimal inflationary models. The contributions of all other interactions are suppressed relative to what we find here by at least one additional power of $H/\Mp$ or $\slrl$. 

How can this be possible? The main difference between our calculation and earlier ones is that for the earlier calculations the division between system and environment was done in such a way that the environmental operator $\cO_\ssB$ appearing in \pref{LintFactors} contains all scalar or all tensor operators. These turn out to be slow-roll suppressed relative to the $\gamma (\partial \zeta)^2$ interaction of \pref{LintSample} where one of the $\zeta$'s is assigned to $\cO_\ssA$ while $\gamma_{\mu\nu}$ and the other $\zeta$ are assigned to $\cO_\ssB$. The dominant contribution comes neither from short-wavelength tensor or scalar environments separately; the dominant contribution requires {\it both} scalar and tensor environments taken together. In that sense the tensor modes can be regarded as being catalysts whose presence enhances the efficiency with which the scalar environment decoheres long-wavelength scalar perturbations. 

Assuming an initially pure state that agrees in the remote past with the Bunch-Davies vacuum, our complete perturbative result for the purity -- {\it c.f.}~eq.~\pref{LindbladPurity5int} below -- is
\be \label{PertResult}
  \gamma = \prod_\bmk \gamma_\bmk \quad \hbox{where} \quad
  \gamma_\bmk(z)    
  \simeq      1  - \frac{32}{45\pi^2} \left( \frac{H^2}{\Mp^2} \right)   \frac{k}{k_\UV} \left( \frac{a H}{k_\UV}\right)^2    \,,
\ee  
where $\bmk$ is the comoving momentum, $k = |\bmk|$ and $k_\UV$ is the co-moving scale that separates the system from the environment so $k < k_\UV$ for the system (though \pref{PertResult} assumes $k \ll k_\UV$). Here $a$ is the usual cosmological scale factor and $H = \dot a/a$ is its associated Hubble scale (which is approximately constant during inflation). Notice the absence of any factor of $\slrl = - \dot H/H^2$.

Expression \pref{PertResult} confirms many of the features that had emerged from earlier studies of the contributions of sub-dominant interactions. First, we find that the perturbative purity evolution shows `secular growth' in the sense that the small factor $\slrl (H^2/\Mp^2)$ appears multiplied by a function of time $(a H)^2 \propto H^2 e^{2Ht}$ that diverges in the late-time limit. This shows that perturbation theory in $H/\Mp$ and $\slrl$ must break down at sufficiently late times.\footnote{This is indeed the generic situation for perturbation theory in quantum mechanics: once one divides the Hamiltonian into large and small parts, $H = H_0 + H_{\rm int}$, there is always a time for which $e^{-i H_{\rm int} t}$ is not well-approximated by $1 - i H_{\rm int} t$ no matter how small $H_{\rm int}$ is relative to $H_0$.} 

But all is not lost because we show how it is also possible to resum the perturbative result and acquire reliable predictions in the later-time regime for which straight-up perturbation theory naively fails. The result we find in this way is
\be \label{ReSumResult}
  \gamma = \prod_\bmk \gamma_\bmk \quad \hbox{where} \quad
  \gamma_\bmk(z)  \simeq \frac{1}{\sqrt{1 + \Xi_k}} \quad \hbox{with} \quad
   \Xi_k \simeq   \frac{64}{45\pi^2} \left( \frac{H^2}{\Mp^2} \right)   \frac{k}{k_\UV} \left( \frac{a H}{k_\UV}\right)^2   \,,
\ee  
Unlike \pref{PertResult}, this expression remains well-behaved even once $\Xi_k$ is not small. Eq.~\pref{ReSumResult} very closely resembles one first found in \cite{Burgess:2022nwu}, but we are here able to put it on a more precise footing because we have access to the full leading contribution.

The key to the above resummation lies in the observation that the evolution of the reduced system density matrix becomes {\it time-local} (not necessarily Markovian, see footnote \ref{fn:nonMarkovian}) and Gaussian in the late-time deep super-Hubble regime (for which $k/a \ll H$). It is Gaussian in the sense that the density matrix factorizes for all times (at leading order) into a product of Gaussian density matrices $\varrho_\bmk$ for each mode. It is time-local in the sense that the evolution of the reduced density matrix becomes well-approximated by a Lindblad-like equation -- {\it c.f.}~eq.~\pref{eq:lindbladlikeform} below,
\begin{equation}
\label{LindbladForm}
\partial_{\eta} \varrho_{\bmk} = -i\Bigl[ \scrH_{\rm eff}(\eta), \varrho_{\bmk}(\eta)\Bigr] +\sum_{r,s=1}^2 {\rm h}^{s r}_\bmk\left[ {O}_{\bmk,s}\varrho_{\bmk} {O}^{\dag}_{\bmk,r}-\frac{1}{2}\left\{ {O}^{\dag}_{\bmk,r} {O}_{\bmk,s},\varrho_{\bmk}\right\}\right] \,,
\end{equation}
where $O_1 = \zeta_\bmk$ and $O_1 = \mfp_\bmk$ are the field modes and their canonical momenta and the effective Hamiltonian $\scrH_{\rm eff}$ and the coupling matrix\footnote{Notice that the determinant is $\det \mfh_\bmk = - |g_\bmk|^2 < 0$ and so at least one of the eigenvalues of $\mfh$ is negative. This need not mean \pref{CouplingMatrix} is unphysical. The Gorini-Kossakowski-Sudarshan-Lindblad theorem \cite{Lindblad:1975ef,Gorini:1976cm} applies to quantum dynamical semigroups generated by a time-independent Lindbladian, for which $\mfh$ must be positive semidefinite. For master equations with time-dependent coefficients (such as the one we find), positivity of the instantaneous matrix $\mfh(\eta)$ at all times $\eta$ is instead sufficient to ensure the related property of ``CP-divisibility'', namely that every intermediate propagator $\Phi(\eta_1,\eta_2)$ translating states via $\varrho(\eta_2)=\Phi(\eta_1,\eta_2)[ \varrho(\eta_1) ]$ is completely positive \cite{Breuer:2009pku,Rivas:2010pzm,Plenio:2014nyj,Breuer:2015zlm}. We instead regard the present dynamics as genuinely non-Markovian (and not necessarily CP-divisible), allowing the instantaneous matrix $\mathfrak{h}(\eta)$ to acquire negative eigenvalues. This interpretation is consistent with the exact Gaussian oscillator analyses of Di\'osi and Ferialdi \cite{Diosi:2014upa,Ferialdi:2017dnt}, where time-local master equations with temporarily negative rates reflect non-Markovian memory rather than a breakdown of physicality.\label{fn:nonMarkovian}} 
\be \label{CouplingMatrix}
  \mfh_\bmk = \Bigl( {\rm h}^{rs}_\bmk \Bigr) = \left( \begin{matrix} f_\bmk(t) & g_\bmk(t) \cr g^{\ast}_\bmk(t) & 0 \end{matrix} \right) \,,
\ee
completely calculable in terms of known functions $f_\bmk(t)$ and $g_\bmk(t)$. Notice that the evolution \pref{LindbladForm} is consistent with Gaussianity because its right-hand side is at most quadratic in the operators $\zeta_\bmk$ and $\mfp_\bmk$.

We are able to compute the leading late-time form of the couplings ${\rm h}^{rs}(t)$ explicitly and use them to compute the late-time evolution of the general equal-time correlators $\Sigma_{ij}(t) = \Tr[ \tfrac12 \{ O_i \,, O_j \} \varrho_\bmk]$, verifying explicitly that they satisfy the late-time constraints required on general grounds \cite{Assassi:2012et} (and in particular do not modify the leading prediction for the amplitude of primordial fluctuations). These constraints amount to an upper limit to how fast the functions $f_\bmk(t)$ and $g_\bmk(t)$ can grow at late times and we show explicitly why this is true and what this implies for the late-time evolution of $\Sigma_{ij}(t)$. The purity is related to $\Sigma_{ij}$ by $\gamma = (4 \det \boldsymbol{\Sigma})^{-1/2}$ and so these results also dictate its late-time evolution, leading to the prediction \pref{ReSumResult}. 
 
We present our results as follows. We start in \S\ref{sec:OpenEFT} with a short review of the open system of metric fluctuations in vanilla inflationary models, with some extra details given in Appendix \ref{App:sec:OpenEFT}. This is followed in \S\ref{sec:NoSloRollSuppression} by an identification of the interactions that produce decoherence that is not slow-roll suppressed and a calculation of the environmental correlators that are required in order to obtain the Lindblad-like form \pref{LindbladForm} and compute its coefficients. Extra details about this calculation are also provided in Appendix \ref{sec:Integrals}. Many of the arguments of \S\ref{sec:OpenEFT} and \pref{sec:NoSloRollSuppression} are compressed versions of arguments in \cite{Burgess:2022nwu} and \cite{Burgess:2025dwm} and so the interested reader can also consult these for more explicit versions. Finally \S\ref{ssec:TransportCoeffs} sets up and solves the general late-time evolution in the time-local and Gaussian limit and by doing so derives the resummed expression \pref{ReSumResult} for the late-time purity (with details provided in Appendix \ref{sec:Transport}). Our main conclusions are summarized in \S\ref{sec:Conclusions}.

\section{Open system of metric modes}
\label{sec:OpenEFT}

This section briefly summarizes those parts of the general calculational setup developed in \cite{Burgess:2022nwu, Burgess:2025dwm} that are needed for the present calculation. For readers who are double-parked Appendix \ref{sec:OpenEFT} contains a quick summary of the main results needed from these references.

\subsection{Semiclassical description}

The system of interest is the simplest `single-clock' inflationary models, with the metric $g_{\mu\nu}$ coupled to a real scalar inflaton field $\varphi$ through the action\footnote{With deep regret we use MTW conventions \cite{MTW} (wistfully gazing at \cite{Weinberg:1972kfs}).}
\begin{equation}
\label{actionstart}
S = \int \exd^4 x\; \sqrt{ - g} \bigg[\tfrac12  \Mp^2 \, R
    - \tfrac{1}{2} g^{\mu\nu} \, \partial_{\mu} \varphi \,
    \partial_{\nu} \varphi  - V(\varphi) \bigg]
\end{equation}
where $R$ is the Ricci scalar and $V(\varphi)$ is the potential energy of the inflaton $\varphi$. This action contains the minimal interactions that are required for the model's success and our goal is to determine the decoherence rate these imply. The intuition is that decoherence will proceed faster than this in a realistic model with more degrees of freedom and interactions, most of which are stronger than the gravitational strength interactions in \pref{actionstart}.  

To this end we study the standard evolution of semiclassical\footnote{We emphasize we follow standard semiclassical practice here and quantize both scalar field and metric, since in some circles `semiclassical' is perversely taken to mean `only quantize matter fields and not the metric'.} fluctuations about a homogeneous spatially flat configuration where $\varphi = \phi(t)$ and the spacetime geometry is described by an FRW metric
\begin{equation}
\label{metric}
\exd  s^2 = - \exd  t^2 + a^2(t) \, \exd  \bm{x}^2 = a^2(\eta)
  \left( - \exd  \eta^2 + \exd \bm{x}^2 \right) \,,
\end{equation}
where $t$ and $\eta$ respectively denote cosmic and conformal time, being related by $\exd t = a \, \exd \eta$. Throughout the paper we denote differentiation with respect to $t$ by overdots and differentiation with respect to $\eta$ by primes. In the strict de Sitter limit the scale factor is given by
\begin{equation}
\label{eq:defscalefactor}
a = e^{H_\ssI t} =-\frac{1}{H_\ssI \eta}\qquad \hbox{(de Sitter)}\,,
\end{equation}
for constant $H_\ssI$, with $-\infty < t < \infty$ corresponding to $- \infty < \eta < 0$. We only ask the evolution to be near-de Sitter, in the sense that the Hubble scale $H = \dot a/a$ satisfies $\slrl =-{\dot{H}}/{H^2} \ll 1$, so $H(t) \simeq H_\ssI$ is only approximately constant. The $\phi$ field then satisfies 
\be\label{phidotvseps}
\dot{\phi}^2=2H^2\Mp^2\slrl \,, 
\ee
and so its kinetic energy is small compared to its potential energy.  

\subsubsection{Fluctuations}

Fluctuations about this background are described by expanding the scalar field and metric
\begin{eqnarray} \label{ADMmetric}
\varphi = \phi(t) + \delta \varphi \quad\hbox{and} \quad
  \exd  s^2 = - N^2 \exd  t^2 + h_{ij} \big( \exd  x^i + N^i \exd  t \big)
  \big( \exd  x^j + N^j \exd  t \big) \,,
\end{eqnarray}
where the fluctuations $\delta \varphi$, $h_{ij}$, $N$ and $N^i$ are all functions of both $\bmx$ and $t$ (or $\eta$). We follow standard arguments and pick a gauge to fix time and spatial reparametrizations, after which the action (\ref{actionstart}) ends up leaving a single physical scalar degree of freedom plus the two tensor modes describing gravitational waves. Following \cite{Maldacena:2002vr} we write
\begin{equation} \label{metricsplit}
     h_{ij} = a^2 e^{2\zeta} \hat h_{ij} \quad\hbox{with}\quad
     \hat h_{ij} = \delta_{ij} + \gamma_{ij} + \tfrac12 \,
     \delta^{kl} \gamma_{ik} \gamma_{lj} + \cdots \,,
\end{equation}
where $\det \hat h_{ij} = 1$ and $\delta^{ij} \partial_i \gamma_{jk} = \delta^{ij} \gamma_{ij} = 0$. The lapse $N$ and shift $N^i$ are determined by solving the energy and momentum constraints and the remaining two scalar functions $\delta \varphi$ and $\zeta$ are not independent -- a coordinate transformation can be used to switch between $\delta \varphi = 0$ (co-moving gauge) or $\zeta = 0$ (spatially-flat gauge). 

The leading (quadratic) part of the action governing the propagation of fluctuations has the form (see for example \cite{Kodama:1985bj, Mukhanov:1990me, Maldacena:2002vr})
\be 
\label{freescalaraction}
  ^{(2)}S =  \int \exd  \eta \; \exd^3 \bm{x}\; a^2 \Mp^2  \bigg\{
  {\slrl}\Bigl[ (\zeta')^2 -  (\partial \zeta )^2 \Bigr] + \tfrac18 \Bigl[( \gamma_{ij}' )^2 - \partial_k \gamma_{ij} \, \partial_k \gamma_{ij} \Bigr] \bigg\} \,,
\ee
where $(\partial \zeta)^2 = \delta^{ij} \partial_i \zeta \, \partial_j \zeta$. For some purposes (such as power-counting how small parameters enter into calculations) it is useful to canonically normalize the fluctuation fields so that their correlators are order unity. For the scalar modes this leads to the standard Mukhanov-Sasaki variable~\cite{Mukhanov:1981xt,Kodama:1985bj} 
\begin{equation}
\label{eq:defzeta}
v(\eta, {\bm x}) := \mfz_s \, \zeta(\eta,{\bm x}) \qquad \hbox{where} \quad \mfz_s := a\Mp \sqrt{2\slrl } \,,
\end{equation}
in terms of which the scalar part of the quadratic action \pref{freescalaraction} is given in \pref{app:freescalaractionv} (see for example \cite{Kodama:1985bj, Mukhanov:1990me, Maldacena:2002vr}). For tensor modes the analogous canonical field is given by \pref{app:eq:defvij}.

Quantization proceeds by going to momentum space\footnote{For later convenience we box normalize momentum states in a fixed comoving volume.} and imposing the canonical quantization conditions between the fields and their momenta (for more details see Appendix \ref{sec:OpenEFT}). For instance, for the tensor modes and the field $\zeta$ the canonical momenta are $\mfp_{\bm{k}} = \delta S/\delta \zeta'_{\bm{k}}$ and $\pi_{ij} = \delta S/\delta {\gamma^{ij}_\bmk}'$ and the normalized mode decompositions become
\be
\label{eq:scalarmodedecomp}
\zeta(\bmx ,\eta)   =  \frac{1}{\sqrt{V}} \sum_{\bmk} \zeta_{\bmk}(\eta)\, e^{i \bmk\cdot\vec{x}} 
\quad \hbox{with} \quad
\zeta_{\bmk}(\eta) =  \hat u_k(\eta) \, c_{\bmk}+\hat u_k^*(\eta) \, c^{\dagger}_{-\bmk} 
\ee
and
\be \label{eq:tensormodedecomp}
\gamma_{i j}(\bmx ,\eta)  =   \frac{1}{\sqrt{V}} \sum_{\bmk,\sigma} \epsilon_{i j}(\hat{\bmk},\sigma) h_{\bmk\sigma}(\eta) \, e^{i \bmk\cdot\bmx }
\quad \hbox{with} \quad
 h_{\bmk\sigma}(\eta) = w_k(\eta) \, b_{\bmk\sigma} + s_{\sigma} w_k^*(\eta) \, b^\dagger_{-\bmk\sigma} \,,
\ee  
where $\sigma =+,\times$ denote the two graviton spin states and $\epsilon_{i j}(\hat{\bmk},\sigma)$ are their polarization tensors (whose explicit form depends on the direction $\hat{\bmk}$ of the graviton's momentum). The quantity $s_{\sigma}=\pm 1$ is the polarization-dependent sign appearing in the relation $\epsilon_{i j}(- \hat{\bmk},\sigma) =s_{\sigma} \, \epsilon_{i j}(\hat{\bmk},\sigma)$, whose detailed form does not matter in what follows.

In our applications we work to leading order in an expansion in powers of the small slow-roll parameter so it suffices to know expressions for the mode functions in the de Sitter limit, in which case $\hat u_k(\eta)$ and $w_k(\eta)$ just differ in normalization:  
\be \label{ModeExpressions}
\hat u_k(\eta) = \frac{u_k(\eta)}{\mfz_s} = \frac{i H}{\Mp} \frac{1}{\sqrt{4 k^3 \slrl}} (1+i k\eta) e^{-i k \eta}
\quad \hbox{and} \quad
w_k (\eta) = \frac{u_k(\eta)}{\mfz_t} = \frac{i H}{\Mp} \sqrt{\frac{2}{k^3}}(1+i k\eta) e^{-i k \eta} \,.
\ee

\subsection{The system and the environment}
\label{sec:SysEnv}

\begin{figure}
\centering
  \includegraphics[width=0.65\linewidth]{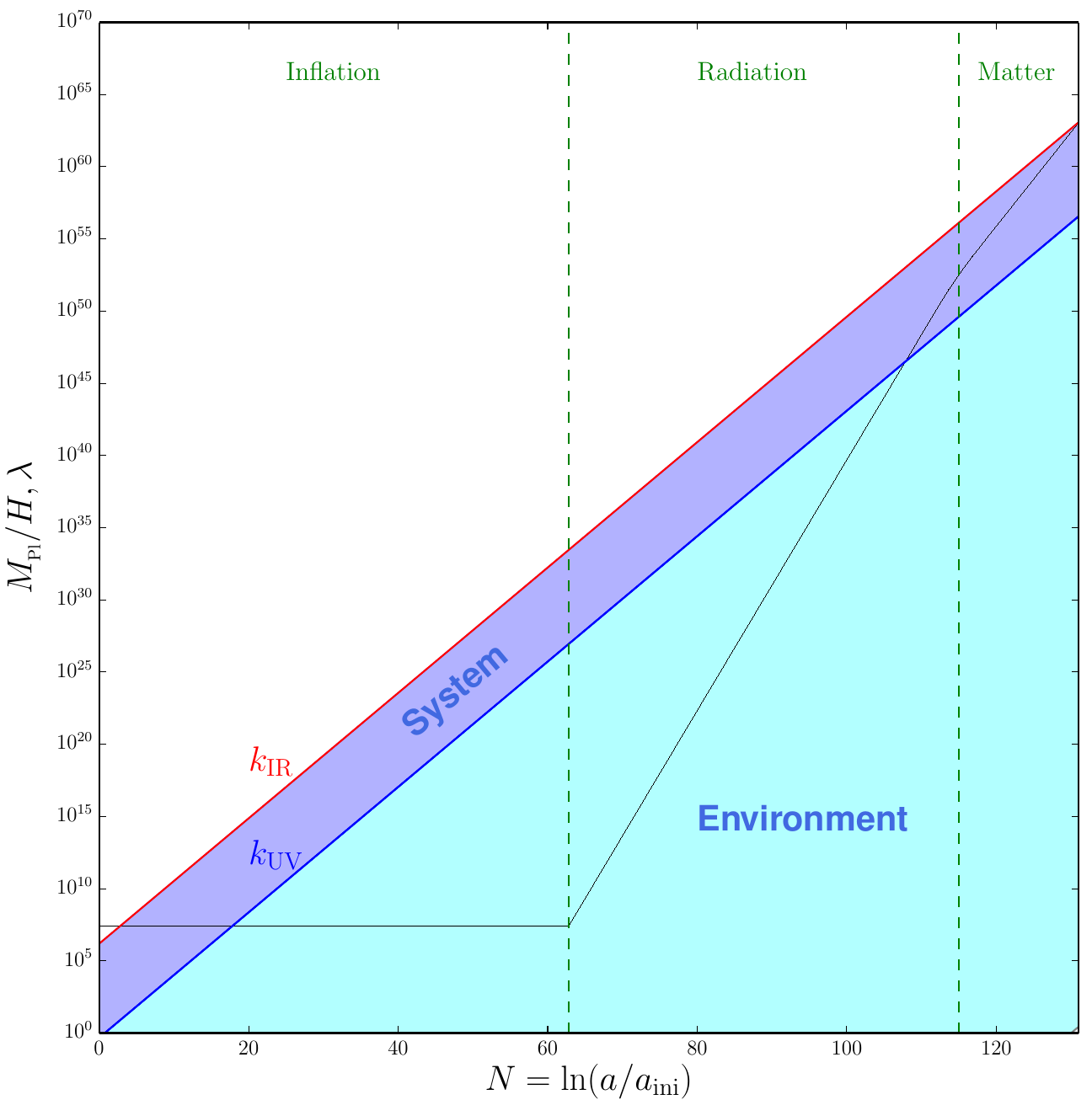}
  \caption{\small A sketch of the domain of the system and environment modes described in the text. The black line denotes the Hubble radius and the coloured lines stand for the mode wavelengths. The vertical axis represents length scales -- such as physical wavelengths or the Hubble length -- and the horizontal axis represents scale factor, which can be taken as a proxy for time. Vertical dashed lines denote the boundary between the epochs of inflation, radiation domination and matter domination. The co-moving scales between $k_{\IR}$ and $k_{\UV}$ are the ones that cross the horizon while we are here to watch, both of which are outside the Hubble radius at the end of inflation. We define the environment to be all scalar and tensor modes for which $k>k_\UV$. (Figure taken from \cite{Burgess:2022nwu}.)
   }
\label{fig:defsystem}
\end{figure}

For decoherence calculations we divide the problem into a `system' (whose properties we measure) and an `environment' (whose properties we do not). In cosmology there is some freedom of choice in how to make this split. Fig.~\ref{fig:defsystem} provides a cartoon of scales within inflationary cosmology that helps understand the choice we make here (following \cite{Burgess:2022nwu, Burgess:2025dwm}).

At present only scalar fluctuations have been observed although tensor modes are also actively being sought. Only a particular range of wavelengths that were once super-Hubble are accessible to us at present, though more and more become available the longer we wait. We define the environment to be all scalar and tensor modes whose comoving wavelengths are large enough that they became sub-Hubble in the late universe too early to play a role in observed primordial fluctuations: $k_{\rm env} > k_\UV$.

In practice we do not measure very long-wavelength modes and have not yet observed tensor modes at any cosmologically interesting scale, so it is tempting to also include these into the environment. We do not do so here for two reasons. First, we do not integrate out extremely long-wavelength modes -- those with $k < k_\IR$ in the figure -- since there are expected to modify the effective background cosmology rather than affect the decoherence rate we ultimately compute.\footnote{This is an assumption that should be checked, though if decoherence were to arise from these modes it presumably would add to -- and so increase -- the decoherence rate we find here, which we regard in any case as a lower bound on the total rate in an realistic framework due to the existence of other particles interactions that can also add decoherence.}

A second reason to focus only on a short-wavelength mode is a technical one: general arguments ensure in this case that the leading evolution of system modes relevant to decoherence is purely gaussian, in a way that goes beyond the naive use of perturbation theory. The reason for this is described in detail in   \cite{Burgess:2022nwu, Burgess:2025dwm}, but is also summarized briefly in \S\ref{App:GaussianProperty} of Appendix \ref{sec:OpenEFT}. 

We therefore write the position-space field 
\be\label{vdecomp}
  \zeta(\eta,\bmx) = \zeta^\sys(\eta,\bmx) + \zeta^\env(\eta,\bmx)
\ee
with 
\begin{equation}
\label{vclassicalsplit}
\zeta^{\sys}(\eta,\bm{x})  : = 
 \frac{1}{\sqrt{V}}\sum_{k<k_{UV}}\; \;
\zeta_{\bm{k}}(\eta) e^{i \bm{k}\cdot\bm{x}} \,,
\end{equation}
and similarly for the tensor field $\gamma_{ij}(\eta,\bmx) = \gamma^\sys_{ij}(\eta,\bmx) + \gamma^\env_{ij}(\eta,\bmx)$. Under this decomposition the free Hamiltonian splits into a piece that evolves system and environment separately,
\begin{equation}
  {\scrH}_0(\eta) = {\scrH}_{\sys}(\eta) \otimes {\mathcal{I}}_{\env}
  + {\mathcal{I}}_{\sys} \otimes {\scrH}_{\env}(\eta)
\end{equation}
with $\scrH_\sys$ and $\scrH_\env$ both given by \pref{freeHclassical} with the momentum range appropriately restricted. The initial state -- the Bunch-Davies vacuum -- also conveniently factorizes with this choice of system/environment split, since $|\Omega \rangle = |0 \rangle_\sys \otimes |0\rangle_\env$, where
\begin{equation}
\label{vacuumsplitting}
| 0 \rangle_{\sys} := \bigotimes_{k<k_{\UV} } | 0_{\bm{k}} \rangle
\quad  \mathrm{and} \quad | 0 \rangle_{\env} :=
\bigotimes_{ k>k_{\UV} } | 0_{\bm{k}} \rangle 
\end{equation}
with $| 0_{\bm{k}} \rangle$ defined by ${c}_{\bm{k}}(\eta_{\mathrm{in}}) | 0_{\bm{k}} \rangle = 0$ for all $\bm{k}$.

The plan is to predict the evolution of the system's reduced density matrix, ${\varrho}(\eta)$, obtained by tracing out the environmental degrees of freedom: 
\begin{equation} \label{reducedSpic}
{\varrho}(\eta) := \Trenv \Bigl[ {\rho}(\eta) \Bigr] \,,
\end{equation}
where $\rho$ is the density matrix for the full system. Knowledge of $\varrho(\eta)$ suffices for computing the time evolution of observables that do not depend on environmental degrees of freedom (and for computing how the system decoheres). The evolution is computed starting from the Bunch-Davies initial  state, for which each mode is initially uncorrelated in the remote past:
\begin{equation} \label{rhofactor}
{\varrho}(\eta_{\rm in}) = \bigotimes_{k < k_{\UV} }
{\varrho}_{\bm{k}}(\eta_{\rm in})  \,,
\end{equation}
where $\varrho_\bmk(\eta_\in) = | 0_\bmk \rangle \, \langle 0_\bmk |$ and we take $\eta_\in \to - \infty$. The gaussian character of cubic interactions -- a key feature when the environment is restricted to short-wavelengths (see \S\ref{App:GaussianProperty}) -- is that the product structure seen in \pref{rhofactor} remains preserved (at leading order) for all times: $\varrho(\eta) = \otimes_\bmk \varrho_\bmk(\eta)$, so decoherence can be computed mode-by-mode.

This state does not evolve at all unless cubic and higher interactions are included. Evolution including these interactions is computed perturbatively, working within the interaction picture. Within the full theory this implies states evolve according to
\begin{equation} \label{INTpicVN}
  \frac{\partial {\rho}}{\partial \eta}
  = - i \Bigl[ {\scrH}_{\mathrm{int}}(\eta) , {\rho}(\eta) \Bigr] \,,
\end{equation}
where in practice $\scrH_{\rm int}$ is built from the interactions cubic and higher in the fluctuations.  

In general the evolution of the system's reduced density matrix can be very complicated in the presence of interactions with the environment, in general being nonlocal in time. A very general expression for this evolution can be derived in perturbation theory, called the Nakajima-Zwanzig form \cite{Nakajima:1958pnl, Zwanzig:1960gvu} (see \S\ref{app:sec:SysEnv} of Appendix \ref{sec:OpenEFT} for a very brief summary of the result found in \cite{Burgess:2025dwm} for cosmological applications). Long story short: for super-Hubble modes during inflation it happens that the complicated evolution simplifies, leaving an  approximately time-local description of late-time super-Hubble evolution (see \S\ref{app:sec:markstrat} for a summary).  

\section{Scalar Decoherence without Slow-roll Suppression}
\label{sec:NoSloRollSuppression}

This section uses the formalism of the previous section to identify a source of decoherence for scalar modes that is bigger than any of the partial calculations proposed to date \cite{Nelson:2016kjm, Burgess:2022nwu, Burgess:2025dwm}. The contribution we find here is larger than those considered previously for scalar modes because it is {\it not} slow-roll suppressed (and so is similar in size as the decoherence of tensor modes).

\subsection{Dominant decohering interactions}

As argued in \cite{Burgess:2022nwu, Burgess:2025dwm} the leading contribution to decoherence (in powers of $H/\Mp$) arises at second-order in $\scrH_{\rm int}$ and so is dominated by interactions coming from the piece of the action that is cubic in fluctuations. A complete list of interactions that are cubic in $\zeta$ and $\gamma_{ij}$ are given in \cite{Maldacena:2002vr}. 

Decoherence is driven only by those terms in the Hamiltonian that connect the system and environment and these are obtained by inserting the decomposition (\ref{vdecomp}) (and its tensor counterpart) into any particular cubic interaction of interest. Denoting the fields collectively by $f = \{\zeta, \gamma_{ij} \}$ and writing $f = f_\sys + f_\env$ gives interactions of the schematic form $f_{\sys}^3$, $f_{\sys}^2 f_{\env}$, $f_{\sys} f_{\env}^2$ and $f_{\env}^3$. Of these, only the cross terms $f_{\sys}^2 f_{\env}$ and $f_{\sys} f_{\env}^2$ couple the system to the environment and of these momentum conservation suppresses the $f_{\sys}^2 f_{\env}$ interactions because it is impossible to sum two small momenta to get a large one. We seek only cubic interactions where two fields are regarded as environmental and one as a system field. 

Now comes the part where our current story deviates from earlier calculations. Ref.~\cite{Burgess:2025dwm} computes the contribution to the decoherence of scalar fluctuations using all of the cubic $\zeta^3$ interactions, so for which both the system and environment fields are scalars. Ref.~\cite{Burgess:2022nwu} computes a particular contribution to the decoherence of scalar modes due to a $\zeta \gamma^2$ interaction where both of the tensor fields are environmental and the scalar is a system field. Both of these contributions have the same size and turn out to be slow-roll suppressed, leading to a decoherence rate proportional to $\slrl (H/\Mp)^2$. 

Ref.~\cite{Burgess:2022nwu} also computes the contribution of a $\zeta^2 \gamma$ interaction to the decoherence of tensor fluctuations due to a scalar environment. Like the $\gamma^3$ interactions this interaction is not slow-roll suppressed, leading to the expectation that the decoherence rate for tensor modes should be order $(H/\Mp)^2$ (without slow-roll suppression). The numerical coefficient coming from the $\gamma^3$ interactions remains a work in progress. In what follows we compute the contribution of the $\zeta^2 \gamma$ interaction to the decoherence of {\it scalar} fluctuations, showing that scalar fluctuations also decohere with a rate of order $(H/\Mp)^2$ that is not slow-roll suppressed, but only because of their unsupressed cubic couplings to tensor modes. 

The unique cubic interaction that contributes to scalar decoherence {\it without} slow-roll suppression is
\be \label{SintUnsuppressed}
S_{\rm int} = \int \exd\eta \, \exd^3 x\ \slrl \Mp^2 a^2  \gamma^{i j} \, \partial_i\zeta \, \partial_j  \zeta= \tfrac12  \int \exd\eta \, \exd^3 x\   \gamma^{i j} \, \partial_iv \, \partial_j  v \,,
\ee
which we evaluate with $\gamma_{i j}$ and one of the $\partial \zeta$'s regarded as environment fields.
The lack of slow-roll suppression is most transparent when the interactions are expressed in terms of the variable $v$ (because their correlation functions are $\slrl$-independent). 

\subsection{Leading order Lindblad-like Coefficients}
\label{sec:leadLiind}

We next compute the contribution to scalar decoherence using the interaction Hamiltonian density implied by \pref{SintUnsuppressed}:
\be  \label{BigHam}
\scrH_{\rm int}  = G(\eta) \, \partial_i \zeta^{\sys} \left(\gamma^{i p}_{\env} \partial_p \zeta^{\env}\right) 
\ee
with $G(\eta)=-2 \slrl \Mp^2 a^2(\eta)$. Other ways of assigning the fields to system and environment do not contribute to leading order, for the reasons summarized above. 

This interaction falls into the category of interactions \pref{app:HintRdef0} for which the discussion in Appendix \ref{App:sec:OpenEFT} applies. Repeating the steps given in the appendix leads to the Nakajima-Zwanzig equation \pref{app:NZ}, which for super-Hubble modes is approximated by a time-local Lindblad-like equation \pref{app:eq:lindbladlikeform}. For the interaction \pref{BigHam} this leads to the following Lindblad-like equation for the evolution of the reduced density matrix:
\be
\label{eq:Lindblad}
\partial_{\eta} \varrho(\eta)=-\int_{\eta_{\rm in}}^{\eta} \exd\eta^{\prime} \int \exd^3 x \, \exd^3 x^{\prime}\ \left\langle\Bigl[\scrH_{\rm int}(\bmx ,\eta), \scrH_{\rm int}(\bmx^{\prime},\eta^{\prime}) \varrho(\eta^{\prime})\Bigr] \right\rangle_\env+{\rm c.c.}
\ee
where the expectation value is taken over environmental variables only. 

Explicitly isolating the system fields this becomes
\be \label{OurEvovarrho}
\partial_{\eta} \varrho(\eta)=-\sum_{\bmk} \int_{\eta_{\rm in}}^{\eta} \exd\eta^{\prime}\,  G(\eta)\, G(\eta^{\prime}) \,\Bigl[\zeta^\sys_{\bmk }(\eta), \zeta^\sys_{-\bmk }(\eta^{\prime})\varrho(\eta^{\prime})\Bigr]\mathcal{D}_{\bmk }(\eta,\eta^{\prime})+{\rm c.c.},
\ee
where the environmental correlator to be computed is 
\be 
\label{eq:corrfun}
\mathcal{D}_{\bmk}(\eta,\eta^{\prime})  := k_i k_j  
  \int \exd^3x \, \exd^3 x' \, e^{-i \bmk \cdot (\bmx + \bmx')} \left\langle \left[ \gamma^{i a}_{\env}(\bmx,\eta) \partial_a \zeta^{\env}(\bmx,\eta)\right]\left[ \gamma^{j b}_{\env}(\bmx^{\prime},\eta^{\prime}) \partial_b\zeta^{\env}(\bmx^{\prime},\eta^{\prime})\right]\right\rangle_\env \,.
\ee 
Because \pref{BigHam} is linear in the system field the evolution \pref{OurEvovarrho} to which it gives rise is automatically quadratic in these fields, and so preserves the gaussian property of the initial state (for the reasons described in more detail in \S\ref{App:GaussianProperty}). 

The expectation value in \pref{eq:corrfun} can be computed in the interaction picture using Wick's theorem, and the time-evolution of $\zeta_{\bmk}$ can be made explicit in the interaction picture to relate $\zeta_{\bmk}(\eta^{\prime})$ to $\zeta_{\bmk}(\eta)$ (and the same for its conjugate momentum, though this turns out not to contribute), along the lines described in \S\ref{app:sec:markstrat}. The difference between the density matrix $\varrho(\eta^{\prime})$ and $\varrho(\eta)$ can also be neglected to leading order (again see \S\ref{app:sec:markstrat}). Finally, having the right-hand-side of \pref{OurEvovarrho} be quadratic in system fields also allows the density matrix, $\varrho_\bmk$, for each mode number $\bmk$ to be evolved separately: $\partial_\eta \varrho = \sum_\bmk \partial_\eta \varrho_\bmk$, with
\be
\label{eq:markovianlind}
\partial_{\eta} \varrho_\bmk (\eta)=-  \Bigl[\zeta_{\bmk}(\eta), \zeta_{-\bmk}(\eta)\varrho_\bmk(\eta)\Bigr] G(\eta) \int_{\eta_{\rm in}}^{\eta} \exd\eta^{\prime}\ G(\eta') \, k_i k_j \, \mathcal{C}_{\bmk}^{i j}(\eta,\eta^{\prime})+{\rm c.c.}
\ee
where
\bea
\label{eq:correlation}
\mathcal{C}_{\bmk}^{i j}(\eta,\eta^{\prime})&=&\tfrac{1}{2}  \mathcal{W}_k(\eta,\eta^{\prime}) \int   \frac{\exd^3 p}{(2\pi)^3}\ \frac{\exd^3 q}{(2\pi)^3}\ \left[(2\pi)^3\delta^{(3)}\left(\bmp+\bmq-\bmk\right)\right] \Pi^{iajb}(\hat\bmq)  p_a p_b \nn\\
&&\qquad\qquad\qquad\qquad\qquad\qquad\qquad\qquad  \times \left[\hat u_{\bmp}(\eta) \, w_{\bmp}^*(\eta^{\prime}) \right] \left[ w_{\bmq}(\eta) \, \hat u_{\bmq}^*(\eta^{\prime}) \right] \,.
\eea
Here $\mathcal{W}_k(\eta,\eta^{\prime})$ is the conversion kernel relating $\zeta_{\bmk}(\eta^{\prime})$ to $\zeta_{\bmk}(\eta)$ that is given explicitly in \pref{app:eq:etaprimeToeta2} while $\hat u$ and $w$ are the mode functions for $\zeta$ and $\gamma_{ij}$ that are given explicitly for the de Sitter limit by \pref{ModeExpressions}. 

The tensor operator $\Pi^{iajb}(\hat\bmq)$ appearing in \pref{eq:correlation} is defined by the polarization sum
\be
\label{eq:projectiondef}
\sum_{\sigma} \epsilon_{i a}(\hat{\bmq},\sigma) \epsilon^*_{j b}(\hat{\bmq},\sigma) =: \tfrac{1}{2} \, \Pi_{iajb}(\hat{\bmq}) \,,
\ee
which evaluates to
\be 
\label{eq:projection}
 \Pi^{iajb}(\hat{\bmq})  =  \Pi^{ij}(\hat{\bmq}) \Pi^{ab}(\hat{\bmq})  + \Pi^{ib}(\hat{\bmq}) \Pi^{ja}(\hat{\bmq}) - \Pi^{ia}(\hat{\bmq}) \Pi^{jb}(\hat{\bmq}) \,,
 \ee 
 where
\be 
\label{eq:projection2}
\Pi_{ab}(\hat{\bmq}) :=  \delta_{a b} - \hat{q}_{a} \hat{q}_{b} =  \delta_{a b} - \frac{q_a q_b}{\bmq^2} \,,
\ee
projects onto the space perpendicular to $\bmq$. Notice that these definitions imply the polarization-dependent quantity appearing in \pref{eq:correlation} can be simplified using 
\be
  \Pi^{iajb}(\hat{\bmq}) p_a p_b = \Pi^{ij}(\hat{\bmq}) \, \Pi^{ab}(\hat{\bmq}) \, p_a p_b = \Pi^{ij}(\hat{\bmq}) \, \Bigl[ \bmp \cdot \bmp - (\bmp \cdot \hat{\bmq})^2 \Bigr] \,.
\ee

The resulting integrals are evaluated in detail in Appendix \ref{sec:Integrals}, leading to the time-local evolution described by \pref{app:eq:lindbladlikeform} at leading order for the reduced density matrix for super-Hubble modes. This allows \pref{eq:markovianlind} to be written in the Lindblad-like form\footnote{This is Lindblad-{\it like} in the sense that it has the familiar operator structure from \cite{Lindblad:1975ef,Gorini:1976cm}, however the matrix $ h_\bmk$ is not restricted to be constant nor positive semi-definite. It is well known that all local master equations (including the generally non-Markovian one we derive here) can be expressed in this form \cite{Hall:2014oiy}.}
\begin{equation}
\label{eq:lindbladlikeform}
\partial_{\eta} \varrho_{\bmk} = -i\Bigl[ \scrH_{\rm eff}(\eta), \varrho_{\bmk}(\eta)\Bigr] +\sum_{r,s=1}^2 h^{s r}_\bmk\left[\mathcal{O}_{\bmk,s}\varrho_{\bmk}\mathcal{O}^{\dag}_{\bmk,r}-\frac{1}{2}\left\{\mathcal{O}^{\dag}_{\bmk,r}\mathcal{O}_{\bmk,s},\varrho_{\bmk}\right\}\right] \,,
\end{equation}
with
\be \label{OiandZdefs}
 \mathcal{O}_{1} := \frac{ \zeta}{Z} = z \sqrt{k}\; v \qquad \hbox{and} \qquad
 \mathcal{O}_{2} := Z \mfp \qquad \hbox{with} \qquad
 Z^2 := \frac{H^2}{2\slrl k^3 \Mp^2}  = \frac{1}{\mfz_s^2 z^2 k} \,,
\ee 
where on de Sitter space $\mfz_s = -\sqrt{2\slrl } \Mp/(H \eta) = \sqrt{2\slrl} k \Mp/(Hz)$ is as defined in \pref{eq:defzeta} and $0 < z < \infty$ is defined by $z = -k\eta = k/(aH)$. Since the Liouville-like $\scrH_{\rm eff}$ term does not contribute to decoherence it is largely ignored in what follows. 

The matrix of couplings $h^{sr}_\bmk$ appearing in \pref{eq:lindbladlikeform} is given at leading order in slow roll by 
\begin{equation}
\label{eq:decohmatrixhnosr}
\left(\begin{array}{cc}h_\bmk^{11}  & h_\bmk^{12}  \\ h_\bmk^{21}  & h_\bmk^{22}  \end{array}\right)  = k\left(\frac{H}{\Mp}\right)^2\left(\begin{array}{cc}{ \rm Re}[\widetilde{\cJ}^{\zeta\zeta}_{k0}] & \widetilde{\cJ}^{\zeta\mfp *}_{k0}  + \slrl \widetilde{\cJ}^{\mfp\zeta}_{k0} \\ \widetilde{\cJ}^{\zeta\mfp}_{k0} + \slrl \widetilde{\cJ}^{\mfp\zeta *}_{k0} & 4 \slrl {\rm Re}[\widetilde{\cJ}^{\mfp\mfp}_{k0}] \end{array}\right) \,,
\end{equation}
which reduces at leading order in slow roll to
\begin{equation}
\label{eq:decohmatrixh}
\left(\begin{array}{cc}h_\bmk^{11}  & h_\bmk^{12}  \\ h_\bmk^{21}  & h_\bmk^{22}  \end{array}\right)  = k\left(\frac{H}{\Mp}\right)^2\left(\begin{array}{cc}{ \rm Re}[\widetilde{\cJ}^{\zeta\zeta}_{k0}] & \widetilde{\cJ}^{\zeta\mfp *}_{k0}  \\ \widetilde{\cJ}^{\zeta\mfp}_{k0} & 0 \end{array}\right) \,.
\end{equation}
Here the $\widetilde{\cJ}^{\mfa \mfb}_{kn}$ are integrals obtained when evaluating the environmental correlators of $\cO_\mfa$ and $\cO_\mfb$, as defined by eqs.~\pref{app:eq:pvkernels} through \pref{app:curlyJdimless}. 

Direct calculation -- see Appendix \ref{sec:Integrals} for details -- gives the following leading asymptotic forms for these integrals in the $z \ll 1$ ({\it i.e.}~super-Hubble) limit during inflation
\be \label{Jzz}
\hbox{Re } \widetilde \cJ^{\zeta \zeta} \simeq \frac{32}{45\kappa^3 z^3} \left(\frac{2}{\pi^2} \right)\Bigl[1 +\cO(z) \Bigr] \,, \quad
\hbox{Im } \widetilde \cJ^{\zeta \zeta} \simeq - \frac{8}{5\kappa^2 z^2} \left(\frac{2}{\pi^2} \right)\Bigl[1 +\cO(z) \Bigr] \,,
\ee
and 
\be \label{Jzp}
\hbox{Re } \widetilde \cJ^{\zeta \mfp} \simeq  \frac{2}{25\kappa^5 z^2} \left(\frac{2}{\pi^2} \right) \Bigl[1 +\cO(z) \Bigr] \,, \quad
\hbox{Im } \widetilde \cJ^{\zeta \mfp} \simeq - \frac{2z}{15\kappa^2} \left(\frac{2}{\pi^2} \right) \Bigl[1 +\cO(z) \Bigr]\,,
\ee
where $\kappa = \kUV\slash k \gg 1$ and $\kappa z = - k_\UV \eta =: z_\UV$. For these expressions subleading terms involve at least one additional positive power of $z$. Notice that these expressions ensure that for small $z$ the dominant eigenvalue of the coupling matrix $h^{\mfa\mfb}_\bmk$ is positive.

The other integral kernels $\widetilde \cJ^{\mfp \zeta}$ and $\widetilde \cJ^{\mfp\mfp}$ appearing in \pref{eq:decohmatrixhnosr} only do so at subdominant order in $\slrl$, and so can be ignored at leading order. We have nonetheless computed their leading contributions to verify their subdominance, finding 
\be \label{Jpz}
\hbox{Re } \widetilde \cJ^{\mfp \zeta} \simeq \frac{c_{pzr}}{\kappa^3 } z^0 +\cdots \,, \quad
\hbox{Im } \widetilde \cJ^{\mfp \zeta} \simeq \frac{c_{pzi}}{\kappa^2 } z+\cdots \,,
\ee
and
\be \label{Jpp}
\hbox{Re } \widetilde \cJ^{\mfp \mfp} \simeq \frac{\pi^2 z}{128} +\cdots \,, \quad
\hbox{Im } \widetilde \cJ^{\mfp \mfp} \simeq \frac{\pi z}{64} \log(2 \kappa) +\cdots \,.
\ee
We do not quote explicitly the values for the coefficients $c_{pzr}$ and $c_{pzi}$ because these are divergent. This need not be a problem for two reasons. First, these divergent coefficients only appear in slow-roll suppressed contributions and so do not matter at all for the leading result in slow roll that is our current focus. But they need not be problems even once the terms subdominant in slow roll are sought because they compete with -- and can cancel -- the divergent contributions of other interactions not described above. (This is why divergences are difficult to treat systematically until all interactions relevant to a given order are included in the calculation.) A calculation of these subdominant terms is underway and we report elsewhere on their implications for subdominant contributions to super-Hubble evolution \cite{ToAppear}.  

\subsection{Implications for the purity}
\label{sec:PurityCalc}

The final step is to compute the implications of \pref{eq:lindbladlikeform} for the evolution of the purity of the system's state, defined by \pref{PurityDef}. This discussion closely parallels the discussion in \cite{Burgess:2022nwu, Burgess:2025dwm}. 

The evolution simplifies given the uncorrelated initial state \ref{rhofactor} and the gaussian nature of the evolution -- which allows each mode to be evolved separately as in \pref{eq:lindbladlikeform} -- leading to
\be \label{PurityProd}
    \gamma = \prod_\bmk  \gamma_\bmk \quad \hbox{where} \quad
   \gamma_\bmk =  \hbox{Tr}_{\rm sys} \left(\varrho_\bmk^2 \right)
\ee
and so
\be \label{PurityRateSum}
  \frac{ \partial_\eta\gamma}{\gamma} = \sum_\bmk \frac{\partial_\eta \gamma_\bmk }{ \gamma_\bmk }  
\ee
with
\be \label{LindbladPurity}
  \partial_\eta \gamma_\bmk := 2 \hbox{Tr}_{\rm sys} \left( \varrho_\bmk \, \partial_\eta \varrho_\bmk \right) 
  \simeq 2 \sum_{r,s=1}^2 h^{s r}_\bmk\left[ \hbox{Tr}_{\rm sys} \Bigl(  \varrho_{\bmk} \, \mathcal{O}^{\dag}_{\bmk,r}\varrho_{\bmk} \,  \mathcal{O}_{\bmk,s} \Bigr) - \hbox{Tr}_{\rm sys} \Bigl( \varrho_\bmk^2 \, \mathcal{O}^{\dag}_{\bmk,r}\mathcal{O}_{\bmk,s} \Bigr) \right]\,.
\ee

\subsubsection{Perturbative evolution}

We start by evaluating the right-hand side of \pref{LindbladPurity} within straight-up perturbation theory.  In this case it suffices on the right-hand side to use the lowest-order expression: the initial (pure) value for the system state: $\varrho_{\bmk\,{\rm in}} = | \Omega \rangle \, \langle  \Omega |$. Choosing this to be the Bunch-Davies state, $|\Omega \rangle$, and the initial time to be $\eta_{\rm in} \to - \infty$ provides several simplifications. First, it implies $\gamma_{\bmk\,{\rm in}} = 1$ since the initial state is pure. Second, it implies
\be
   \hbox{Tr}_{\rm sys} \Bigl(  \varrho_{\bmk} \, \mathcal{O}^{\dag}_{\bmk,r}\varrho_{\bmk} \,  \mathcal{O}_{\bmk,s} \Bigr) = \langle  \, \mathcal{O}^{\dag}_{\bmk,r}  \rangle \, \langle   \mathcal{O}_{\bmk,s} \rangle \,,
\ee
where $\langle \cdots \rangle := \langle \Omega | \cdots | \Omega \rangle$. This expectation value vanishes in the Bunch-Davies state for the operators appearing in \pref{LindbladPurity}. Third, it allows us to use $\varrho_\bmk^2 = \varrho_\bmk$ in the remaining contribution, leading to 
\be  \label{LindbladPurity2}
  \partial_\eta \gamma_\bmk   \simeq  - 2 \sum_{r,s=1}^2 h^{s r}_\bmk  \; \Bigl\langle  \mathcal{O}^{\dag}_{\bmk,r}\mathcal{O}_{\bmk,s} \Bigr\rangle  \,.
\ee

Using the operators $\cO_r$ from \pref{OiandZdefs} and using \pref{eq:decohmatrixh} for the coefficients $h^{s r}$ gives the evolution of the purity in terms of the dimensionless integrals $\widetilde{\cJ}^{rs}_{k0}$:
\bea \label{LindbladPurity3}
\partial_z \gamma_\bmk &\simeq& 
\frac{2H^2}{M_p^2}  \bigg[ \mathrm{Re}[\widetilde{\mathcal{J}}^{\zeta \zeta}_{k0}] \,  Z^{-2} \Bigl\langle \zeta_{\bmk} \zeta_{-\bmk} \Bigr\rangle +  \mathrm{Re}[\widetilde{\mathcal{J}}^{\zeta \mfp}_{k0}]  \,    \left( \Bigl\langle \mfp_{-\bmk}\zeta_{\bmk} \Bigr\rangle + \Bigl\langle   \zeta_{-\bmk} \mfp_{\bmk} \Bigr\rangle \right)  \\
&& \qquad\qquad\qquad \qquad\qquad\qquad \qquad\qquad\qquad + i \, \mathrm{Im}[ \widetilde{\mathcal{J}}^{\zeta\mfp}_{k0}  ]  \left( \Bigl\langle   \zeta_{-\bmk} \mfp_{\bmk} \Bigr\rangle - \Bigl\langle \mfp_{-\bmk}\zeta_{\bmk} \Bigr\rangle \right)  
\bigg] \,, \nn
\eea 
where all operators are evaluated at $\eta$ (or $z = - k \eta$) and $Z^{-2} =2 \epsilon_1 k^3 M_p^2/H^2$, as in \pref{OiandZdefs}. 

The leading system correlators can be evaluated using the Bunch-Davies mode functions given in \pref{ModeExpressions}, 
\be
   \frac{2}{Z^2}\Bigl\langle  \zeta_{-\bmk} \zeta_{\bmk} \Bigr\rangle = \frac{2}{Z^2}| \hat u_k |^2  
   = (1+z^2) \,, \qquad 2Z^2\Bigl\langle  \mfp_{-\bmk} \mfp_{\bmk} \Bigr\rangle = 2Z^2\mfz_s^4 | \hat u'_k |^2  
   = \frac{1}{z^2} \,,   
\ee
and
\be
  \Bigl\langle \zeta_{-\bmk} \mfp_{\bmk} \Bigr\rangle = \mfz_s^2 \hat u_k \, (\hat u'_k)^* = -\frac{1}{2z}\,(1+iz) = \Bigl\langle  \mfp_{-\bmk} \zeta_{\bmk} \Bigr\rangle^* \quad,  
\ee
and using these in \pref{LindbladPurity3} finally gives
\be \label{LindbladPurity4}
\partial_z \gamma_\bmk  \simeq   
\frac{2H^2}{M_p^2}  \left[ \frac{1}{2} (1+z^2) \mathrm{Re}[\widetilde{\mathcal{J}}^{\zeta \zeta}_{k0}]  - \frac{1}{z} \, \mathrm{Re}[\widetilde{\mathcal{J}}^{\zeta \mfp}_{k0}]  + \mathrm{Im}[\widetilde{\mathcal{J}}^{\zeta\mfp}_{k0}  ]  
  \right] \,. 
\ee

Inserting expressions \pref{Jzz} and \pref{Jzp} for the $\widetilde \cJ^{rs}_{k0}$ coefficients in the small $z = -k\eta$ ({\it i.e.}~super-Hubble) limit -- into \pref{LindbladPurity4} gives
\be
\partial_z \gamma_\bmk  \simeq 
\frac{4H^2}{\pi^2 M_p^2}  \left[ \frac{1}{2} (1+z^2)  \frac{32}{45\kappa^3 z^3} - \frac{1}{z} \left( \frac{2}{25\kappa^5 z^2}\right)  -  \frac{2z}{15\kappa^2}  
 \right] \,,
\ee
showing that the terms involving the real part of $\widetilde \cJ^{\zeta\zeta}_{k0}$ and $\widetilde \cJ^{\zeta\mfp}_{k0}$ dominate, giving a contribution that scales as $z^{-3}$ for small $z$, with all other terms down compared to this by additional positive powers of $z$. 

Using  $z_\UV := \kappa z = - k_\UV \eta$ we find the leading result 
\be   \label{LindbladPurity5}
\partial_z \gamma_\bmk   \simeq  \frac{8H^2}{5\pi^2z_\UV^3 \Mp^2} \Bigl(\frac{8 }{9 }-\frac{1}{ 5\kappa^2  }\Bigr) \Bigl[ 1 + \cO(z) \Bigr]   
 \,,
\ee  
in the super-Hubble regime $0 < z = - k\eta = {k}/{aH} \ll 1$. Notice that the $\widetilde \cJ^{\zeta\zeta}_{k0}$ term dominates for large $\kappa$ ({\it i.e.}~when $k_\UV \gg k$). The perturbative evolution of purity evaluated for super-Hubble modes during inflation grows with time (since $z_\UV \propto \eta$) at late times and is strictly negative, corresponding to a monotonically decreasing purity.  

The result \pref{LindbladPurity5} may be integrated term by term  to give the perturbative expression for the purity of a state in the super-Hubble limit:
\be    \label{LindbladPurity5int}
 \gamma_\bmk(z)    
 \simeq  1  - \frac{32H^2}{45\pi^2\kappa^3 z^2\Mp^2}  + \cdots =   1  - \frac{32}{45\pi^2} \left( \frac{H^2}{\Mp^2} \right)   \frac{k}{k_\UV} \left( \frac{a H}{k_\UV}\right)^2  + \cdots \,,
\ee  
where we specialize to the initial condition $\gamma_\bmk \to 1$ as $z_{\rm in} \to \infty$, as appropriate if the state approaches the Bunch-Davies vacuum in the remote past. Notice that this result is ultraviolet finite, as must be the leading order contribution to the purity. It applies for all $k \ll k_\UV$ provided the time is late enough that all such modes have left the Hubble scale, with linearity in $k$ showing that decoherence is larger for modes closer in wavelength to the wavelength of the environment. Finally, $\gamma_\bmk - 1$ grows quadratically with $a \propto  e^{Ht}$, and so the prediction \pref{LindbladPurity5int} eventually breaks down as the correction to $\gamma_\bmk$ moves beyond the domain of validity of perturbative methods.

\subsubsection{Secular growth}
\label{sec:Observables}

From the point of view of late-time predictions the key feature of the results \pref{LindbladPurity5} and \pref{LindbladPurity5int} is their divergence as $z \to 0$, since this reveals the late-time secular growth \cite{Fleming:2011irq,Kaplanek:2019dqu,Burgess:2022rdo,Lampert:2025cus} that often plagues near-de Sitter cosmological calculations \cite{Tsamis:2005hd, Burgess:2009bs}. This growth prevents the straight-up perturbative result from being trusted once $z$ is small enough that
\be
  z^2 \sim \lambda := \frac{H^2}{4\pi^2\Mp^2\kappa^3} \,,
\ee
where we use $\lambda$ to denote the core underlying perturbative expansion parameter that justifies the semiclassical approximation \cite{Burgess:2003jk, Burgess:2009ea, Adshead:2017srh}.

The observed size of primordial fluctuations fixes the size $H^2/(8\pi^2 \Mp^2 \slrl) \sim 10^{-10}$ during inflation and the absence of observable primordial tensor fluctuations implies $\slrl \lsim 10^{-2}$ and these together imply $\lambda \lsim 10^{-12}/\kappa^3$ or so. The apparent transition to $\gamma_\bmk < 0$ seen in \pref{LindbladPurity5int} is due to the late-time breakdown of perturbation theory that occurs once $z^2 \sim  \lambda$, which for $\lambda \sim 10^{-12}/\kappa^3$ implies $z \sim 10^{-6}/\kappa^{3/2}$. Since $z = - k \eta = k/(aH)$ and $a \propto e^{H t}$ during inflation we see that $z = 10^{-6}/\kappa^{3/2}$ occurs around $14+\frac32 \log \kappa$ $e$-foldings of expansion after horizon crossing. 

Reliable predictions for the purity at the end of inflation 50 $e$-foldings after horizon exit require a better treatment of the late-time regime, and this is where having the evolution be gaussian and time-local at late times helps, by providing a way to resum the perturbative result to give a reliable prediction at very late times. The resummation (described below) puts a similar result in \cite{Burgess:2022nwu}) on a firmer foundation and gives results that work to all orders in $\lambda/z^2$.  Importantly, by the end of inflation $\gamma_\bmk$ is extremely small and so the prediction of efficient decoherence from then on relies on the absence of a mechanism for recoherence (see however \cite{Colas:2022kfu,Colas:2024xjy,Cielo:2025ibc}).

\section{Late-time Gaussian \& Time Local Behaviour}
\label{ssec:TransportCoeffs}

The secular growth (and the resulting need for resummation) of the purity at late times during inflation is a special case of the more general phenomena of secular growth (and resummation) in inflationary correlation functions at late times. 

Following \cite{Mulryne:2009kh,Mulryne:2010rp,Mulryne:2013uka,Burgess:2022nwu,Colas:2022hlq,Werth:2023pfl} we therefore pause in this section to broaden our discussion to more general equal-time correlators, of the form
\begin{equation}
\Sigma_{ij} = \mathrm{Tr}\big[ \tfrac{1}{2} \{ O_{i}, O_{j} \} \varrho \big] \,.
\end{equation}
Our interest is specifically in the cases where $O_i$ represent the hermitian fields 
\be
    \tilde \zeta_\bmk = \frac{1}{\sqrt 2}(\zeta^\ssR_\bmk + \zeta^\ssI_\bmk) \qquad
    \tilde \mfp_\bmk = \frac{1}{\sqrt2}(\mfp^\ssR_\bmk + \mfp^\ssI_\bmk) 
\ee
where superscripts $R$ and $I$ denote the real and imaginary parts of the fields, normalized such that $\zeta_\bmk = \frac{1}{\sqrt 2}(\zeta^\ssR_\bmk + i \zeta^\ssI_\bmk)$ and similarly for $\mfp_\bmk$. $\zeta^\ssR_\bmk$ and $\zeta^\ssI_\bmk$ can be obtained from $\tilde \zeta_\bmk$ by taking its symmetric and antisymmetric parts under $\bmk \to - \bmk$ (see \S\ref{App:GaussianProperty} for more details). 

For these operators the desired correlators form a two-by-two matrix
\be  \label{SigmaGaussian}
\boldsymbol{\Sigma}_\bmk = \left[ \begin{matrix} \Sigma_{11} & \Sigma_{12} \\  \Sigma_{12} & \Sigma_{22} \end{matrix} \right] =  \left[ \begin{matrix} \mathrm{Tr}\big[ \tilde \zeta_\bmk \tilde \zeta_\bmk \varrho_\bmk \big] & \tfrac12\mathrm{Tr}\big[ \{ \tilde \zeta_\bmk, \tilde \mfp_\bmk \}  \varrho_\bmk \big] \\  \tfrac12\mathrm{Tr}\big[ \{ \tilde \zeta_\bmk, \tilde \mfp_\bmk \}  \varrho_\bmk \big]  & \mathrm{Tr}\big[ \tilde \mfp_\bmk\tilde \mfp_\bmk \varrho_\bmk \big] \end{matrix} \right] \,.
\ee
 In the special case of free minimally coupled massless fields in near-de Sitter space prepared in the Bunch-Davies vacuum these correlators evaluate (see \S\ref{ssecApp:GaussianEvo}) to     
\be \label{LeadingBDSigmaij}
\Sigma_{11}(k,\eta)  \to  \frac{H^2 \left(\eta ^2 k^2+1\right)}{4 k^3 M_p^2 \epsilon _1}  \,, \qquad
\Sigma_{12}(k,\eta) \to \frac{1}{2 k \eta}   \quad \hbox{and} \quad
\Sigma_{22}(k,\eta) \to \frac{k \epsilon _1 M_p^2}{\eta ^2 H^2}  \,.
\ee
 
\subsection{Transport equations}
\label{sec:Transport}

Much can be said about the evolution of the equal-time correlators using only that the evolution is time-local and Gaussian. Here time-local means the system's state evolves through the Lindblad-like equation
\bea \label{LindbladAgain}
\partial_{\eta} \varrho_{\bmk} &=& -i\Bigl[ \scrH_{\rm eff}(\eta), \varrho_{\bmk}(\eta)\Bigr] +\sum_{r,s=1}^2 \mathrm{h}^{s r}_\bmk\left[  O _{\bmk,s}\varrho_{\bmk} O^{\dag}_{\bmk,r}-\frac{1}{2}\left\{ O^{\dag}_{\bmk,r} O_{\bmk,s},\varrho_{\bmk}\right\}\right] \nn\\
&=& -i\Bigl[ \scrH_{\rm eff}(\eta), \varrho_{\bmk}(\eta)\Bigr] +\sum_{r,s=1}^2 h^{s r}_\bmk\left[\mathcal{O}_{\bmk,s}\varrho_{\bmk}\mathcal{O}^{\dag}_{\bmk,r}-\frac{1}{2}\left\{\mathcal{O}^{\dag}_{\bmk,r}\mathcal{O}_{\bmk,s},\varrho_{\bmk}\right\}\right]   \,,
\eea
where $O_1 = \tilde \zeta_\bmk$ and $O_2 =\tilde  \mfp_\bmk$ imply
\begin{equation} \label{Kosmatv1}
\left[ \begin{matrix} \mathrm{h}^{11}_\bmk & \mathrm{h}^{12}_\bmk \\ \mathrm{h}^{21}_\bmk & \mathrm{h}^{22}_\bmk \end{matrix} \right]  = \left[ \begin{matrix} 2\mathrm{Re}[\mathcal{J}^{\zeta \zeta}_{k0} ] & \mathcal{J}^{\zeta \mfp \ast}_{k0} + \mathcal{J}^{\mfp \zeta}_{k 0}  \\ \mathcal{J}^{\zeta \mfp}_{k0} + \mathcal{J}^{\mfp \zeta \ast}_{k 0}  & 2 \mathrm{Re}[\mathcal{J}^{\mfp \mfp}_{k0} ] \end{matrix} \right]_{\rm leading} \,,
\end{equation}
while the rescaled variables $\mathcal{O}_{1} := Z^{-1} \tilde\zeta_\bmk$ and $\mathcal{O}_{2} := Z \tilde \mfp_\bmk$ with $Z := \frac{1}{\sqrt{2\slrl k^3}} \left( \frac{H}{\Mp}\right)$ instead give
\begin{equation} \label{Kossmat}
\left[ \begin{matrix} h_{\bmk}^{11} & h_{\bmk}^{12} \\ h_{\bmk}^{21} & h_{\bmk}^{22} \end{matrix} \right] = k\left(\frac{H}{\Mp}\right)^2  \left[ \begin{matrix}  \mathrm{Re}[\widetilde{\mathcal{J}}^{\zeta \zeta}_{k0} ] & \widetilde{\mathcal{J}}^{\zeta \mfp \ast}_{k0}+\slrl \widetilde{\mathcal{J}}^{\mfp \zeta }_{k0}   \\  \widetilde{\mathcal{J}}^{\zeta \mfp}_{k0}+ \slrl \widetilde{\mathcal{J}}^{\mfp \zeta \ast}_{k0}  &  4 \slrl \mathrm{Re}[\widetilde{\mathcal{J}}^{\mfp \mfp}_{k0} ] \end{matrix} \right]_{\rm leading} \,.
\end{equation}
Here the subscript `leading' means to use the leading terms in the small-$z$ expansion -- such as given in \pref{Jzz} through \pref{Jpp} -- since this expansion is required in order for the evolution to be time-local (in the sense that it is well-described by \pref{LindbladAgain}). 

The detailed form of the effective Hamiltonian $\scrH_{\rm eff}$ appearing in these equations is not required for the applications to decoherence discussed here, though in the interaction picture it does {\it not} include the free Hamiltonian (\ref{freeHclassical}) since only interactions -- like those in eq.~\pref{app:HeffCorr} coming from cubic self-interactions, for instance -- are included. Notice that in \pref{Kossmat} we include for generality subdominant terms in the slow-roll expansion since we hope to apply this evolution beyond the leading order regime, though this is not crucial.

\subsubsection{Evolution of $\boldsymbol{\Sigma}$}

An evolution equation for $\boldsymbol{\Sigma}$ is found by directly differentiating the definition \pref{SigmaGaussian} 
\be
   \partial_{\eta} \Sigma_{ij} = \mathrm{Tr}[ \tfrac12 \{\partial_\eta O_{i}, O_j\} \varrho ] + \mathrm{Tr}[ \tfrac12 \{ O_{i} ,  \partial_\eta O_j \} \varrho ] + \mathrm{Tr}[\tfrac12 \{ O_{i} , O_j \}  \partial_\eta \varrho ] \,, 
\ee
and using \pref{LindbladAgain} to evaluate $\partial_\eta \varrho$. Time derivatives of the fields $\tilde \zeta_\bmk$ and $\tilde \mfp_\bmk$ are evaluated using their evolution equation, which in the interaction picture comes purely from the free Hamiltonian $\mathscr{H}_0$ given in (\ref{freeHclassical})), leading to the Klein-Gordon equation $\zeta''_{\bmk} + ({2a'}/{a}) \zeta'_{\bmk} + k^2 \zeta_{\bmk} = 0$, or equivalently
\be
\partial_\eta \zeta_{\bmk} = {\mfp_{\bmk}}/{\mfz_s^2} \qquad \mathrm{and} \qquad \partial_\eta \mfp_\bmk =   - \mfz_s^2 k^2 \zeta_{\bmk}  \,,
\ee
where $\mfz_s = a\Mp \sqrt{2\slrl } \simeq - \sqrt{2\slrl} \, \Mp/(H \eta)$ is defined in \pref{eq:defzeta}. This calculation is performed explicitly in \S\ref{Appssec:Transport}, with the result  
\begin{small}
\be \label{matTrans3}
\frac{\partial}{ \partial \eta } \left[ \begin{matrix} \Sigma_{11} \\ \Sigma_{12} \\ \Sigma_{22} \end{matrix}  \right] =\left[ \begin{matrix} 4 \mathrm{Im}[\mathcal{J}^{\mfp \zeta}_{k0} ]  & 2 \mfz_s^{-2} + 4 \mathrm{Im}[\mathcal{J}^{\mfp \mfp}_{k0} ] & 0 \\
 - \mfz_s^2 k^2 - 2 \mathrm{Im}[\mathcal{J}^{\zeta \zeta}_{k0} ] & 2\mathrm{Im}[-\mathcal{J}^{\zeta \mfp}_{k0}  + \mathcal{J}^{\mfp \zeta}_{k0} ] & \mfz_s^{-2} + 2 \mathrm{Im}[\mathcal{J}^{\mfp \mfp}_{k0} ] \\
 0 & -2 \mfz_s^{2} k^2 - 4 \mathrm{Im}[\mathcal{J}^{\zeta \zeta}_{k0} ] & - 4 \mathrm{Im}[\mathcal{J}^{\zeta \mfp}_{k0} ] \end{matrix}  \right] \left[ \begin{matrix} \Sigma_{11} \\ \Sigma_{12} \\ \Sigma_{22} \end{matrix}  \right] + \left[ \begin{matrix} 2 \mathrm{Re}[\mathcal{J}^{\mfp \mfp}_{k0} ]  \\ - \mathrm{Re}[\mathcal{J}^{\zeta \mfp}_{k0}  + \mathcal{J}^{\mfp \zeta}_{k0} ] \\   2 \mathrm{Re}[\mathcal{J}^{\zeta \zeta}_{k0}]   \end{matrix}  \right] 
\ee
\end{small}\ignorespaces

Notice that turning off the interactions implies that the free correlators $\sigma_{ij}$ must satisfy
\begin{small}
\be \label{matTrans3z0}
\frac{\partial}{ \partial \eta } \left[ \begin{matrix} \sigma_{11} \\ \sigma_{12} \\ \sigma_{22} \end{matrix}  \right] =\left[ \begin{matrix} 0 & 2 \mfz_s^{-2}   & 0 \\
 - \mfz_s^2 k^2   & 0 & \mfz_s^{-2}   \\
 0 & -2 \mfz_s^{2} k^2   &0\end{matrix}  \right] \left[ \begin{matrix} \sigma_{11} \\ \sigma_{12} \\ \sigma_{22} \end{matrix}  \right]  
\ee
\end{small}\ignorespaces
which for $\mfz_s =- c/\eta$ implies
\be \label{BDlatetime}
  \partial_\eta \sigma_{11} = \frac{2 \eta^2 \, \sigma_{12}}{c^2} \,, \quad
  \partial_\eta \sigma_{22} = - \frac{2 c^2 k^2 \sigma_{12}}{\eta^2} \quad \hbox{and} \quad
  \partial_\eta \sigma_{12} = - \frac{c^2k^2 \sigma_{11}}{\eta^2} + \frac{\eta^2 \sigma_{22}}{c^2} \,.
\ee
These are satisfied in particular by the free massless Bunch-Davies correlators given in \pref{FreeSigma}:
\be \label{BDleadingform}
\sigma_{11}(\eta)   = \frac{H^2 \left(\eta ^2 k^2+1\right)}{4 k^3 M_p^2 \epsilon _1} \,, \quad
\sigma_{12}(\eta)  = \frac{1}{2 k \eta}  \quad \hbox{and} \quad
\sigma_{22}(\eta) = \frac{k \epsilon _1 M_p^2}{\eta ^2 H^2}  \,,
\ee
because $c^2 = {2 \epsilon _1 M_p^2}/{ H^2}$.

\subsubsection{Asymptotic forms}

The Gaussian time-local behaviour described above provides an insight into late-time deep super-Hubble behaviour once the asymptotic forms \pref{Jzz} through \pref{Jpp} are used in the $\eta \to 0$ limit. To identify the late-time limit we write $\widetilde\cJ^{\mfa\mfb}_{k0} \simeq j^{\mfa\mfb}/z^p$ for the leading small-$z$ form, where the coefficients $j^{\mfa\mfb} = \cR^{\mfa\mfb}+ i \cI^{\mfa\mfb}$ can be read off explicitly from \pref{Jzz} through \pref{Jpp}. Using these in eqs.~\pref{app:curlyJdimless} imply the asymptotic forms for the $\cJ^{\mfa\mfb}_{k0}$ are
\bea\label{app:curlyJdimless1}
    \mathcal{J}^{\zeta \zeta}_{k0}(\eta, -\infty) &=& \slrl k^4\widetilde{\mathcal{J}}^{\zeta \zeta}_{k0}(z) \simeq \slrl k^4 \left[ \frac{\cR^{\zeta\zeta}}{z^3} + \frac{i \cI^{\zeta\zeta}}{z^2} \right] \nn\\
  \mathcal{J}^{\zeta \mfp}_{k0}(\eta, -\infty)  &=& \frac{ k H^2}{M_p^2} \widetilde{\mathcal{J}}^{\zeta \mfp}_{k0}(z) \simeq  \frac{ k H^2}{M_p^2} \left[ \frac{\cR^{\zeta \mfp}}{z^2} + i \cI^{\zeta\mfp} z  \right]   \\
 \mathcal{J}^{\mfp \zeta}_{k0}(\eta, -\infty) &=&  \frac{ \slrl k H^2} {M_p^2} \widetilde{\mathcal{J}}^{\mfp \zeta}_{k0}(z)\simeq  \frac{ \slrl k H^2} {M_p^2} \Bigl[ \cR^{\mfp\zeta} + i \cI^{\mfp\zeta} z  \Bigr]  \nn\\
\mathcal{J}^{\mfp \mfp}_{k0}(\eta, -\infty) &=&  \frac{H^4}{ k^2 M_p^4} \Bigl[ \cR^{\mfp\mfp} z + i \cI^{\mfp\mfp} z  \Bigr]  \,. \nn
\eea

Using these asymptotic limits in the evolution equation \pref{matTrans3} gives the dominant small-$\eta$ evolution
\be \label{Sig11Evo}
  \partial_\eta \Sigma_{11} \simeq \frac{ 4 \slrl k H^2 z \cI^{\mfp \zeta}}{\Mp^2} \, \Sigma_{11} +   \frac{H^2z}{k^2 \Mp^2} \left[ \frac{z}{\slrl} + \frac{4H^2 \cI^{\mfp\mfp}}{\Mp^2}\right]   \Sigma_{12} +   \frac{2H^4 \cR^{\mfp\mfp} z}{ k^2 M_p^4}  
\ee
\be \label{Sig22Evo}
  \partial_\eta \Sigma_{22} \simeq -  \frac{4  k H^2  z\cI^{\zeta\mfp}  }{M_p^2}  \Sigma_{22} - \frac{4\slrl k^4}{z^2}   \left[  \frac{ \Mp^2}{H^2}  +  \cI^{\zeta\zeta} \right]   \Sigma_{12} +  \frac{ 2 \slrl k^4 \cR^{\zeta\zeta}}{z^3}  
\ee
and
\bea \label{Sig12Evo}
\partial_\eta \Sigma_{12} &\simeq& -\frac{ 2\slrl k^4}{z^2}\left[  \frac{ \Mp^2}{H^2} + \cI^{\zeta \zeta} \right] \Sigma_{11} + \frac{2 k H^2z}{\Mp^2} \Bigl[  \slrl  \cI^{\mfp\zeta}    -     \cI^{\zeta\mfp}     \Bigr] \Sigma_{12} \\
&& \qquad\qquad + \frac{H^2z}{k^2\Mp^2} \left[ \frac{z}{2\slrl} +  \frac{2H^2}{ M_p^2}  \cI^{\mfp\mfp}   \right]  \Sigma_{22}  -\frac{kH^2}{\Mp^2} \left[ \frac{  \cR^{\zeta \mfp}}{z^2}   -  \slrl  \cR^{\mfp\zeta} \right]   \,.\nn
\eea
Notice that the real parts of the correlators, $\cR^{\mfa\mfb}$, act as source terms for the $\Sigma_{ij}$ correlators while the imaginary parts, $\cI^{\mfa\mfb}$, contribute to their mixing amongst themselves as time evolves. 

Although the Bunch-Davies free contributions to these equations are often subdominant in $z$ they are enhanced relative to the interaction terms by a factor of $H^2/\Mp^2$. This means the Bunch-Davies behaviour can dominate for long periods of time but it eventually becomes swamped by secular effects if one waits long enough. The beauty of the evolution equation \pref{matTrans3} is that its validity relies only on the underlying evolution being Gaussian and time-local and so it can be used to understand the evolution at times when perturbation theory starts to break down.\footnote{It should be borne in mind when using \pref{matTrans3} that we have not computed the contributions of quartic self-interactions to field evolution and although the arguments of \cite{Burgess:2022nwu} show why this does not matter when evolving the purity, it can matter when computing the correlators $\Sigma_{ij}$ more generally.} 

One might worry that secular growth and the breakdown of perturbation theory might undermine the validity of standard calculations \cite{Mukhanov:1981xt, Starobinsky:1982ee,  Hawking:1982cz, Mukhanov:1988jd} for the size of primordial perturbations in the later universe, but these are protected by general consistency arguments that protect super-Hubble long-wavelength perturbations \cite{Assassi:2012et}. These arguments can be made very explicit using eqs.~\pref{Sig11Evo} through \pref{Sig12Evo}, by examining the leading small-$z$ behaviour in the regime where the Bunch-Davies contributions no longer dominate. 

To see how, notice that the evolution implied by \pref{Sig11Evo} through \pref{Sig12Evo} is consistent with the leading small-$z$ behaviour taking the form
\be
  \Sigma_{11} \simeq s_{11} z^0 \,, \qquad \Sigma_{12} \simeq \frac{s_{12}}{z} \qquad \hbox{and} \qquad \Sigma_{22} \simeq \frac{s_{22}}{z^2} \,,
\ee
which is precisely the form also found in the non-interacting Bunch-Davies result \pref{BDlatetime}. This asymptotic form is consistent even in the presence of interactions because using it in the evolution equations implies the small-$z$ expansion of $\partial_z \Sigma_{ij} = - \partial_\eta \Sigma_{ij}/k$ becomes
\be \label{S11Assymp}
  \partial_z \Sigma_{11} 
  \simeq  -   \frac{4H^4 \cI^{\mfp\mfp}}{k^3\Mp^4}  s_{12} + \cO(z)
\ee
\be\label{S22Assymp}
 \partial_z \Sigma_{22}   \simeq  \frac{4\slrl k^3}{z^3}  \left[   \left(  \frac{ \Mp^2}{H^2}  +  \cI^{\zeta\zeta} \right)   s_{12}  - \tfrac12 \, \cR^{\zeta\zeta}   \right] + \cO(z^{-1}) 
\ee
and
\be\label{S12Assymp}
  \partial_z \Sigma_{12}    \simeq     \left[ 2\slrl k^3 \left( \frac{ \Mp^2}{H^2} + \cI^{\zeta \zeta} \right) s_{11}  + \frac{H^2}{\Mp^2}  \cR^{\zeta \mfp}\right] \frac{1}{z^2} + \cO(z^{-1})  \,, 
\ee 
and so the derivatives grow no faster as $z \to 0$ than does the assumed asymptotic form:
\be
   \partial_z \Sigma_{11} \propto z^p \;\;\hbox{with $p>0$} \,, \qquad \partial_z \Sigma_{12} \simeq - \frac{s_{12}}{z^2} \qquad
   \hbox{and} \qquad   \partial_z \Sigma_{22}\simeq -\frac{2 s_{22}}{z^3}  \,.
\ee

In particular, eq.~\pref{S11Assymp} implies interactions contribute only to terms in $\Sigma_{11}$ that vanish at least linearly with $z$ as $z \to 0$, and so leave untouched the leading Bunch-Davies result seen in \pref{BDlatetime}:
\be \label{s11ans}
   s_{11} =  \frac{H^2 }{4 k^3 M_p^2 \epsilon _1} \,.
\ee
Equating the leading powers of $1/z$ in \pref{S22Assymp} and \pref{S12Assymp} then also identifies how the interactions modify the leading coefficients, $s_{12}$ and $s_{22}$, with
\be \label{s12ans}
 s_{12} = - \left[ 2\slrl k^3 \left( \frac{ \Mp^2}{H^2} + \cI^{\zeta \zeta} \right) s_{11}  + \frac{H^2}{\Mp^2}  \cR^{\zeta \mfp} \right] = - \left[ \tfrac12 + \frac{H^2}{\Mp^2}  \Bigl(   \cR^{\zeta \mfp} +  \tfrac12\, \cI^{\zeta \zeta}    \Bigr) \right]  \,,
\ee 
and
\be \label{s22ans}
   s_{22} =   \slrl k^3  \left[ \cR^{\zeta\zeta}   -2  \left(  \frac{ \Mp^2}{H^2}  +  \cI^{\zeta\zeta} \right)   s_{12}   \right]  =  \slrl k^3  \left[\frac{ \Mp^2}{H^2} + \cR^{\zeta\zeta} + 2(\cR^{\zeta\mfp} + \cI^{\zeta\zeta}) + \cO\left( \frac{H^2}{\Mp^2} \right)  \right] \,.
\ee
Notice that the leading terms in powers of $H^2/\Mp^2$ of these expressions agree with eqs.~\pref{LeadingBDSigmaij} and \pref{BDleadingform} for the Bunch-Davies correlators.

The above consistency check is important in practice because it provides a constraint on how quickly the integrals $\widetilde{\cJ}^{\mfa\mfb}_{k0}$ can grow as $z \to 0$. If these integrals grow too strongly for small $z$ the above arguments can lead to asymptotic forms for $\Sigma_{ij}$ that are inconsistent with general constraints on late-time de Sitter behaviour of correlators.\footnote{This is indeed how we have found errors in our earlier calculations.}

\subsection{Implications for the purity}

We now return to our main line of development and focus on the evolution of the purity, starting with the general evolution allowed in the time-local and Gaussian limit. The purity turns out to be given in terms of the correlation matrix $\boldsymbol{\Sigma}_\bmk$ built using $\tilde \zeta_\bmk$, $\tilde \mfp_\bmk$ and $\varrho_\bmk$ by the expression\footnote{As a check notice that the Bunch Davies expressions \pref{BDleadingform} imply $\det \boldsymbol{\Sigma} = \frac14$.}
\be
    \gamma_\bmk = \frac{1}{\sqrt{4 \det \boldsymbol{\Sigma}_\bmk}} \,
\ee
and so we require the rate of change of the determinant of the covariance matrix, which a simple calculation using \pref{matTrans3} reveals to be
\begin{equation}
\frac{\partial \det \boldsymbol{\Sigma}}{\partial \eta} = 4 \mathrm{Im}[ \mathcal{J}^{\mfp \zeta}_{k0} -\mathcal{J}^{\zeta \mfp}_{k0} ] \det \boldsymbol{\Sigma} + 2\Big( \mathrm{Re}[\mathcal{J}^{\zeta \zeta}_{k0} ] \Sigma_{11} + \mathrm{Re}[\mathcal{J}^{\zeta \mfp}_{k0} + \mathcal{J}^{\mfp \zeta}_{k0}] \Sigma_{12} + \mathrm{Re}[\mathcal{J}^{\mfp \mfp}_{k0} ] \Sigma_{22} \Big)
\end{equation}
and so
\bea
 \frac{\partial_\eta \gamma_\bmk}{\gamma_\bmk}  &=& - \tfrac12 \, \frac{\partial_\eta \det \boldsymbol{\Sigma}_\bmk}{\det \boldsymbol{\Sigma}_\bmk} \nn\\
 &=& 2 \mathrm{Im}[ \mathcal{J}^{\zeta \mfp}_{k0} - \mathcal{J}^{\mfp \zeta}_{k0}  ]  - 4 \gamma^2_\bmk \Bigl[ \mathrm{Re}[\mathcal{J}^{\zeta \zeta}_{k0} ] \Sigma_{11} + \mathrm{Re}[\mathcal{J}^{\zeta \mfp}_{k0} + \mathcal{J}^{\mfp \zeta}_{k0}] \Sigma_{12} + \mathrm{Re}[\mathcal{J}^{\mfp \mfp}_{k0} ] \Sigma_{22} \Bigr]\,.
\eea

This last equation allows a very general conclusion whose validity relies only on the time-locality and Gaussianity of the evolution. Rewriting it as  
\be
  \partial_\eta \gamma_\bmk = 2 \mathrm{Im}[ \mathcal{J}^{\zeta \mfp}_{k0} - \mathcal{J}^{\mfp \zeta}_{k0}  ] \gamma_\bmk - 4 \gamma^3_\bmk \Bigl[ \mathrm{Re}[\mathcal{J}^{\zeta \zeta}_{k0} ] \Sigma_{11} + \mathrm{Re}[\mathcal{J}^{\zeta \mfp}_{k0} + \mathcal{J}^{\mfp \zeta}_{k0}] \Sigma_{12} + \mathrm{Re}[\mathcal{J}^{\mfp \mfp}_{k0} ] \Sigma_{22} \Bigr]\,,
\ee
shows that $\gamma_\bmk = 0$ is a fixed point of the evolution (barring any unexpected singular behaviour in the $\cJ^{\mfa\mfb}_{k0}$ integrals or correlators $\Sigma_{ij}$). Consequently once the state becomes very classical it tends to stay that way - perhaps the starting point for an inflationary {\it no-recoherence} theorem. 

To extract more information we next use the small-$z$ expansion using the leading form of the integrals $\cJ^{\mfa\mfb}_{k0}$ given in \pref{app:curlyJdimless1}, to identify the late-time form for the purity evolution. Changing variables from $\eta$ to $z = - k \eta$,
\bea \label{mastergammaevo}
  \partial_z \gamma_\bmk  &\simeq&    \frac{ 4 \gamma^3_\bmk}{z^3} \left[\slrl k^3   \cR^{\zeta\zeta} s_{11} + \frac{  H^2}{M_p^2}   \cR^{\zeta \mfp}  s_{12}   \right] + \cO(z^{-2}) \nn\\
 &=& \frac{  \gamma^3_\bmk}{z^3} \left( \frac{H^2}{\Mp^2} \right) \left[  \cR^{\zeta\zeta} - 2  \cR^{\zeta \mfp}  + \cO\left(\frac{H^2}{\Mp^2} \right) \right] + \cO(z^{-2})   \,,
\eea
where the last equality uses \pref{s11ans} and \pref{s12ans} for $s_{11}$ and $s_{12}$. Notice that for small $z$ only the real parts of the two integrals $\cJ^{\zeta\zeta}_{k0}$ and $\cJ^{\zeta\mfp}_{k0}$ and the leading parts of $\Sigma_{11}$ and $\Sigma_{12}$ play a role.  $\cJ^{\zeta\zeta}_{k0}$ and $\cJ^{\zeta\mfp}_{k0}$ are also the only two kernels that appear in the leading slow-roll limit, though this was not used in deriving \pref{mastergammaevo}.

If we use the explicit expressions for $\cR^{\zeta\zeta}$ and $\cR^{\zeta \mfp}$ given in \pref{Jzz} and \pref{Jzp} we find
\be
 \cR^{\zeta\zeta} = \frac{32}{45 \kappa^3} \left( \frac{2}{\pi^2} \right) \quad \hbox{and} \quad
 \cR^{\zeta\mfp} = \frac{2}{25 \kappa^5} \left( \frac{2}{\pi^2} \right) \,,
\ee 
and so
\be
 \cR^{\zeta\zeta} - 2 \cR^{\zeta \mfp} = \frac{4}{5\kappa^3} \left( \frac{2}{\pi^2} \right) \left( \frac89 - \frac{1}{5 \kappa^2} \right) \,.
\ee 
In the large-$\kappa$ limit (when $k_\UV \gg k$) the $\cR^{\zeta \mfp}$ contribution also becomes negligible, leaving only the contribution coming from $\cR^{\zeta\zeta}$. 

\subsection{Late-time resummation}

Working perturbatively and starting from the Bunch-Davies vacuum (whose initial condition is $\gamma_\bmk(z = \infty) \to 1$ we can use $\gamma_\bmk \simeq 1$ on the right-hand side of \pref{mastergammaevo}, leading back to our perturbative result \pref{LindbladPurity5}, including the singular $z^{-3}$ behaviour that signals secular growth. The beauty of the derivation of \pref{mastergammaevo} is that its domain of validity goes beyond straight-up perturbation theory in $H^2/\Mp^2$, since it is valid for any situation for which the evolution is time-local and Gaussian. But the justification of these conditions relies on the small-$z$ limit, which gets better and better the later in time one works. 

We can consequently solve for $\gamma_\bmk$ at late times by integrating \pref{mastergammaevo}, which can be done explicitly regardless of the particular values of $\cR^{\mfa\mfb}$. The equation to be solved has the form 
\be
   \gamma'(z) = \frac{ \mfc \gamma^3}{z^3} \,, 
\ee
where the constant $\mfc = (\cR^{\zeta\zeta} - 2 \cR^{\zeta \mfp})(H/\Mp)^2$ is small, but need not be smaller than $z$. This has as its general solution
\be
   \gamma(z) = \frac{z}{\sqrt{\mfc +  \cC z^2}} \,,
\ee
where $\cC > 0$ is the integration constant, whose value can be determined from the boundary condition 
\be
   \gamma_\infty = \lim_{z \to \infty} \gamma(z) = \frac{1}{\sqrt{\cC}} \,.
\ee
For the Bunch-Davies system we choose $\cC = 1$, leading to
\be
  \gamma_\bmk(z) = \frac{z}{\sqrt{\mfc + z^2}} \qquad \hbox{where} \quad \mfc =   (\cR^{\zeta\zeta} - 2 \cR^{\zeta \mfp}) \frac{H^2}{\Mp^2} \,.
\ee

This looks very similar to the expressions obtained in \cite{Burgess:2022nwu}, who argued that the resummed result of a related purity calculation could be written as
\be \label{ResummedResult}
   \gamma_\bmk(\eta) = \frac{1}{\sqrt{1 + \Xi_k(\eta)}}   
\ee
where $\Xi_k(\eta)$ was perturbatively calculable. Our derivation here shows why resummation gives a similar result for the calculation given here of the leading-in-slow-roll late-time inflationary purity evolution, where
\be \label{XiResult}
   \Xi_k(\eta) = \frac{\mfc}{z^2} = \frac{\mfc}{k^2 \eta^2} =(\cR^{\zeta\zeta} - 2 \cR^{\zeta \mfp}) \frac{H^2}{\Mp^2} \left( \frac{aH}{k} \right)^2 \,.
\ee
This returns a sensible expression for $\gamma_\bmk$ for any positive $\Xi_k$ (even if it diverges, as it does at late times, which in this case just corresponds to $\gamma_\bmk \to 0$).  

\section{Lessons learnt}
\label{sec:Conclusions}
 
In this paper we report what we believe to be the first full calculation of the the leading decoherence of long-wavelength scalar-metric fluctuations due to their gravitational interactions with short wavelength scalar and tensor metric perturbations. We do so within a minimal single-clock inflationary framework in which we are able to explicitly track system-environment interactions into the remote past to make contact with the Bunch-Davies initial conditions. For applications to primordial fluctuations we specialize to the late-time regime when the decohering long-wavelength modes are super-Hubble, for which we find several simplifications, leading to the late-time super-Hubble result -- {\it c.f.}~eq.~\pref{LindbladPurity5int}, 
\be    \label{LindbladPurityRecap}
 \gamma_\bmk(a)  =  1  - \frac{32}{45\pi^2} \left( \frac{H^2}{\Mp^2} \right)   \frac{k}{k_\UV} \left( \frac{a H}{k_\UV}\right)^2  + \cdots 
 \quad \hbox{for} \quad k \ll aH \,.
\ee  
As discussed in the main text, obtaining this expression requires careful handling of the ordering of limit, with care taken to properly project onto the Bunch Davies vacuum at early times.

What is new in this calculation is the observation that at leading order in slow roll one must evolve both tensor and scalar environments {\it simultaneously}. Interactions that couple scalar system fluctuations to an environment that is pure scalar or pure tensor are all systematically slow-roll suppressed, and an implicit restriction to these two cases is why earlier calculations \cite{Nelson:2016kjm, Burgess:2022nwu,Lopez:2025arw, Burgess:2025dwm, DaddiHammou:2022itk} find a slow-roll suppressed result. The absence of slow-roll suppression makes the decoherence rate for scalar and tensor fluctuations the same order of magnitude. Our calculation is complete inasmuch as it includes all of the gravitational self-interactions that can contribute to lowest order in the semiclassical expansion. This allows us to identify systematically which interactions decohere most efficiently, which parts of the environment are responsible and when the decoherence occurs.  

Our calculation reveals several characteristic properties of gravitation-mediated decoherence within minimal inflationary models, some but not all of which are also shared by partial calculations of the subdominant slow-roll suppressed rate:
\begin{itemize}
\item As always, semiclassical perturbation theory organizes the calculation into powers of $H^2/M_p^2$, where $H$ is the inflationary Hubble scale, consistent with general EFT power-counting arguments \cite{Adshead:2017srh, Burgess:2009ea} (for a textbook review see \cite{Burgess:2020tbq}). For minimal inflationary models the leading nontrivial decoherence contribution arises at order $H^2/M_p^2$ and is mediated by interactions cubic in the fields. Choosing the environment to be short-wavelength modes then ensures that momentum conservation implies the effective evolution of the purity of long-wavelength system modes at this order is gaussian. This in turn implies that it can be understood mode-by-mode without mode-mixing. 
\item The result \pref{LindbladPurityRecap} is not universal in the sense that it depends explicitly on the details of the split between system and environment (through the value of $k_\UV$). 
\item Eq.~\pref{LindbladPurityRecap} shows that decoherence for super-Hubble modes during inflation has a small amplitude (because of the weakness of gravitational interactions), although the lack of slow-roll suppression makes it slight larger than previously anticipated. The observed amplitude of primordial fluctuations implies the prefactor $H^2/(8\pi^2\Mp^2) \simeq 2 \times 10^{-9}\slrl$ and so since the non-observation of primordial tensor fluctuations implies $\slrl \lsim 10^{-2}$ we infer $32H^2/(45\pi^2\Mp^2) \simeq 1 \times 10^{-8} \slrl \lsim 1 \times 10^{-10}$.  
\item Primordial purity grows very quickly with time, with eq.~\pref{ResummedResult} showing that $\Xi_\bmk  \propto \eta^{-2}$ and so $\partial_\eta \Xi_\bmk \propto \eta^{-3} \propto a^3$, much as found in earlier partial calculations like the ones in \cite{Nelson:2016kjm, Burgess:2022nwu}, but more slowly than found in other partial calculations like v1 of \cite{Burgess:2025dwm}. 
\item The leading behaviour for super-Hubble modes is time-local, with deviations from the time-local limit controlled by powers of $k/(aH)$ (much as found in earlier partial calculations), with the leading deviations suppressed by a single power of $k/(aH)$.
\item The time-local and Gaussian nature of the leading system behaviour for super-Hubble modes allows the late-time evolution to be resummed in a way that allows predictions to be extended to times much later than the point where perturbative methods naively fail -- for the reasons argued in \cite{Burgess:2015ajz, Burgess:2022nwu, Burgess:2024eng}. This is crucial for making reliable predictions in practice because for most inflationary modes 50 $e$-foldings suffice to cause naive perturbation theory to fail. 
\item Similar to \cite{Burgess:2022nwu} -- but unlike, say, \cite{Nelson:2016kjm} -- we find the leading contributions to decoherence to be UV finite, despite UV divergences arising in many intermediate steps. Although this in retrospect was a happy accident for \cite{Burgess:2022nwu}, it is expected for the calculation presented here because the decoherence calculation arises at loop order in the general semiclassical $H^2/M_p^2$ expansion \cite{Burgess:2003jk, Burgess:2009ea, Adshead:2017srh}. UV finiteness is required at leading order because general arguments ensure that UV divergences can be absorbed into the renormalization of counterterms in the effective lagrangian, but decoherence can never arise due to a particular choice for a value of a coupling in the lagrangian. Calculations are in progress to see how UV divergences arise and are renormalized within the purity at subdominant order in $\slrl$.  
\end{itemize}
 
There is much more to be learned from explicit calculations of quantum coherence in cosmology, even if they only tell us that quantum effects are irretrievably erased in primordial perturbations by the time they return to the sub-Hubble regime. In the end what is useful is to know how decoherence depends on the parameters of any underlying model, since this is what determines what we learn from any future attempts to measure quantum behaviour. 

\section*{Acknowledgements}
We thank Rafael Bravo Guerraty, Sebastian Cespedes, Thomas Colas, Francescopaolo Lopez, Enrico Pajer and Jason Pollack for helpful conversations. CB's research was partially supported by funds from the Natural Sciences and Engineering Research Council (NSERC) of Canada. Research at the Perimeter Institute is supported in part by the Government of Canada through NSERC and by the Province of Ontario through MRI.

\appendix
%
\addtocontents{toc}{\setcounter{tocdepth}{1}}

\section{Open system of Primordial Fluctuations}
\label{App:sec:OpenEFT}

This section briefly summarizes the calculational framework, following closely on the description given in \cite{Burgess:2022nwu, Burgess:2025dwm}. 

\subsection{Semiclassical setup}

The system consists of vanilla `single-clock' inflationary models, with the metric $g_{\mu\nu}$ and inflaton $\varphi$ coupled through the action
\begin{equation}
\label{app:actionstart}
S = \int \exd^4 x\; \sqrt{ - g} \bigg[\tfrac12  \Mp^2 \, \cR
    - \tfrac{1}{2} g^{\mu\nu} \, \partial_{\mu} \varphi \,
    \partial_{\nu} \varphi  - V(\varphi) \bigg]
\end{equation}
where $\cR$ is the metric's Ricci scalar and $V(\varphi)$ is the inflationary potential. We study the properties predicted by this action for fluctuations about a homogeneous spatially flat FRW geometry.

The background configuration is $\varphi = \phi(t)$ and $\exd  s^2 = a^2(\eta) \left( - \exd  \eta^2 + \exd \bm{x}^2 \right)$, with near-de Sitter slow roll assumed so $\slrl =-{\dot{H}}/{H^2} \ll 1$ and
\be\label{app:phidotvseps}
\dot{\phi}^2 \simeq 2H^2\Mp^2\slrl \,.
\ee

Fluctuations about this background are described by expanding the scalar field and metric
\begin{eqnarray} \label{app:ADMmetric}
\varphi = \phi(t) + \delta \varphi(t,\bm{x}) \quad\hbox{and} \quad
  \exd  s^2 = - N^2 \exd  t^2 + h_{ij} \big( \exd  x^i + N^i \exd  t \big)
  \big( \exd  x^j + N^j \exd  t \big) \,,
\end{eqnarray}
and picking a gauge to fix time and spatial reparametrizations. Standard arguments show that using this expansion in the action (\ref{actionstart}) ends up leaving a single physical scalar degree of freedom plus the two tensor modes describing gravitational waves, which we write as
\begin{equation} \label{app:metricsplit}
     h_{ij} = a^2 e^{2\zeta} \hat h_{ij} \quad\hbox{with}\quad
     \hat h_{ij} = \delta_{ij} + \gamma_{ij} + \tfrac12 \,
     \delta^{kl} \gamma_{ik} \gamma_{lj} + \cdots \,,
\end{equation}
where $\det \hat h_{ij} = 1$ and $\delta^{ij} \partial_i \gamma_{jk} = \delta^{ij} \gamma_{ij} = 0$ and the lapse $N$ and shift $N^i$ are determined by solving the energy and momentum constraints. The scalar perturbation that remains depends on the gauge choice. The single degree of freedom is conveniently represented either by $\delta \varphi$ or $\zeta$, depending on whether we use comoving gauge ($\delta \varphi = 0$) or spatially flat gauge ($\zeta = 0$). 

The action is expanded about the background and the expansion starts at quadratic order in the fluctuations because the background is chosen to solve the classical field equations. The leading (quadratic) order describes the `free' evolution of fluctuations that interact only with the background while cubic and higher orders describe the interactions of fluctuations with one other. These interactions are treated perturbatively below, but part of the story is identifying the small parameters that justify a perturbative treatment.

\subsubsection{Free evolution about a fixed background}

Consider first the fluctuations' free evolution. Working in the co-moving gauge and restricting to the part of the action quadratic in fluctuations gives (see for example \cite{Kodama:1985bj, Mukhanov:1990me, Maldacena:2002vr}) ${}^2S = {}^2 S_s + {}^2 S_t$ where the action for the tensor fluctuations is
\be \label{app:freetensoraction}
 {}^2 S_t = \tfrac18\Mp^2 \int \exd  t \; \exd^3 \bm{x}\; \Bigl(
    a^3 \dot{\gamma}_{ij} \dot{\gamma}^{ij} - a \, \partial_k \gamma_{ij} \, \partial^k \gamma^{ij} \Bigr) = \tfrac18 \Mp^2\int \exd  \eta \; \exd^3 \bm{x}\; a^2\Bigl[
(\gamma'_{ij})^2  - ( \partial_k \gamma_{ij} )^2 \Bigr] \,,
\ee
with spatial indices raised and lowered using $\delta_{ij}$ and the second equality changing to conformal time. The action for the scalar fluctuations similarly is
\be 
\label{app:freescalaraction}
  {}^{(2)}S_s = \int \exd  t \; \exd^3 \bm{x}\;
  \frac{\dot{\phi}^2}{2H^2}\bigg[ a^3 \dot{\zeta}^2
    - a (\partial \zeta )^2 \bigg]   = \int \exd  \eta \; \exd^3 \bm{x}\;
  {\slrl} \Mp^2 a^2\bigg[ (\zeta')^2 -  (\partial \zeta )^2 \bigg]   \,,
\ee
where the second equality uses \pref{app:phidotvseps} to eliminate $\dot\phi^2/H^2$ in favour of the slow-roll parameter $\slrl$ and $(\partial \zeta)^2$ denotes $\delta^{ij} \partial_i \zeta \, \partial_j \zeta$. For evolution in conformal time the momenta conjugate to $\zeta$ and $\gamma_{ij}$ as obtained from \pref{app:freetensoraction} and \pref{app:freescalaraction} are
\be \label{app:pzeta0}
  \pi_{ij} = \frac{\delta S}{\delta \gamma'_{ij}} = \tfrac14 \Mp^2 a^2 \gamma'_{ij} \quad \hbox{and} \quad
   \mfp = \frac{\delta S }{ \delta \zeta'} = 2{\slrl}\Mp^2 a^2\zeta' \,. 
\ee

When identifying the small parameters that control perturbation theory it is useful to define canonical variables whose correlation functions are independent of the small control parameters. For tensors these are
\begin{equation}
\label{app:eq:defvij}
v_{ij}(\eta, {\bm x}) := \mfz_t \, \gamma_{ij}(\eta,{\bm x}) \qquad \hbox{where} \quad \mfz_t := \tfrac12 a\Mp  \,,
\end{equation}
and for scalars these are the Mukhanov-Sasaki variable~\cite{Mukhanov:1981xt,Kodama:1985bj} 
\begin{equation}
\label{app:eq:defzeta}
v(\eta, {\bm x}) := \mfz_s \, \zeta(\eta,{\bm x}) \qquad \hbox{where} \quad \mfz_s := a\Mp \sqrt{2\slrl } \,,
\end{equation}
in terms of which the quadratic action \pref{app:freescalaraction} becomes (see for example \cite{Kodama:1985bj, Mukhanov:1990me, Maldacena:2002vr})
\be
\label{app:freescalaractionv}
  ^{(2)}S_s   = \tfrac{1}{2}
\int{\exd  \eta\; \exd^3 \bm{x} \;
\bigg[\left(Dv\right)^2
- (\partial v )^2
  \bigg]}  
 = \tfrac{1}{2}
\int{\exd  \eta\; \exd^3 \bm{x} \;
\bigg[\left(v^\prime\right)^2
-(\partial v )^2
+\frac{\mfz''}{\mfz } v^2 \bigg]} + \hbox{s.t.}\,,
\ee
where 
\begin{equation} \label{app:Ddef}
 Dv := v' - \frac{\mfz' v}{\mfz} 
\end{equation}
and the second equality performs an integration by parts, for which the corresponding surface terms are denoted by `s.t.'. (See \cite{Burgess:2025dwm} for a more detailed discussion of issues associated with the integration by parts.) 

Quantization of this free system goes through using standard methods. For the canonical scalar variable we have
\begin{equation}
\label{eq:v:Fourier}
v(\eta,\bm{x}) =\frac{1}{\sqrt{V}}\sum_{\bm{k}}\; 
v_{\bm{k}}(\eta) \, e^{i \bm{k}\cdot\bm{x}}\,,
\end{equation}
where $v(\eta,\bm{x}) = v^\dagger(\eta,\bm{x})$ implies $ v_{-\bm{k}}(\eta) = v_{\bm{k}}^\dagger(\eta)$ and a similar expression holds for the conjugate momentum $p_{\bm{k}} = \delta S/\delta v'_{\bm{k}}$. Time evolution is made explicit by expanding $v_{\bm{k}}(\eta)$ in terms of mode functions $u_{{k}}(\eta)$
\begin{equation} \label{app:vhatk}
 v_{\bm{k}}(\eta) = u_{{k}}(\eta)\, {c}_{\bm{k}}
  + u^{\ast}_{{k}}(\eta)\, {c}^{\dagger}_{-\bm{k}} \,,
\end{equation}
where the $u_k(\eta) \, e^{i \bmk \cdot \bmx}$ is a complete basis of solutions to the linear field equation for $v$ obtained from the action \pref{app:freescalaractionv}, subject to an appropriate initial condition. For instance, for massless states on a de Sitter background satisfying the Bunch-Davies initial condition this implies
\be \label{app:modev}
  u_k(\eta) = \frac{1}{\sqrt{2k}} \left( 1 - \frac{i}{k\eta} \right) e^{-i k \eta} \,,
\ee
where the modes are normalized using $u_{{k}} u_{{k}}^{*\prime}-u_{{k}}^* u_{{k}}'=i$. This normalization ensures the equal-time commutation relations imply 
\be \label{app:ccstarcommutator}
     [{c}_{\bm{k}}, {c}^{\dagger}_{\bm{q}}] = \delta_{\bm{k},\bm{q}} \,.
\ee

For tensor modes an identical argument implies the canonical field has the expansion
\be \label{app:eq:tensormodedecomp}
v_{i j}(\bmx ,\eta)  =   \frac{1}{\sqrt{V}} \sum_{\bmk,\sigma} \epsilon_{i j}(\hat{\bmk},\sigma) v_{\bmk\sigma}(\eta) \, e^{i \bmk\cdot\bmx }
\quad \hbox{with} \quad
 v_{\bmk\sigma}(\eta) = u_k(\eta) \, b_{\bmk\sigma} + s_{\sigma} u_k^*(\eta) \, b^\dagger_{-\bmk\sigma} \,,
\ee  
where $\sigma = +, \times$ denotes the two spin states, whose polarization tensors $\epsilon_{ij}(\hat{\bmk},\sigma)$ satisfy $\epsilon_{ij}(-\hat{\bmk},\sigma) = s_\sigma \, \epsilon_{ij}(\hat{\bmk},\sigma)$ for a sign $s_\sigma = \pm 1$. Because the equation of motion for $v_{\bmk\sigma}$ is the same as for $v_\bmk$ (for massless scalars when $\slrl$ is constant) the mode functions appearing in \pref{app:vhatk} and \pref{app:eq:tensormodedecomp} are identical, and the operators $b_{\bmk\sigma}$ satisfy the analog of \pref{app:ccstarcommutator}:
\be \label{app:aastarcommutator}
     [{b}_{\bm{k}\sigma}, {b}^{\dagger}_{\bm{q}\lambda}] = \delta_{\bm{k},\bm{q}} \, \delta_{\sigma \lambda} \,.
\ee
Specializing to de Sitter backgrounds and converting back to the variables $\zeta$ and $\gamma_{ij}$ leads to expressions \pref{ModeExpressions} of the main text.

Free quantum evolution is described in the usual way by the free Hamiltonian obtained from the quadratic action. For instance for scalars 
\begin{equation}
\label{freeHclassical}
\scrH_0(\eta)  :=  \tfrac{1}{2}\sum_{\bm{k}} \;
\Bigl[ p_{\bm{k}}(\eta)  p_{-\bm{k}}(\eta)
+ \omega^2(\bm{k},\eta)  v_{\bm{k}}(\eta)  v_{-\bm{k}} (\eta) \Bigr] \,,
\end{equation}
where the time-dependent frequency is
\begin{equation}
\label{omegadef}
\omega^2(\bm{k},\eta) := k^2 - \frac{\mfz^{\prime\prime}}{\mfz} \ .
\end{equation}   

\subsubsection{Interactions}

When canonical variables are used each additional power of $v$ or $v_{ij}$ comes together with a power of $\Mp^{-1}$ and so for $n \geq 2$ a term involving $n$ powers of fluctuations is proportional to $\Mp^{2-n}$. This implies all terms cubic and higher are suppressed by at least one power of $\Mp^{-1}$, with the minimal suppression coming from cubic interactions. Here $\Mp^{-1}$ is implicitly always compared with another scale, either associated with the background (like $H$) or the fluctuations (like $k/a$). It is the small size of $H/\Mp$ and/or $k/(a\Mp)$ that justifies treating these interactions perturbatively. 

Time evolution is then treated perturbatively using the interaction picture (for which the fluctuation fields satisfy their free equations of motion and so remain described by \pref{app:eq:tensormodedecomp}). The density matrix, $\rho$, describing the state of the system then satisfies
\begin{equation} \label{app:INTpicVN}
  \frac{\partial {\rho}}{\partial \eta}
  = - i \Bigl[ {\scrH}_{\mathrm{int}}(\eta) , {\rho}(\eta) \Bigr] \,,
\end{equation}
where in practice $\scrH_{\rm int}$ is built from the interactions contained when expanding the action \pref{app:actionstart} beyond quadratic order in the fluctuations.

For inflationary applications initial conditions are usually chosen in the remote past to be in the Bunch Davies vacuum \cite{Bunch:1978yq}: ${\rho}(\eta_{\mathrm{in}}) 
= | \Omega \rangle \langle \Omega |$, where $|\Omega \rangle$ is defined by 
\be
   c_\bmk | \Omega \rangle = b_{\bmk\sigma} |\Omega \rangle = 0 \,.
\ee

\subsection{System and environment}
\label{app:sec:SysEnv}

As described in the main text, we follow \cite{Burgess:2022nwu, Burgess:2025dwm} and divide the Hilbert space of states into a `system' and `environment' variables using comoving momenta, with long-wavelength modes making up the system and the environment comprising short-wavelength modes. We assume the co-moving scale $k_\UV$ separating system from environment does not evolve in time. These choices are meant to capture how only a limited range of wavelengths are seen by later observers, as sketched in Fig.~\ref{fig:defsystem}. 

We wish to compute the evolution of the system's reduced density matrix, ${\varrho}(\eta)$, obtained by tracing out the environmental degrees of freedom: 
\begin{equation} \label{app:reducedSpic}
{\varrho}(\eta) := \Trenv \Bigl[ {\rho}(\eta) \Bigr] \,,
\end{equation}
where $\rho$ is the density matrix for the full system, subject to the initial condition that initial state is the Bunch-Davies vacuum in the remote past. In particular
\begin{equation} \label{app:rhofactor}
{\varrho}(\eta_{\rm in}) = \bigotimes_{k < k_{\UV} }
{\varrho}_{\bm{k}}(\eta_{\rm in})  \,,
\end{equation}
where $\eta_{\rm in} \to - \infty$.

The evolution of the reduced density matrix $\varrho$ is in principle obtained by taking the trace of \pref{app:INTpicVN}, but (as described at length in \cite{Burgess:2022nwu, Burgess:2025dwm} -- see also the review \cite{Burgess:2022rdo}) this is not immediately useful because the trace of the right-hand side of the equation involves both $\varrho$ and the state of the environmental degrees of freedom. A more useful expression is obtained by solving for the evolution of the environment as a function of $\varrho$ and then substituting this back into the evolution equation for $\varrho$, but this does so at the expense of introducing nonlocality in time since $\partial_\eta \varrho$ depends on the entire past history $\varrho(\eta')$ for $\eta' < \eta$. The linearity of \pref{app:INTpicVN} allows this to be done in great generality within perturbation theory in $\scrH_{\rm int}$, with a result called the Nakajima-Zwanzig equation \cite{Nakajima:1958pnl, Zwanzig:1960gvu}.

The result is reasonably simple if the interaction Hamiltonian of has the general form
\begin{equation}
\label{app:HintRdef0}
      {\scrH}_{\mathrm{int}}(\eta) =   \int\exd^3 \bm{x}\;
     \cS_a (\eta, \bm{x}) \otimes \cE^a(\eta,\bm{x}) \,,
\end{equation}
(with an implied sum on $a$), as is the case for our cosmological applications, where $\cS_a$ involves only the system fields and $\cE^a$ involves only environment fields. In this case the Nakajima-Zwanzig evolution equation at second order in $\scrH_{\rm int}$ turns out to be \cite{Burgess:2022nwu, Burgess:2025dwm}
\bea
\label{app:NZ}
\frac{\partial {\varrho}}{\partial \eta} &\simeq&
- i  \scrE^a(\eta) \int \exd^3\bm{x} \;
 \Bigl[  \cS_a(\eta,\bm{x}),  {\varrho}(\eta) \Bigr] \nn\\
 &&\qquad -  \int \exd^3\bm{x} \int \exd^3\bm{x}' \int_{\eta_{\mathrm{in} } }^{\eta} \exd  \eta'  \; \bigg\lbrace \Bigl[  \cS_a(\eta,\bm{x}) ,
  \cS_b(\eta',\bm{x}')  {\varrho}(\eta') \Bigr] C^{ab}(\eta, \eta';
\bm{x} - \bm{x}') \nn\\
&& \qquad\qquad\qquad\qquad 
+ \Bigl[  {\varrho}(\eta') \cS_b(\eta',\bm{x}')  ,  \cS_a(\eta,\bm{x}) \Bigr]
C^{ba}(\eta', \eta;\bm{x} - \bm{x}') \bigg\rbrace 
\eea
where
\begin{equation}
\label{app:R_1pt_def}
\scrE^a(\eta) := \langle 0_{\env} | \cE^a(\eta,\bm{x}) | 0_{\env} \rangle
\end{equation}
and
\begin{equation}
\label{app:2pt_text}
C^{ab}(\eta,\eta' ; \bm{x} - \bm{x}') =  \langle 0_{\env} |
\Bigl[ \cE^a(\eta,\bm{x}) - \mathscr{E}^a(\eta) \Bigr]
\Bigl[ \cE^b(\eta',\bm{x}') - \mathscr{E}^b(\eta')  \Bigr] | 0_{\env} \rangle \,,
\end{equation}
evaluated in the Bunch-Davies state. So far as decoherence is concerned, we can omit the `tadpole' part (or $\scrE$-dependent first line) of the NZ equation \pref{app:NZ}, since it does not contribute at all to the decoherence. 

\subsection{Gaussian property}
\label{App:GaussianProperty}

As noticed in \cite{Burgess:2022nwu, Burgess:2025dwm} two important simplifications follow at leading order in the interactions from choosing the environment to consist only of short-wavelength modes. These properties follow from the fact that the leading interactions are cubic in the fields (at least for those that contribute to decoherence). This, together with momentum conservation, implies that the important cubic interactions always involve only one system field and two envirnomental fields. They only appear in this way because momentum conservation requires the three momenta in the interaction (one for each field) to form the sides of a triangle, but there are no triangles with two short and one long side. 
 
As a result the evolution found using these interactions is always gaussian in the system variables. It therefore does not mix modes with different values of $\bm{k}$, similar to free evolution, leaving the reduced density matrix with different momenta uncorrelated,
\be
  \varrho = \bigotimes_{\bm{k}}  \varrho_{\bm{k}} \,,
\ee
just as was true for the initial Bunch-Davies state. The reduced density matrix for each $\bm{k}$ can therefore be evolved separately, with decoherence governed to leading order by
\begin{equation}
\label{app:NZmodesrhok}
\frac{\partial \varrho_{\bm{k}}}{\partial \eta}
 = -   \int_{ \eta_{\mathrm{in}} }^{\eta} \exd  \eta' \;
\bigg\lbrace G(\eta) \, G(\eta') \Bigl[ \cS_{a{\bm{k}}}(\eta) ,
  \cS_{b{-\bm{k}}}(\eta')  \varrho_\bmk(\eta') \Bigr]
\mathscr{C}^{ab}_{\bm{k}}(\eta,\eta') + \hbox{h.c.}
\bigg\rbrace  \,,
\end{equation}
where
\begin{equation}
\label{app:CR_FT}
C^{ab}(\eta, \eta' ; \bm{y}) =  \frac{1}{V}\sum_{\bm{k}}\; 
\mathscr{C}^{ab}_{\bm{k}}(\eta,\eta') e^{i \bm{k} \cdot \bm{y} } \,,
\end{equation}
and 
\begin{equation}
\label{app:eq:v:FourierS}
\cS_a(\eta,\bm{x}) =\frac{1}{\sqrt{V}}\sum_{\bm{k}}\; 
\cS_{a{\bm{k}}}(\eta) \, e^{i \bm{k}\cdot\bm{x}}\,.
\end{equation}

Gaussianity allows a very explicit representation of the reduced density matrix. 
To see how, it is convenient to start by adopting the canonical field variable $v$ (rather than $\zeta$, say) and to  follow \cite{Martin:2018zbe} by breaking its Fourier components $v_{-\bm{k}}$ into real and imaginary parts 
\begin{equation} \label{app:alphaRI}
v_{\bm{k}}(\eta) =: \tfrac{1}{\sqrt2} \Bigl[ v^\ssR_{\bm{k}}(\eta)
+i \, v^\ssI_{\bm{k}}(\eta) \Bigr] \,,
\end{equation}
for which $v_{-\bm{k}} = v^\dagger_{\bm{k}}$ implies $v^{\ssR}_{\bm{k}} = v^{\ssR}_{-\bm{k}}$ while $v^{\ssI}_{\bm{k}} =
-v^{\ssI}_{-\bm{k}}$, and implies there is an independent mode $\bmk$ in the half-space $\mathbb{R}^{+3}$ \cite{Lesgourgues:1996jc,Kiefer:1998jk,Martin:2018zbe}. These Hermitian variables are convenient as they satisfy
$[\,v^{\alpha}_{\mathbf{k}},p^{\beta}_{\mathbf{q}}\,]
=
i(2\pi)^3\delta(\mathbf{k}-\mathbf{q})\delta^{\alpha\beta}
$ with $\alpha,\beta=\{R,I\}$ and evolve separately under linear evolution provided that the physics is isotropic (as is true here).

This makes it convenient to treat the system as if it were a single real field, $\w_{\bm{k}} = \w^\dagger_{\bm{k}}$ for {\it all} $\bm{k}$ and then identify $\sqrt2 \; v^\ssR_{\bm{k}} = \w_{\bm{k}}+\w_{-\bm{k}}$ and $\sqrt2 \; v^\ssI_{\bm{k}} = \w_{\bm{k}}-\w_{-\bm{k}}$ respectively as its even and odd parts under reflections of $\bm{k}$, since this simplifies the notation by allowing us to use real variables but drop the superscripts `R' and `I' on the fields. A similar story also
applies for the canonical momentum field, whose real Fourier components we similarly denote $\p_{\bm{k}}$.

In terms of these variables the Gaussian nature of the reduced density matrix boils down to the statement that its matrix elements can be written
\be 
\label{app:eq:rho:Gaussian}
\left\langle \w_{\bm{k},1} \right\vert \varrho_{\bm{k}}
\left\vert \w_{\bm{k},2} \right\rangle =
\sqrt{\frac{ \mathrm{Re} \, \mfa_k -  \mfc_k }{\pi}} \; \exp\left(-\frac{\mfa_k}{2}
  \, \w^2_{\bm{k},1}  -\frac{\mfa^*_k}{2}
 \, \w^2_{\bm{k},2} 
+ \mfc_k \, \w_{\bm{k},1} \w_{\bm{k},2} \right)\,,
\ee
for some choice of time-dependent functions $\mfa_k(\eta)$ and $\mfc_k(\eta)$. As written, this state is normalised to satisfy $\mathrm{Tr}(\varrho_{\bm{k}}) = 1$ and the requirement $\varrho_{\bm{k}}^\dagger = \varrho_{\bm{k}}$ further implies $\mfc_k$ is real. From this point of view the purpose of the Nakajima-Zwanzig equation is to determine the time-dependence of the functions $\mfa_k(\eta)$ and $\mfc_k(\eta)$. 

Because the state is Gaussian the coefficients $\mfa_k$ and $\mfc_k$ are completely determined by the equal-time two-point functions, through the formulae
\bea \label{app:Pcorrdefs}
&&\left\langle  \w_{\bm{k}} \w_{\bm{k}'}  \right\rangle
= P_{vv}(k) \, \delta_{\bm{k},\bm{k}'} \,, \qquad
 \left\langle  \p_{\bm{k}}  \p_{\bm{k}'}   \right\rangle
 = P_{pp}(k) \, \delta_{\bm{k}.\bm{k}'}\qquad \hbox{and} \qquad
 \left\langle \w_{\bm{k}} {\p}_{\bm{k}'}  \right\rangle
 =  \left[P_{v p}(k)+\frac{i}{2}\right] \, 
\delta_{\bm{k},\bm{k}'}  \,,
\eea
where the quantities $P_{vv}(k)$, $P_{vp}(k)$ and $P_{pp}(k)$ are given by
\be 
P_{vv}(k)=\frac{1}{2 \left[\mathrm{Re} \, \mfa_k -\mfc_k\right]}\, ,\quad
P_{vp}(k)=-\frac{\mathrm{Im} \,\mfa_k}
{2 \left[\mathrm{Re} \, \mfa_k -\mfc_k\right]} \quad\hbox{and} \quad
P_{pp}(k)=\frac{\left\vert \mfa_k\right\vert^2-\mfc_k^2}
{2 \left[\mathrm{Re} \, \mfa_k -\mfc_k\right]}\, .
\ee
In terms of these the original complex correlators are given by
\be
  \left \langle v_\bmk^\dagger v_\bmk \right\rangle = P_{vv}(k) + P_{vv}(-k) \,, \qquad
  \left \langle p_\bmk^\dagger p_\bmk \right\rangle = P_{pp}(k) + P_{pp}(-k) \,,
\ee
and so on. The state's purity is given in terms of these by \cite{Serafini:2003ke, Grain:2019vnq, Colas:2021llj,Martin:2021znx}
\be 
\label{app:purityk}
\gamma_{\bm{k}}(\eta) :=
\mathrm{Tr}\left[ {\varrho}^2_{\bm{k}} (\eta)\right]   = \sqrt{\frac{\mathrm{Re} \, \mfa_k -\mfc_k}{\mathrm{Re} \, \mfa_k +\mfc_k} } \,.
\ee
  
\subsection{Super-Hubble evolution}
\label{app:sec:markstrat}

We next compute the implications of the Nakajima-Zwanzig equation \pref{app:NZmodesrhok} for the evolution of $\varrho_{\bm{k}}(\eta)$, focussing on the super-Hubble regime $|k \eta| \ll 1$. We start by making the dependence on $\eta'$ of the right-hand side of the Nakajima-Zwanzig equation as explicit as possible. 

\subsubsection{Field evolution}
\label{app:subsec:convfieldvars}

Identifying the $\eta'$ dependence hidden in the field variables is a simple matter since we work in the interaction picture where the fields evolve through the free Hamiltonian \eqref{freeHclassical}. For scalar field this is captured by expanding the fields at $\eta'$ in terms of the creation and annihilation operators $c_{\bm{k}}$ using \pref{eq:scalarmodedecomp} and then re-expressing the $c_{\bm{k}}$'s in terms of the fields evaluated at $\eta$ (again using \pref{eq:scalarmodedecomp}) (see \cite{Burgess:2025dwm}). An identical argument goes through for the tensor modes, using \pref{eq:tensormodedecomp}.

For $\zeta_\bmk(\eta)$ and its canonical momentum $\mfp_\bmk(\eta)$ this leads to \cite{Burgess:2025dwm}
\begin{eqnarray}
\label{app:eq:etaprimeToeta}
\zeta_{-\bmk }(\eta^{\prime})  &=&  \cW_k (\eta^{\prime}, \eta) \, \zeta_{-\bmk }(\eta)+\cX_k (\eta^{\prime}, \eta) \, \mfp_{-\bmk }(\eta)\,,\nn\\
\mfp_{-\bmk }(\eta^{\prime})  &=&  \cZ_k (\eta^{\prime}, \eta) \, \zeta_{-\bmk }(\eta)+\cY_k (\eta^{\prime}, \eta) \, \mfp_{-\bmk }(\eta) \,,
\end{eqnarray}
with 
\begin{eqnarray}
\label{app:eq:etaprimeToeta2}
 \cW_k (\eta^{\prime}, \eta) &:=&   
  i\ \mfz_s^2(\eta) \Bigl[ \hat u_k^*(\eta^{\prime}) \hat u_k^{\prime}(\eta)- \hat u_k(\eta^{\prime}) \hat u_k^{* \prime}(\eta)\Bigr] \nn \\
 \cX_k (\eta^{\prime}, \eta) &:=&  
   i  \Bigl[ \hat u_k(\eta^{\prime}) \hat u_k^{*}(\eta)- \hat u_k^*(\eta^{\prime}) \hat u_k(\eta)\Bigr]\\
 \cY_k (\eta^{\prime}, \eta) &:=&   
   i\ \mfz_s^2(\eta')  \Bigl[ \hat u' _k(\eta^{\prime}) \hat u_k^{*}(\eta)-\hat u_k^{*\prime}(\eta^{\prime}) \hat u_k(\eta)\Bigr]\nn\\
 \cZ_k (\eta^{\prime}, \eta) &:=&
i\  \mfz_s^2(\eta) \,\mfz_s^2(\eta^{\prime}) \ \Bigl[ \hat u_k^{*\prime}(\eta^{\prime}) \hat u_k^{\prime}(\eta)- \hat u_k^{\prime}(\eta^{\prime}) \hat u_k^{*\prime}(\eta)\Bigr]\,.\nn
\end{eqnarray}

\subsubsection{Density matrix evolution}
\label{app:subsec:convreddens}

The $\eta^{\prime}$ dependence in $\varrho(\eta^{\prime})$ can be made explicit by Taylor expanding $\varrho(\eta^{\prime})$ around $\eta^{\prime}=\eta$,
\begin{equation} \label{app:TaylorRho}
\varrho(\eta^{\prime})=\sum_{n=0}^{\infty} \frac{\partial_{\eta}^n \varrho(\eta)}{n!} (\eta^{\prime}-\eta)^n \,,
\end{equation}
and we seek information about the size of these derivatives in order to know how many of these terms must be kept to any given order in the small expansion parameters.

This can seem at first sight to be overkill because the time derivatives of $\varrho$ on the right-hand side of equations like \pref{app:eq:TCL2a} are themselves proportional to further powers of the semiclassical expansion parameter $\lambda := (H/\Mp)$, and so in principle contribute at higher than leading order in perturbation theory. This overkill is deliberate, however, because at late time perturbation theory itself breaks down and we wish to determine whether other control parameters can step in on which calculations can rely. We find that some do, at least for super-Hubble modes.

\subsubsection{TCL2 evolution}

Using these in eq.~\pref{app:NZmodesrhok} leads to the `TCL2' result:\footnote{TCL2 denotes the form of the TCL equation at second-order in perturbation theory.} 
\bea
\label{app:eq:TCL2a}
\partial_{\eta} \varrho_{\bmk }(\eta) &=& -\sum_{n=0}^{\infty} \frac{1}{n!} \Biggl \{  \mathcal{J}^{\zeta \zeta}_{kn}(\eta;\eta_{\rm{in}})\Bigl[\zeta_{\bmk}(\eta),\ \zeta_{-\bmk}(\eta)\ \partial^n_{\eta}\varrho(\eta)\Bigr] + \mathcal{J}^{\zeta \mfp}_{kn}(\eta;\eta_{\rm{in}})\Bigl[\zeta_{\bmk}(\eta),\ \mfp_{-\bmk}(\eta)\ \partial^n_{\eta}\varrho(\eta)\Bigr] \nn\\
&&\; \vphantom{\frac12} +\mathcal{J}^{\mfp \zeta}_{k n}(\eta;\eta_{\in})\Bigl[\mfp_{\bmk}(\eta),\ \zeta_{-\bmk}(\eta)\ \partial^n_{\eta}\varrho(\eta)\Bigr] +\mathcal{J}^{\mfp \mfp}_{k n}(\eta;\eta_{\in})\Bigl[\mfp_{\bmk}(\eta),\ \mfp_{-\bmk}(\eta)\ \partial^n_{\eta}\varrho(\eta)\Bigr] +{\rm h.c.}\Biggr\},\nn\\
\eea
where  
\be 
\label{app:eq:pvkernels}
 \mathcal{J}^{\mfa \mfb}_{k n}(\eta;\eta_{\in}) =    \int_{\eta_{\in}}^{\eta} \exd\eta^{\prime}\   \, \mathcal{T}^{\mfa\mfb}_{k}(\eta^{\prime},\eta)   \, (\eta^{\prime}-\eta)^n \,,
\ee 
where $\mfa$ and $\mfb$ denote either $\zeta$ or $\mfp$.

The correlators appearing here are obtained by decomposing the system operators defined in \pref{app:HintRdef0} as linear combinations of $\zeta$, $\partial_i \zeta$ and its canonical momenta $\mfp$ and $\partial_i \mfp$: 
\be
   \cS_a = K_a^\zeta \zeta + K_a^{\zeta i} \partial_i \zeta + K^{\mfp}_a \mfp + K^{\mfp i}_a \, \partial_i \mfp \,,
\ee
and then defining 
\be
   \cE^\zeta = K_a^\zeta \cE^a \, \quad \cE^{\zeta i} = K_a^{\zeta i} \cE^a \,, \quad \cE^{\mfp} = K^{\mfp}_a \cE^a \quad \hbox{and} \quad
   \cE^{\mfp i} =  K^{\mfp i}_a \, \cE^a \,.
\ee

In terms of these
\begin{equation}
\label{app:eq:zetazeta}
\begin{split}
\mathcal{T}_k^{\zeta\zeta}(\eta,\eta^{\prime}) &=\cW_k(\eta,\eta^{\prime})\left\langle \left[\cE^{\zeta}+\cE^{\zeta i} \left(-i\hat{\bmk}_i k\right)\right]_{\eta} \left[\cE^{\zeta}+\cE^{\zeta i} \left(i\hat{\bmk}_i k\right)\right]_{\eta^{\prime}} \right\rangle\\
&\qquad\qquad + \cZ_k(\eta,\eta^{\prime}) \left\langle \left[\cE^{\zeta}+\cE^{\zeta i} \left(-i\hat{\bmk}_i k\right)\right]_{\eta} \left[ \cE^{\mfp}+\cE^{\mfp i} \left(-i\frac{\hat{\bmk}_i }{k}\right)\right]_{\eta^{\prime}} \right\rangle ,
\end{split}
\end{equation}
\begin{equation}
\label{app:eq:zetap}
\begin{split}
\mathcal{T}_k^{\zeta\mfp}(\eta,\eta^{\prime}) &=\cX_k(\eta,\eta^{\prime})\left\langle \left[\cE^{\zeta}+\cE^{\zeta i} \left(-i\hat{\bmk}_i k\right)\right]_{\eta} \left[\cE^{\zeta}+\cE^{\zeta i} \left(i\hat{\bmk}_i k\right)\right]_{\eta^{\prime}} \right\rangle\\
&\qquad\qquad + \cY_k(\eta,\eta^{\prime}) \left\langle \left[ \cE^{\zeta}+\cE^{\zeta i} \left(-i\hat{\bmk}_i k\right)\right]_{\eta} \left[\cE^{\mfp}+\cE^{\mfp i} \left(-i\frac{\hat{\bmk}_i }{k}\right)\right]_{\eta^{\prime}} \right\rangle ,
\end{split}
\end{equation}
\begin{equation}
\label{app:eq:pzeta}
\begin{split}
\mathcal{T}_k^{\mfp\zeta}(\eta,\eta^{\prime}) &=\cW_k(\eta,\eta^{\prime})\left\langle \left[\cE^{\mfp}+\cE^{\mfp i} \left(i\frac{\hat{\bmk}_i }{k}\right)\right]_{\eta}  \left[\cE^{\zeta}+\cE^{\zeta i} \left(i\hat{\bmk}_i k\right)\right]_{\eta^{\prime}} \right\rangle\\
&\qquad\qquad + \cZ_k(\eta,\eta^{\prime}) \left\langle\left[\cE^{\mfp}+\cE^{\mfp i} \left(i\frac{\hat{\bmk}_i }{k}\right)\right]_{\eta}\left[\cE^{\mfp}+\cE^{\mfp i} \left(-i\frac{\hat{\bmk}_i }{k}\right)\right]_{\eta^{\prime}} \right\rangle ,
\end{split}
\end{equation}
and
\begin{equation}
\label{app:eq:pp}
\begin{split}
\mathcal{T}^{\mfp\mfp}_k(\eta,\eta^{\prime}) &= \cY_k(\eta,\eta^{\prime}) \left\langle \left[\cE^{\mfp}+\cE^{\mfp i} \left(i\frac{\hat{\bmk}_i }{k}\right)\right]_{\eta} \left[\cE^{\mfp}+\cE^{\mfp i} \left(-i\frac{\hat{\bmk}_i }{k}\right)\right]_{\eta^{\prime}} \right\rangle\\
&\qquad\qquad + \cX_k(\eta,\eta^{\prime}) \left\langle \left[\cE^{\mfp}+\cE^{\mfp i} \left(i\frac{\hat{\bmk}_i }{k}\right)\right]_{\eta} \left[\cE^{\zeta}+\cE^{\zeta i} \left(i\hat{\bmk}_i k\right)\right]_{\eta^{\prime}}  \right\rangle.
\end{split}
\end{equation}
 
Ref.~\cite{Burgess:2025dwm} shows that for scalar environments in the superhorizon limit -- where $z=-k\eta\ll1$ -- we can neglect the effects of the kernels $\mathcal{J}^{rs}_{k n}(\eta;\eta_{\rm{in}})$ for all $r$ and $s$ and for all $n\geq 1$ because these are all suppressed by at least one power of $z$ compared to the leading result $\mathcal{J}^{rs}_{k \, 0}(\eta;\eta_{\rm{in}})$. Ref.~\cite{Burgess:2025dwm} also finds that $\mathcal{J}^{\zeta\zeta}_{k \, 0}(\eta;\eta_{\rm{in}})$ is less suppressed than the other three $\mathcal{J}^{rs}_{k \, 0}(\eta;\eta_{\rm{in}})$ by at least one power of $z$. The leading forms for all four of the $\mathcal{J}^{rs}_{k \, 0}(\eta;\eta_{\rm{in}})$ turn out to compete in the predictions for the purity evolution.

This leaves the four integrals $\cJ^{\zeta\zeta}_{k0}$, $\cJ^{\zeta\mfp}_{k0}$, $\cJ^{\mfp\zeta}_{k0}$ and $\cJ^{\mfp\mfp}_{k0}$ as the dominant coefficients  respectively multiplying the terms $\left[\zeta_{\bmk}(\eta), \zeta_{-\bmk}(\eta)\varrho(\eta)\right]$, $\left[\zeta_{\bmk}(\eta), \mfp_{-\bmk}(\eta)\varrho(\eta)\right]$, $\left[\mfp_{\bmk}(\eta), \zeta_{-\bmk}(\eta) \varrho(\eta)\right]$ and $\left[\mfp_{\bmk}(\eta), \mfp_{-\bmk}(\eta)\varrho(\eta)\right]$ in the Nakajima-Zwanzig equation \pref{app:eq:TCL2a}.

\subsubsection{Time-local evolution}

The leading evolution for super-Hubble modes predicted by \pref{app:eq:TCL2a} turns out to be time-local for super-Hubble modes at late times. This happens for {\it two} reasons. First, at leading order in semiclassical perturbation theory the evolution is time-local since stopping at leading order allows the neglect of all but the $n = 0$ terms in \pref{app:eq:TCL2a}. Second, evolution remains time-local (for super-Hubble modes) even once secular effects make perturbation theory alone break down at late times because small $z = - k \eta$ then protects it.

The dependence on the small parameters of the problem is most clearly seen by rescaling
\be \label{app:OiandZdefs}
 \mathcal{O}_{1} := Z^{-1} \zeta \qquad \hbox{and} \qquad
 \mathcal{O}_{2} := Z \mfp \qquad \hbox{with} \qquad
 Z := \frac{1}{\sqrt{2\slrl k^3}} \left( \frac{H}{\Mp}\right) \,,
\ee 
and defining:
\begin{equation}
\label{app:curlyJdimless}
\begin{aligned}[c]
\widetilde{\mathcal{J}}^{\zeta\zeta}_{k0}(z)
&:=
\lim_{\eta_{\mathrm{in}}\to-\infty}
\frac{1}{\epsilon_1 k^4}
\mathcal{J}^{\zeta\zeta}_{k0}(\eta,\eta_{\mathrm{in}})
\\
\widetilde{\mathcal{J}}^{\mfp\zeta}_{k0}(z)
&:=
\lim_{\eta_{\mathrm{in}}\to-\infty}
\frac{M_p^2}{\slrl k H^2}
\mathcal{J}^{\mfp\zeta}_{k0}(\eta,\eta_{\mathrm{in}})
\end{aligned}
\qquad\qquad
\begin{aligned}[c]
\widetilde{\mathcal{J}}^{\zeta\mfp}_{k0}(z)
&:=
\lim_{\eta_{\mathrm{in}}\to-\infty}
\frac{M_p^2}{k H^2}
\mathcal{J}^{\zeta\mfp}_{k0}(\eta,\eta_{\mathrm{in}})
\\
\widetilde{\mathcal{J}}^{\mfp\mfp}_{k0}(z)
&:=
\lim_{\eta_{\mathrm{in}}\to-\infty}
\frac{k^2 M_p^4}{H^4}
\mathcal{J}^{\mfp\mfp}_{k0}(\eta,\eta_{\mathrm{in}})
\end{aligned}
\end{equation}
In terms of these the leading contributions to eq~\pref{app:eq:TCL2a} take the Lindblad{\it-like} form \cite{Burgess:2025dwm}
\begin{equation}
\label{app:eq:lindbladlikeform}
\partial_{\eta} \varrho_{\bmk} = -i\Bigl[ \scrH_{\rm eff}(\eta), \varrho_{\bmk}(\eta)\Bigr] +\sum_{r,s=1}^2 h^{s r}_\bmk\left[\mathcal{O}_{\bmk,s}\varrho_{\bmk}\mathcal{O}^{\dag}_{\bmk,r}-\frac{1}{2}\left\{\mathcal{O}^{\dag}_{\bmk,r}\mathcal{O}_{\bmk,s},\varrho_{\bmk}\right\}\right] \,,
\end{equation}
where $\scrH_{\rm eff}$ contains both the environmental average of the interaction Hamiltonian and the parts of the second-order contributions that have Liouville form ({\it i.e.}~the commutator of something with $\varrho$), given by
\begin{eqnarray}
\scrH_{\rm eff} = \mathrm{Im}[ \mathcal{J}^{\zeta \zeta}_{k0}] \zeta^2 - \mathrm{Im}[\mathcal{J}^{\zeta \mfp}_{k0} + \mathcal{J}^{\mfp \zeta}_{k0}]  \tfrac{1}{2} \{\zeta,\mfp \} + \mathrm{Im}[\mathcal{J}^{\mfp \mfp}_{k0}] \mfp^2 \, .
\end{eqnarray} 
Because $\scrH_{\rm eff}$ contributes only to Hamiltonian evolution it plays no role in the decoherence calculations described in the main text.

The matrix of couplings appearing in \pref{app:eq:lindbladlikeform} is given explicitly by
\begin{equation}
\label{app:eq:decohmatrixh}
\left(\begin{array}{cc}h_\bmk^{11}  & h_\bmk^{12}  \\ h_\bmk^{21}  & h_\bmk^{22}  \end{array}\right)  = k\left(\frac{H}{\Mp}\right)^2\left(\begin{array}{cc}{ \rm Re}[\widetilde{\cJ}^{\zeta\zeta}_{k0}] & \widetilde{\cJ}^{\zeta\mfp *}_{k0} + \slrl \widetilde{\cJ}^{\mfp\zeta}_{k0} \\ \widetilde{\cJ}^{\zeta\mfp}_{k0} + \slrl \widetilde{\cJ}^{\mfp\zeta *}_{k0} & 4 \slrl {\rm Re}[\widetilde{\cJ}^{\mfp\mfp}_{k0}] \end{array}\right) \,.
\end{equation}
The eigenvalues of this matrix are not positive in the limit $\slrl \to 0$ but this need not be a worry for the reasons described in footnote 2 above eq.~\pref{CouplingMatrix}.

\section{Evaluating the integrals}
\label{sec:Integrals}

Here we describe in some detail how the integrals relevant to the computation of the Lindblad-like coeffecients are calculated.

There are a number of steps we should take before undertaking the evaluation of both the momentum as well as the time integrals involved in the construction of the various Lindblad-like coefficients appearing in the evolution of the reduced density matrix. Starting from equation eq.\eqref{eq:Lindblad}, we can write the generic Lindblad-like coefficient in the following form:
\be
\label{app:eq:genericLcoeff}
\mathcal{L}_k(\eta)=-\int_{\eta_{\rm in}}^{\eta} \exd \eta^{\prime}\ \hat{G}(\eta) \hat{G}(\eta^{\prime})\ \mathcal{Q}_k(\eta,\eta^{\prime}),
\ee
with
\bea
\label{app:eq:Lcoeffintegrand}
\mathcal{Q}_k(\eta,\eta^{\prime})&=& \int_{q_i>k_{UV}} \prod_{i=1}^2 \frac{\exd^3 q_i}{(2\pi)^3 q_i^3} (2\pi)^3 \delta^{(3)}\big(\vec{k}-\vec{q}_1-\vec{q}_2\big) e^{-i (q_1+q_2)(\eta-\eta^{\prime})} \mathcal{F}(q_1,q_2,\eta,\eta^{\prime}) \mathcal{K}(\eta,\eta^{\prime}),\nonumber\\
 \mathcal{K}(\eta,\eta^{\prime}) &\equiv& \mathcal{K}_+(\eta,\eta^{\prime})e^{i k(\eta-\eta^{\prime})}+\mathcal{K}_-(\eta,\eta^{\prime})e^{-i k(\eta-\eta^{\prime})},
\eea
The delta function comes from projecting out the coefficient corresponding to the wavenumber of interest $\vec{k}$, while the $q_i^3$ denominators, the exponential $e^{-i (q_1+q_2)(\eta-\eta^{\prime})}$ as well as $\mathcal{F}(q_1,q_2,\eta,\eta^{\prime})$ come from the $\zeta$ of $\gamma$ modes (together with whatever ``decoration'' they may have in terms of spatial or temporal derivatives). Finally, $\mathcal{K}(\eta,\eta^{\prime})$ is what we might call the time translation kernel; it takes $\zeta_{\vec{k}}(\eta^{\prime}),\mathfrak{p}_{\vec{k}}(\eta^{\prime})$ into the relevant linear combination of $\zeta_{\vec{k}}(\eta),\mathfrak{p}_{\vec{k}}(\eta)$, which we need to do in order to arrive at the TCL2 form of the Nakajima-Zwanzig equation eq.~\eqref{app:eq:TCL2a}. These are listed in terms of the modes in eqs.~\eqref{app:eq:etaprimeToeta} \& \eqref{app:eq:etaprimeToeta2}. 

\subsection{Environmental Operators}
\label{subsec;envops}

We start by listing the form of the environmental operators we need. We start with the cubic action interaction Hamiltonians:
\bea
\label{app:eq:interactionHs}
H_{\rm int}^{\gamma \zeta \zeta} &=& -\slrl \Mp^2 a^2(\eta)\int \exd^3 x\ \gamma_{i j}\partial_i \zeta \partial_j \zeta\nonumber\\
H_{\rm int}^{\zeta \gamma \gamma} &=& -\frac{\slrl \Mp^2 a^2(\eta)}{8}\int \exd^3 x\ \left\{\zeta\left( \gamma^{\prime}_{i j}\gamma^{\prime}_{i j}+\partial_l \gamma_{i j}\partial_l \gamma_{i j}\right)-2 \gamma_{i j}^{\prime} \partial_i\gamma_{i j} \partial_i \partial^{-2}\zeta^{\prime}\right\}\nonumber\\
H_{\rm int}^{\zeta \zeta \zeta} &=& -\slrl^2 \Mp^2 a^2(\eta) \int \exd^3 x\ \left\{ \zeta \left( \zeta^{\prime 2}+\left(\partial \zeta \right)^2\right) -2\zeta^{\prime} \partial_i \zeta \partial_i \partial^{-2}\zeta^{\prime} \right\}.
\eea
Going through our procedure, separating each field into its system and environment pieces and then using conservation of momentum to only keep the terms with one system field and two environmental ones leads us to the following set of environmental operators together with their couplings
\begin{equation} \label{app:eq:envops}
\begin{aligned}[c]
\cE^{\zeta i}_{\gamma \zeta \zeta}
&= 2 \gamma_{i j}\partial_j \zeta,\\
\cE^{\zeta}_{\zeta \gamma \gamma}
&= \gamma^{\prime}_{i j}\gamma^{\prime}_{i j}
+\partial_l \gamma_{i j}\partial_l \gamma_{i j},\\
\cE^{\mfp i}_{\zeta \gamma \gamma}
&= -2 \gamma^{\prime}_{r s}\partial_i \gamma_{r s},\\
\cE^{\zeta}_{\zeta \zeta \zeta}
&= \zeta^{\prime 2}
+\left(\partial\zeta\right)^2,\\
\cE^{\zeta i}_{\zeta \zeta \zeta}
&= 2\left(
\zeta\partial_i\zeta
-\zeta^{\prime}\partial_i\partial^{-2}\zeta^{\prime}
\right),\\
\cE^{\mfp}_{\zeta \zeta \zeta}
&= 2\left(
\zeta\zeta^{\prime}
-\partial_i\zeta\,\partial_i\partial^{-2}\zeta^{\prime}
\right),\\
\cE^{\mfp i}_{\zeta \zeta \zeta}
&= -2\zeta^{\prime}\partial_i\zeta,
\end{aligned}
\qquad\qquad
\begin{aligned}[c]
\hat{G}^{\zeta i}_{\gamma \zeta \zeta}
&= -\slrl \Mp^2 a^2(\eta),\\
\hat{G}^{\zeta}_{\zeta \gamma \gamma}
&= -\tfrac{1}{8}\slrl \Mp^2 a^2(\eta),\\
\hat{G}^{\mfp i}_{\zeta \gamma \gamma}
&= -\tfrac{1}{8}\slrl \Mp^2 a^2(\eta),\\
\hat{G}^{\zeta}_{\zeta \zeta \zeta}
&= -\slrl^2 \Mp^2 a^2(\eta),\\
\hat{G}^{\zeta i}_{\zeta \zeta \zeta}
&= -\slrl^2 \Mp^2 a^2(\eta),\\
\hat{G}^{\mfp}_{\zeta \zeta \zeta}
&= -\slrl^2 \Mp^2 a^2(\eta),\\
\hat{G}^{\mfp i}_{\zeta \zeta \zeta}
&= -\slrl^2 \Mp^2 a^2(\eta).
\end{aligned}
\end{equation}
Since the environmental assumed to be the Bunch-Davies state for the purposes of the Nakajima-Zwazig equation, we can use Wick's theorem in order to calculate correlation functions of the environmental operators.

\subsection{Scaling}
\label{subsec:scaling}

Our goal in this subsection is to isolate the dependence of each of the Linbdblad coeffcients on the slow-roll parameter $\slrl$, the wavenumber $k$  as well as $H^2\slash \Mp^2$. It is easiest to do this for each coefficient separately, as they each have their own foibles.

There are some common bits of information that can be used for all of the coefficients. First, we will scale $k$ both out of the conformal times as well as the momenta, setting $z=-k\eta,\ z^{\prime}=-k\eta^{\prime}$ and $r_i=q_i\slash k$. Next, since all the couplings contain $a^2(\eta)$ or  $a^2(\eta^{\prime})$, when we rewrite these in terms of $z$, the product of couplings $\hat{G}(\eta) \hat{G}(\eta^{\prime})$ will always yield a factor proportional to $ k^4\slash (z^2 z^{\prime 2})$. More factors of $k$ appear from $d\eta^{\prime}=dz^{\prime}\slash k$ as well as from the momentum delta function: $\delta^{(3)}(\vec{k}-\vec{q}_1-\vec{q}_2)\rightarrow k^{-3} \delta^{(3)}(\hat{k}-\vec{r}_1-\vec{r}_2)$. 
Note that  $d^3 q_i\slash q_i^3$ is invariant under this scaling. We see that the $k^{-4}$ coming from the time integral and the delta function exactly cancels the $k^4$ coming from the couplings. What this means is that the overall power of $k$ appearing in each Lindblad-like coefficient comes from the scaling dimensions of the environmental operators involved as well as from the time translation kernels involved and/or the factors of $\mfz^2$ that might appear. 

The environmental operator mass dimensions are given by:
\be
\label{app:eq:envmassdims}
\left[ \cE^\zeta\right]=2,\ \left[ \cE^{\zeta i} \right]=1,\ \left[ \cE^{\mfp}  \right]=1,\  \left[ \cE^{\mfp i}  \right]=2
\ee

We can write the time translation kernels in dimensionless form as:
\begin{equation}
\label{app:eq:dimlesskerns}
\begin{aligned}[c]
\cW_k(\eta,\eta^{\prime})
&=\frac{\tilde{\cW}(z,z^{\prime})}{z^2}\\
\cX_k(\eta,\eta^{\prime})
&=\frac{1}{2\slrl}\frac{H^2}{\Mp^2}\frac{1}{k^3}
\tilde{\cX}(z,z^{\prime})
\end{aligned}
\qquad\qquad
\begin{aligned}[c]
\frac{\cY_k(\eta,\eta^{\prime})}{\mfz_s^2(\eta')}
&=\frac{1}{2\slrl}\frac{H^2}{\Mp^2}\frac{1}{k^2}
\tilde{\cY}(z,z^{\prime})\\
\frac{\cZ_k(\eta,\eta^{\prime})}{\mfz_s^2(\eta')}
&=\frac{k\,\tilde{\cZ}(z,z^{\prime})}{z^2}
\end{aligned}
\end{equation}
with
\bea
\label{app:eq:dimlesskerns2}
\tilde{\cW}(z,z^{\prime})&=&\frac{z}{2}\left( \left(i+z^{\prime}\right) e^{i (z^{\prime}-z)}+\left(-i+z^{\prime}\right) e^{-i (z^{\prime}-z)}\right)\nonumber\\
\tilde{\cX}(z,z^{\prime})&=&\frac{1}{2}\left( \left(1+i z\right) \left(i+z^{\prime}\right) e^{i (z^{\prime}-z)}+\left(1-i z\right) \left(-i+z^{\prime}\right) e^{-i (z^{\prime}-z)}\right)\nonumber\\
\tilde{\mathcal{Y}}(z,z^{\prime})&=&\frac{z^{\prime}}{2}\left( \left(-i+z\right) e^{i (z^{\prime}-z)}+\left(i+ z\right) e^{-i (z^{\prime}-z)}\right)\nonumber\\
\tilde{\cZ}(z,z^{\prime})&=& \frac{i}{2} z z^{\prime} \left(e^{i (z^{\prime}-z)}-e^{-i (z^{\prime}-z)}\right).
\eea

When we consider the scaling with $H\slash\Mp$, we note the product of couplings $\hat{G}(\eta) \hat{G}(\eta^{\prime})$ is proportional to $\left(H\slash \Mp\right)^{-4}$. But the quantity $\mathcal{F}(q_1,q_2,\eta,\eta^{\prime})$ in eq.~\eqref{app:eq:Lcoeffintegrand} contains four modes, each of which contributes a factor of $H\slash \Mp$, thus canceling what comes from the couplings. The scaling with $H\slash\Mp$ is then dictated by the power appearing in the time translation kernels in eq.~\eqref{app:eq:dimlesskerns}.

The scaling with $\slrl$ has to be done on a case by case basis, since it depends on the scaling in the couplings, in the modes and in the time translation kernels. We can write the scaling as $\slrl^{p_{\rm sr}}$, with
\be
\label{app:slowrollscaling}
p_{\rm sr} = c_{\eta}+c_{\eta^{\prime}}-n_{\zeta}/2-n_{\cX}-n_{\cY}
\ee
where $c_{\eta},\ c_{\eta^{\prime}}$ are the powers of $\slrl$ appearing in the relevant coupling $\hat{G}(\eta),\ \hat{G}(\eta^{\prime})$, respectively, $n_{\zeta}$ is the number of $\zeta$ modes appearing in the expression and $n_{\cX},\ n_{\cY}$ equals $1$ if the time translation kernel is $\cX,\ \cY$, $0$ otherwise. Since we have to match the number of $\zeta$ and $\gamma$ fields in $H_{\rm int}(\eta)$ and $H_{\rm int}(\eta^{\prime})$ to arrive at a non-zero Wick contraction, $c_{\eta}=c_{\eta^{\prime}}\equiv c$, and $p_{\rm sr} = 2 c-n_{\zeta}/2-n_{\cX}-n_{\cY}$.  

Finally, we note that each $\zeta$ mode contributes a factor of $1\slash 2$ to the correlation function while the $\gamma$ modes feed in a factor of $\sqrt{2}$ each. 

\subsubsection{$\mathcal{J}^{\zeta \zeta}_{kn}(\eta;\eta_{\rm{in}})$}
\label{subsub:zetazeta} 
The lowest order contribution in slow-roll to $\mathcal{J}^{\zeta \zeta}_{kn}(\eta;\eta_{\rm{in}})$ comes from 
\be
\label{app:eq:zetagammaenvop}
\hat{G}^{\zeta i}_{\gamma \zeta \zeta}(\eta)\hat{G}^{\zeta i}_{\gamma \zeta \zeta}(\eta^{\prime})\left\langle \cE^{\zeta i}_{\gamma \zeta \zeta} \cE^{\zeta j}_{\gamma \zeta \zeta} \right\rangle=2 \slrl \frac{k^2}{z^2 z^{\prime 2}}\  \left\langle \gamma_{i p}(\eta)\ \gamma_{j q}(\eta^{\prime}) \right\rangle \left\langle \partial_p \zeta(\eta)\ \partial_q \zeta(\eta^{\prime })\right\rangle.
\ee
We used our scaling rules in arriving at the above expression. If we now use the mode decomposition for each of the fields and insert these into eq.~\eqref{app:eq:zetazeta} we can then write:
\be
\label{app:eq:zetazetascaledLcoeff}
\mathcal{J}^{\zeta \zeta}_{k0}(\eta;\eta_{\rm{in}})=\slrl k^4\left( \widetilde{\mathcal{J}}^{\zeta \zeta}_{k0}(z; z_{in})+\mathcal{O}(\slrl)\right)
\ee
with $z_{in}=-k\eta_{\rm{in}}$ and 
\bea
\widetilde{\mathcal{J}}^{\zeta \zeta}_{k0}(z; z_{in})&=&\int_{z(1-i \varepsilon)}^{z_{in}(1-i \varepsilon)} \exd z^{\prime}\ \tilde{\mathcal{T}}^{\zeta \zeta}_k(z,z^{\prime}) \label{app:eq:zzmomint} \\
\tilde{\mathcal{T}}^{\zeta \zeta}_k(z,z^{\prime}) &=& \int_{r_i>\kappa} \prod_{i=1}^2 \frac{\exd^3 r_i}{(2\pi)^3 r_i^3} (2\pi)^3 \delta^{(3)}\left(\hat{k}-\vec{r}_1-\vec{r}_2\right) \frac{e^{i (r_1+r_2)(z-z^{\prime})} }{z^4 z^{\prime 2}}\tilde{\mathcal{F}}(r_1,r_2,z,z^{\prime}) \tilde{\cW}(z,z^{\prime})\nonumber 
\eea
where
\be
\tilde{\mathcal{F}}(r_1,r_2,z,z^{\prime})=\hat{k}_i \hat{k}_j r_{1 p} r_{1 q}  \Pi^{ipjq}(\hat{r}_2)\prod_{a=1,2}(1-i r_a z)(1+i r_a z^{\prime}),
\ee
and as always, $\hat{k}$ is the unit vector along $\vec{k}$. Note that we have already implemented the deformation of the $z^{\prime}$ integration contour to pick out the Bunch-Davies vacuum at early times and that we used eq.~\eqref{app:eq:dimlesskerns2} to replace $\cW_k(\eta,\eta^{\prime})\rightarrow \tilde{\cW}(z,z^{\prime})\slash z^2$, which leads to the factor of $z^4$ in the denominator of eq.~\eqref {app:eq:zzmomint}.

We can simplify the expression for $\tilde{\mathcal{F}}(r_1,r_2,z,z^{\prime})$ by noting two facts. The first is that due to the symmetry of $\hat{k}_i \hat{k}_j r_{1 p} r_{1 q}$ under $i\leftrightarrow j$ and  $p\leftrightarrow q$ we can replace $\Pi^{ipjq}(\hat{r}_2)\rightarrow \Pi^{i j}(\hat{r}_2) \Pi^{p q}(\hat{r}_2)$. Next, thanks to momentum conservation, we can replace $\vec{r}_1\rightarrow \hat{k}-\vec{r}_2$ and then use the fact that $\Pi^{i j}(\hat{r}_2)$ annihilates $\vec{r}_2$ to finally replace
\be
\hat{k}_i \hat{k}_j r_{1 p} r_{1 q}  \Pi^{ipjq}(\hat{r}_2)\rightarrow \hat{k}_i \hat{k}_j \hat{k}_p \hat{k}_q  \Pi^{i j}(\hat{r}_2) \Pi^{p q}(\hat{r}_2)=(1-\mu_2^2)^2
\ee
where $\mu_2 \equiv \hat{k}\cdot \hat{r}_2$. 

We note that the higher order contributions to $\tilde{\mathcal{J}}^{\zeta \zeta}_{k0}(z; z_{in})$ come from the use of $H_{\rm int}^{\zeta \gamma \gamma},\ H_{\rm int}^{\zeta \zeta \zeta}$. 

\subsubsection{$\mathcal{J}^{\zeta \mathfrak{p}}_{kn}(\eta;\eta_{\rm{in}})$}
\label{subsub:zetap}

We can use our result above to quickly generate the leading contribution to $\mathcal{J}^{\zeta \mathfrak{p}}_{kn}(\eta;\eta_{\rm{in}})$. All that needs to be done is to replace $\cW_k(\eta,\eta^{\prime})$ by $\cX_k(\eta,\eta^{\prime})$, so that $n_{\cX}=1$. Writing these in terms of the dimensionless quantities gives
\be
\label{app:eq:zetapscaledLcoeff}
 \mathcal{J}^{\zeta \mathfrak{p}}_{k0}(\eta;\eta_{\rm{in}})=k \left(\frac{H}{\Mp}\right)^2\left( \widetilde{\mathcal{J}}^{\zeta \mathfrak{p}}_{k0}(z; z_{in})+\mathcal{O}(\slrl)\right)
 \ee
 with
 \bea
 \label{app:eq:zpmomint}
\widetilde{\mathcal{J}}^{\zeta \mathfrak{p}}_{k0}(z; z_{in})&=&\int_{z(1-i \varepsilon)}^{z_{in}(1-i \varepsilon)} \exd z^{\prime}\ \tilde{\mathcal{T}}^{\zeta \mathfrak{p}}_k(z,z^{\prime})\\\
\tilde{\mathcal{T}}^{\zeta \mathfrak{p}}_k(z,z^{\prime}) &=& \frac{1}{2}\int_{r_i>\kappa} \prod_{i=1}^2 \frac{\exd^3 r_i}{(2\pi)^3 r_i^3} (2\pi)^3 \delta^{(3)}\left(\hat{k}-\vec{r}_1-\vec{r}_2\right) \, \frac{e^{i (r_1+r_2)(z-z^{\prime})}}{z^2 z^{\prime 2}} \tilde{\mathcal{F}}(r_1,r_2,z,z^{\prime}) \tilde{\mathcal{X}}(z,z^{\prime}).\nonumber
\eea
 Note that the same quantity $\tilde{\mathcal{F}}$ appears both for $\mathcal{J}^{\zeta \mathfrak{p}}_{k0}(\eta;\eta_{\rm{in}})$ as for $\mathcal{J}^{\zeta \zeta}_{k0}(\eta;\eta_{\rm{in}})$. 
 
\subsubsection{$\mathcal{J}^{\mathfrak{p}\zeta }_{k0}(\eta;\eta_{\rm{in}})$}
\label{subsub:pzeta}

 Repeating the by now familiar scaling arguments, we see that we can write
\be
\mathcal{J}^{\mathfrak{p}\zeta }_{k0}(\eta;\eta_{\rm{in}})=\slrl k\left(\frac{H^2}{\Mp^2}\right) \tilde{\mathcal{J}}^{\mathfrak{p}\zeta }_{k0}(z;z_{in})
\ee 
where we have used the fact that in eq.~\eqref{app:eq:pzeta} a factor of $\mfz^{-2}(\eta)$ appears, which gives an overall factor of $(z^2 H^2)\slash (2\slrl k^2\Mp^2)$. Notice that while $H^{\gamma \zeta \zeta}_{\rm int}$ does not contribute to $\tilde{\mathcal{J}}^{\mathfrak{p}\zeta }_{k0}(z;z_{in})$, both $H^{\zeta \zeta \zeta}_{\rm int}$ and $H^{ \zeta \gamma\gamma}_{\rm int}$ do contribute and at the same order in $\slrl$. We write
\be
\tilde{\mathcal{J}}^{ \mathfrak{p}\zeta}_{k0}(z; z_{in})=\int_{z(1-i \varepsilon)}^{z_{in}(1-i \varepsilon)} \exd z^{\prime}\ \left(\tilde{\mathcal{T}}^{ \mathfrak{p}\zeta}_{k \zeta \gamma \gamma}(z,z^{\prime})+\tilde{\mathcal{T}}^{ \mathfrak{p}\zeta}_{k \zeta \zeta \zeta}(z,z^{\prime})\right)
\ee
where we separate the contributions due to each operator. Within each of $\tilde{\mathcal{T}}^{ \mathfrak{p}\zeta}_{k \zeta \zeta \zeta}\ \tilde{\mathcal{T}}^{ \mathfrak{p}\zeta}_{k \zeta \gamma \gamma}$ we further separate the contributions proportional to $\tilde{\cW},\ \tilde{\cZ}$ (which we denote as the direct (D) and indirect (I) pieces respectively, corresponding to whether or not they come directly from an operator proportional to the system $\zeta_S(\eta^{\prime})$ or had to be converted from a system $\mathfrak{p}_S(\eta^{\prime})$):
\begin{eqnarray}\label{app:eq:pzmomint}
\tilde{\mathcal{T}}^{ \mathfrak{p}\zeta}_{k X}(z,z^{\prime}) &=& \int_{r_i>\kappa} \prod_{i=1}^2 \frac{\exd^3 r_i}{(2\pi)^3 r_i^3} (2\pi)^3 \delta^{(3)}\left(\hat{k}-\vec{r}_1-\vec{r}_2\right) \, \frac{e^{i (r_1+r_2)(z-z^{\prime})}}{z^2 z^{\prime 2}}  \nn\\
&& \qquad \times \left(\tilde{\mathcal{F}}_{k X}^D(r_1,r_2,z,z^{\prime}) \tilde{\cW}(z,z^{\prime})+\tilde{\mathcal{F}}_{k X}^I(r_1,r_2,z,z^{\prime}) \tilde{\cZ}(z,z^{\prime})\right)
\end{eqnarray}
where $X=\zeta \gamma \gamma$ or $\zeta \zeta \zeta$. The direct correlation functions that have to be computed are:
\bea
\tilde{\mathcal{F}}_{k \zeta \gamma \gamma}^D(r_1,r_2,z,z^{\prime}) &=& i\hat{\bmk}_i \left\langle \cE^{\mfp i}_{\zeta\gamma\gamma}(\eta)\ \cE^{\zeta}_{\zeta\gamma\gamma}(\eta^{\prime})\right\rangle\nonumber\\
\tilde{\mathcal{F}}_{k \zeta \zeta \zeta}^D(r_1,r_2,z,z^{\prime})&=&\left\langle \left[\cE^{\mfp}_{\zeta \zeta \zeta}+\cE^{\mfp i}_{\zeta \zeta \zeta} \left(i\frac{\hat{\bmk}_i }{k}\right)\right]_{\eta}  \left[\cE^{\zeta}_{\zeta \zeta \zeta}+\cE^{\zeta j}_{\zeta \zeta \zeta} \left(i\hat{\bmk}_j k\right)\right]_{\eta^{\prime}} \right\rangle,
\eea
while the indirect ones are:
\bea
\tilde{\mathcal{F}}_{k \zeta \gamma \gamma}^I(r_1,r_2,z,z^{\prime}) &=& \hat{\bmk}_i  \hat{\bmk}_j\left\langle \cE^{\mfp i}_{\zeta\gamma\gamma}(\eta)\ \cE^{\mfp j}_{\zeta\gamma\gamma}(\eta)\right\rangle\nonumber\\
\tilde{\mathcal{F}}_{k \zeta \zeta \zeta}^I(r_1,r_2,z,z^{\prime})&=&\left\langle \left[\cE^{\mfp}_{\zeta \zeta \zeta}+\cE^{\mfp i}_{\zeta \zeta \zeta} \left(i\frac{\hat{\bmk}_i }{k}\right)\right]_{\eta}  \left[\cE^{\mfp}_{\zeta \zeta \zeta}+\cE^{\mfp j}_{\zeta \zeta \zeta} \left(-i\frac{\hat{\bmk}_j }{k}\right)\right]_{\eta^{\prime}} \right\rangle,
\eea

Computing the correlation functions, we have
\bea
\label{app:eq:gzzDirectIndirect}
\tilde{\mathcal{F}}_{k \zeta \gamma \gamma}^D(r_1,r_2,z,z^{\prime}) &=&\frac{t}{32}\left(r_1 r_2\right)^2 (\hat{k}\cdot\vec{r}_2) z (1-i r_2 z) \left[r_1^2 z^{\prime 2}-\frac{\vec{r}_1\cdot\vec{r}_2}{r_2^2}(1+i r_1 z^{\prime})(1+i r_2 z^{\prime})\right]\\
\tilde{\mathcal{F}}_{k \zeta \gamma \gamma}^I(r_1,r_2,z,z^{\prime}) &=&\frac{t}{32}\left(r_1 r_2\right)^2 (\hat{k}\cdot\vec{r}_2) z z^{\prime}(1-i r_2 z)\left[\left(\hat{k}\cdot\vec{r}_1\right) (1+i r_1 z^{\prime})+\left(\hat{k}\cdot\vec{r}_2\right) \frac{r_1^2}{r_2^2} (1+i r_2 z^{\prime})\right].\nonumber
\eea
with $t\equiv  \Pi^{ipjq}(\hat{r}_1) \Pi^{ipjq}(\hat{r}_2)=1+6 \left(\hat{r}_1\cdot\hat{r}_2\right)^2+ \left(\hat{r}_1\cdot\hat{r}_2\right)^4$.

The direct contributions from the cubic scalar interactions are
\bea
\label{app:eq:zzzDirect}
\tilde{\mathcal{F}}_{k \zeta \zeta \zeta}^{D}(r_1,r_2,z,z^{\prime})&=&  \frac{(r_1 r_2)^2}{8} z \left(1-i r_2 z\right)\left(1-\hat{k}\cdot \vec{r}_2-\frac{\vec{r}_1 \cdot \vec{r}_2}{r_1^2}\right)\times \\
&& \bigg[(r_1 z^{\prime})^2 \bigg(\frac{\hat{k}\cdot \vec{r}_1}{r_1^2}+\frac{\hat{k}\cdot \vec{r}_2}{r_2^2}-1\bigg)+\frac{(1+i r_1 z^{\prime})(1+i r_2 z^{\prime})}{r_2^2}\left(1+\vec{r}_1 \cdot \vec{r}_2\right)\bigg], \nn
\eea
while the indirect ones are given by:
\begin{eqnarray} 
\label{app:eq:zzzIndirect}
\tilde{\mathcal{F}}_{k \zeta \zeta \zeta}^{I}(r_1,r_2,z,z^{\prime}) &=&  \frac{(r_1 r_2)^2}{8} z z^{\prime}(1-i r_2 z)\bigg[ (1+i r_1 z^{\prime}) \bigg( \Big(1-\frac{\vec{r}_1\cdot\vec{r}_2}{r_2^2}\Big)\Big(1-\frac{\vec{r}_1\cdot\vec{r}_2}{r_1^2}-\hat{k}\cdot\vec{r}_2\Big) \qquad \ \  \\
&&\; 
 -{\hat{k}\cdot\vec{r}_2\Big( 1-\hat{k}\cdot\vec{r}_1-\frac{\vec{r}_1\cdot\vec{r}_2}{r_1^2}\Big)} \bigg)  + \frac{r_1^2}{r_2^2}  \left(1+i r_2 z^{\prime}\right) \Big(1-\frac{\vec{r}_1\cdot\vec{r}_2}{r_1^2}-\hat{k}\cdot\vec{r}_2\Big)^2 \bigg] \nn
\end{eqnarray}

\subsubsection{$\mathcal{J}^{\mathfrak{p}\mathfrak{p} }_{k0}(\eta;\eta_{\rm{in}})$}
\label{subsub:pp}

Finally, we arrive at $\mathcal{J}^{\mathfrak{p}\mathfrak{p} }_{k0}(\eta;\eta_{\rm{in}})$. We can use eqs.~\eqref{app:eq:pp} \& \eqref{app:eq:pzeta} to see that we can use the direct and indirect correlators found in eqs.~\eqref{app:eq:gzzDirectIndirect}, \eqref{app:eq:zzzDirect} \& \eqref{app:eq:zzzIndirect} in computing  $\mathcal{J}^{\mathfrak{p}\mathfrak{p} }_{k0}(\eta;\eta_{\rm{in}})$ just by changing the time translation kernels: $\cW_k(\eta,\eta^{\prime})\rightarrow \cX_k(\eta,\eta^{\prime}),\ \cZ_k(\eta,\eta^{\prime})\rightarrow \cY_k(\eta,\eta^{\prime})$. The usual scaling arguments show that 
\be
\mathcal{J}^{\mathfrak{p}\mathfrak{p} }_{k0}(\eta;\eta_{\rm{in}})=\frac{1}{k^2}\frac{H^4}{\Mp^4} \tilde{\mathcal{J}}^{\mathfrak{p}\mathfrak{p} }_{k0}(z;z_{in})
\ee
with
\begin{eqnarray} 
\label{app:eq:ppmomint}
\tilde{\mathcal{J}}^{\mathfrak{p}\mathfrak{p} }_{k0}(z;z_{in}) &=& \frac{1}{2}\int_{r_i>\kappa} \prod_{i=1}^2 \frac{\exd^3 r_i}{(2\pi)^3 r_i^3} (2\pi)^3 \delta^{(3)}\left(\hat{k}-\vec{r}_1-\vec{r}_2\right) \frac{e^{i (r_1+r_2)(z-z^{\prime})}}{ z^{\prime 2}} \nn\\
&& \qquad \times \left(\tilde{\mathcal{F}}_{k X}^D(r_1,r_2,z,z^{\prime}) \tilde{\cX}(z,z^{\prime})+\tilde{\mathcal{F}}_{k X}^I(r_1,r_2,z,z^{\prime}) \tilde{\cY}(z,z^{\prime})\right),
\end{eqnarray}
where as before, $X=\gamma \zeta \zeta,\ \zeta \zeta\zeta$. 

\subsection{The time integral}
\label{subsec:timeint}

Our strategy here will be go first do the time integral
\be
\label{app:eq:generictimeint}
\tilde{\mathcal{J}}^{a b}_{k0}(z; z_{in})=\int_{z(1-i \varepsilon)}^{z_{in}(1-i \varepsilon)} \exd z^{\prime}\ \tilde{\mathcal{T}}^{a b}_k(z,z^{\prime})
\ee
with $\tilde{\mathcal{T}}^{a b}_k(z,z^{\prime})$ given in eqs.~\eqref{app:eq:zzmomint},\eqref{app:eq:zpmomint},\eqref{app:eq:pzmomint} \& \eqref{app:eq:ppmomint}. The reason to do this first is to push the regulator $\varepsilon$ into the momentum integrals. As we will see, it serves a number of purposes there. It places the system into its adiabatic vacuum via an exponential suppression term of the form $\exp(-\delta z_{in}\varepsilon ),\ \delta>0$ (which means that we can take $z_{in}\rightarrow \infty$ in our expressions) and it also acts as a high momentum regulator.

\subsection{Momentum integrals}
\label{subsec:momints}

\subsubsection{Region of Integration}
\label{subsubsec:regionint}

\begin{figure}[h!]
\includegraphics[scale=0.17]{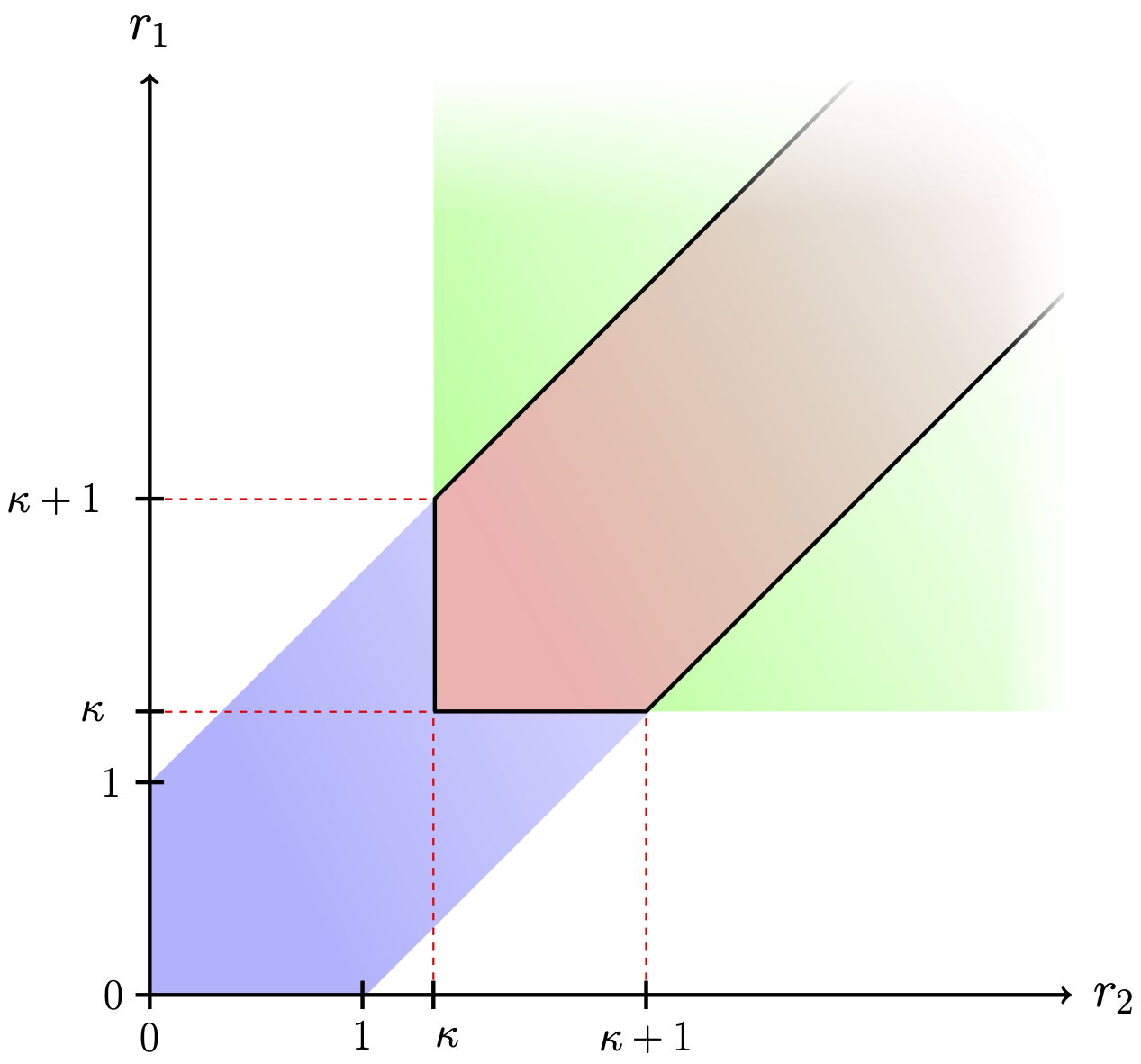}
  \centering
  \caption{The region of integration. The vertical dotted lines are at $r_2=\kappa,\kappa+1$ while the horizontal ones are at $r_1=\kappa,\kappa+1$. The integration region $U$ (in red) is the intersection of the region $r_{1},r_{2} > \kappa$ (in green), and the region between the lines $r_{1} = r_{2} \pm 1$ (in blue).}
 \label{fig:intregionU}
\end{figure}

Next we elucidate what the region of integration is for our calculation. This is driven both by the momentum delta function as well as by the constraint that the momenta are in the environment. In terms of the rescaled momenta, this last constraint reads $r_{1,2}>\kappa$, where $\kappa = \kUV\slash k$. Now turning to the delta function, we can write it as:
\be
\label{eq:sphericaldelta}
\delta^{(3)}\left(\vec{r}_1+\vec{r}_2-\hat{k}\right)=\frac{1}{r_1^2}\ \delta\left(r_1-\left|\hat{k}-\vec{r}_2\right|\right)\delta\left(\cos\theta_{r_1}-\cos\theta_{\hat{k}-\vec{r}_2}\right)\delta\left(\phi_{r_1}-\phi_{\hat{k}-\vec{r}_2}\right).
\ee
We can use this to do the $d\cos\theta_{r_1}$ and $d\phi_{r_1}$ integrals, especially since the integrand does not depend on these variables. We can also cancel out the factor of $r_1^2$ in the $d^3 r_1$ part of the measure. We are then left with 
\be
\int \prod_{a=1}^2 \frac{\exd^3 r_a}{(2\pi)^3 r_a^3}\ (2\pi)^3 \delta^{(3)}\left(\vec{r}_1+\vec{r}_2-\hat{k}\right)  \rightarrow \frac{1}{(2\pi)^2}\int  \frac{ \exd r_2\ r_2^2 \ \exd r_1\ \exd\mu_2 }{(2\pi)^2} \frac{\delta\big(r_1-\sqrt{r_2^2-2 r_2 \mu_2 +1}\big)}{r_1^3 r_2^3},
\ee
where we choose $\hat{k}=\hat{z}$ so that $\mu_2 = \hat{k}\cdot\hat{r}_2=\cos\theta_{r_2}$. We can use the remaining delta function to do the $\mu_2$ integral
\be
\delta\Big(r_1-\sqrt{r_2^2-2 r_2 \mu_2 +1}\Big)=\frac{r_1}{r_2} \delta\left(\mu_2-\mu_2^{(0)}\right),\quad \mu_2^{(0)}=\frac{r_2^2+1-r_1^2}{2 r_2}.
\ee
This gives us the last piece of the information we need about the region of integration: we need $\left |\mu_2^{(0)}\right |\leq 1$. Using the fact that all the momenta are larger than $\kappa\gg1$ this translates to the constraint: $r_2-1\leq r_1\leq r_2+1$. Putting all of these together gives the region $U=\{(r_2,r_1) \, | \,  r_2-1\leq r_1\leq r_2+1, r_{1,2}>\kappa \}$ depicted in Figure \ref{fig:intregionU}. 

The measure is then
\be
\frac{\exd r_2\, \exd r_1}{(2\pi)^2 (r_1 r_2)^2}\; ,
\ee
and we are to replace $\mu_2$ by $\mu_2^{(0)}$ everywhere. The denominator in the measure explains why we took the time to extract a factor of $(r_1 r_2)^2$ from our correlation functions; these will cancel.

\subsubsection{Momentum change of variables}
\label{subsubsec:changevar}

It's useful to do the following change of variables to disentangle the limits of integration:
\be
r_1 = \frac{t+s+1}{2},\quad r_2 = \frac{t-s+1}{2}.
\ee
In terms of $s$ and $t$ the measure factor becomes
\be
\frac{8\  \exd s\ \exd t}{(2\pi)^2(t+s+1)^2 (t-s+1)^2}=\frac{2}{\pi^2} \frac{\exd s\ \exd t}{(t+s+1)^2 (t-s+1)^2}.
\ee
In terms of $r_1$ and $r_2$ the integral could be written as
\be
\int_{\kappa}^{\kappa+1} \exd r_2 \int_{\kappa}^{r_2+1} \exd r_1+\int_{\kappa+1}^{\infty} \exd r_2 \int_{r_2-1}^{r_2+1} \exd r_1
\ee
which becomes
\be
\int_{2\kappa-1}^{2\kappa+2}  \exd t \int_0^1 \exd s+\int_{2\kappa+2}^{\infty}  \exd t \int_{-1}^1 \exd s
\ee
in terms of $s$ and $t$. We will call the first contribution the compact one and the second the non-compact one for obvious reasons.

\subsection{Computing the leading piece of  $\tilde{\mathcal{J}}^{\zeta\zeta }_{k0}(z;z_{in})$}
\label{sec:leadingzz}
Let's apply the above results to compute $\tilde{\mathcal{J}}^{\zeta\zeta }_{k0}(z;z_{in})$. The time integral can be done to arrive at 
\bea
\label{eq:timeintegralpiecets}
&&\tilde{\mathcal{J}}^{\zeta\zeta }_{k0}(z;z_{in})= \left(\frac{1}{z^3}-\frac{i (t+1)}{z^2}+ \frac{s^2-(1+t)^2}{4 z}\right)\times \\ 
&&\left(e^{-i (z^{\prime}-z) t}\bigg[s^2 \bigg(\frac{1}{4 t^2}+\frac{-\frac{1}{4}+\frac{i z^{\prime}}{4}}{t}\bigg)-\frac{1}{4 t^2}+\frac{-\frac{5}{4}-\frac{i z^{\prime}}{4}}{t}+t\left(\frac{1}{4}-\frac{i z^{\prime}}{4}\right)-\frac{i z^{\prime}}{2}-\frac{i}{z^{\prime}}-\frac{3}{4}\bigg]\right .\nn\\
&&\left .+e^{-i (z^{\prime}-z)(t+2)} \bigg[ \frac{\frac{1}{4} i s^2z^{\prime}+\frac{s^2}{4}-\frac{i z^{\prime}}{4}+\frac{5}{4}}{t+2}+\frac{\frac{s^2}{4}-\frac{1}{4}}{(t+2)^2}-\frac{ (t+2)(1 + i z^{\prime})}{4}+\frac{iz^{\prime}}{2}+\frac{i}{z^{\prime}}-\frac{3}{4}\bigg] \right) \nn
\eea
This is then to be combined with the measure as well as the $(1-\mu_2^{(0) 2})^2$ part of the integrand:
\be
\frac{2}{\pi^2} \frac{\exd s\ \exd t}{(t+s+1)^2 (t-s+1)^2}\ \left(\frac{t(t+2)(s^2-1)}{(t-s+1)^2}\right)^2.
\ee

To unravel the early time behavior it's best to do the $s$ momentum integral first. The resulting expression is long, but the relevant feature we'll need is that it is the sum of a (large) number of terms of the form:
\be
e^{-t z_{\rm in}\varepsilon}\frac{t^{q_0}\alpha_0 (z_{\rm in}) \log t+t^{q_2}\alpha_2 (z_{\rm in}) \log (t+2)}{(1+t)^p t^q (2+t)^r}.
\ee
In general we can't do the $t$-momentum integral of this, but it's enough to bound the integral of the absolute value of each term. To do this, we can replace the denominator in the above expression by its value when $t=t_{\rm min}$, i.e. the smallest value of $t$ in the relevant region. In the compact region this is $2\kappa-1$ while it is $2\kappa+2$ in the non compact one. The resulting integrals can then be done and wind up being proportional to $e^{-t z_{\rm in}\varepsilon}$ which shows the exponentially rapid vanishing of the correlator, as expected. Once the $z_{\rm in}\rightarrow \infty$ we can take $\varepsilon$ to zero with impunity.

The strategy outlined for the early time limit is insufficient for the obtaining the late time behavior; a more refined estimate of the momentum integrals is needed. Our strategy here will start with computing the $t$ momentum integrals, which can be done exactly. Since it is in this integral that we need the exponential damping at large $t$, once this integral is done, we can perform the small $z$ expansion of the result, which then allows us to calculate the $s$ integral. Finally, we perform the small $\varepsilon$ expansion. 

For the leading contribution for $\widetilde{\mathcal{J}}_{k 0}^{\zeta \zeta}$, after integrating with respect to $s$ and taking the large $\kappa$ limit, we find  
\be
\widetilde{\mathcal{J}}_{k 0}^{\zeta \zeta}(\eta;\eta_{\rm{in}}) \sim \frac{2}{\pi^2}\left(\frac{32}{45\kappa^3 z^3}-\frac{8 i}{5\kappa^2 z^2}-\frac{4}{45\kappa^3 z}+\frac{16 i}{15}\left(\ln(2\kappa z)+\ln\varepsilon\right)-\frac{16 \pi}{15}\right)
\ee

This result for the leading $z$ behavior of $\widetilde{\mathcal{J}}_{k 0}^{\zeta \zeta}$ passes some useful tests. First, it's positive, as required for the reduced density matrix to actually give probabilities. Second, it scales as inverse powers of $\kappa$. This is as it should be, since if we took $\kappa$ to infinity, \emph{all} modes are kept which means no decoherence should happen. While it's out of the large $\kappa$ regime, if we were to take $\kappa\rightarrow 0$ the Lindblad-like coefficient would blow up which seems consistent with the fact that in this limit \emph{no} modes are kept so decoherence should happen quickly. 

The other Lindblad-like coefficients can also be computed to leading order in $\slrl$ using the same techniques:
\bea
\label{app:eq:leadingLindblad}
\widetilde{\mathcal{J}}^{\zeta \mfp}_{k0}(\eta;\eta_{\rm{in}}) &\sim &\frac{2}{\pi^2}\left(\frac{2}{25\kappa^5 z^2}-\frac{2 i z}{15\kappa^2}+\frac{14}{45 \kappa^3} \right)+\ldots\\
\widetilde{\mathcal{J}}^{\mfp \mfp}_{k0}(\eta;\eta_{\rm{in}})&\sim & z \left(\frac{3 i \pi }{128 \kappa }-\frac{1709}{1280\kappa }+\frac{1}{64} i \pi  \log (2 \kappa)+\frac{\pi ^2}{128}-\frac{379 i \pi}{768}\right)+\ldots
\eea

\section{Transport equations}
\label{sec:Transport}

This appendix derives the transport equations (for the equal-time two-point functions) in the gaussian limit that are used in \S\ref{ssec:TransportCoeffs} of the main text. We do {\it not} here assume the slow-roll dependence found in the main text since our interest is in using the results of this section to resum late-time evolution.

\subsection{Gaussian Lindblad-like evolution}

Our starting point is the $\mathrm{TCL}_2$ equation eq.~(\ref{app:eq:TCL2a}) --- which can be written 
\bea \label{app:TCL2a}
\partial_{\eta} \varrho_{\bmk }(\eta) &=& -  \mathcal{J}^{\zeta \zeta}_{k0}(\eta;\eta_{\rm{in}})\Bigl[\zeta_{\bmk}(\eta),\ \zeta_{-\bmk}(\eta)\ \varrho_{\bmk}(\eta)\Bigr] - \mathcal{J}^{\zeta \mfp}_{k0}(\eta;\eta_{\rm{in}})\Bigl[\zeta_{\bmk}(\eta),\ \mfp_{-\bmk}(\eta)\ \varrho_{\bmk}(\eta)\Bigr] \\
&&\;  - \mathcal{J}^{\mfp \zeta}_{k 0}(\eta;\eta_{\in})\Bigl[\mfp_{\bmk}(\eta),\ \zeta_{-\bmk}(\eta)\ \varrho_{\bmk}(\eta)\Bigr] - \mathcal{J}^{\mfp \mfp}_{k 0}(\eta;\eta_{\in})\Bigl[\mfp_{\bmk}(\eta),\ \mfp_{-\bmk}(\eta)\ \varrho_{\bmk}(\eta)\Bigr] + \mathrm{h.c.}\nn
\eea
where we take $\eta_{\mathrm{in}} \to - \infty$. 

To lighten the notation temporarily, we define
\begin{equation}
\begin{aligned}[c]
\mathcal{J}^{\zeta \zeta}_{k0}(\eta; -\infty)
&= A + ia, \qquad
\mathcal{J}^{\zeta \mfp}_{k0}(\eta; -\infty)
= B + ib,\\[0.5ex]
\mathcal{J}^{\mfp \zeta}_{k0}(\eta; -\infty)
&= C + ic, \qquad
\mathcal{J}^{\mfp \mfp}_{k0}(\eta; -\infty)
= D + id.
\end{aligned}
\end{equation}
and keep in mind that $\mathcal{J}^{\zeta \mfp}_{k0} \neq \mathcal{J}^{\mfp \zeta}_{k 0}$ in general. In terms of these eq.~\pref{app:TCL2a} is 
\bea
\partial_{\eta} \varrho_{\bmk }(\eta) & = & - \big( A + i a \big) \Bigl[\zeta_{\bmk} ,\ \zeta_{-\bmk} \varrho_{\bmk} \Bigr] - \big( A - i a \big) \Bigl[ \varrho_{\bmk} \zeta_{\bmk}, \zeta_{-\bmk} \Bigr] \\
& \ & - ( B + i b ) \Bigl[ \zeta_{\bmk},\ \mfp_{-\bmk}\varrho_{\bmk} \Bigr] - ( B - i b ) \Bigl[ \varrho_{\bmk} \mfp_{\bmk} , \zeta_{-\bmk} \Bigr] \notag \\
& \ & - (C + i c) \Bigl[\mfp_{\bmk},\ \zeta_{-\bmk} \varrho_{\bmk} \Bigr]  - (C - i c) \Bigl[ \varrho_{\bmk} \zeta_{\bmk} ,\mfp_{-\bmk}\Bigr] \nn \\
&\ & - ( D + i d ) \Bigl[\mfp_{\bmk},\ \mfp_{-\bmk} \varrho_{\bmk}\Bigr] - ( D - i d ) \Bigl[ \varrho_{\bmk}\mfp_{\bmk}, \mfp_{-\bmk} \Bigr] \nn
\eea
which uses $\zeta^{\dagger}_{\bmk} = \zeta_{-\bmk}$ and $\mfp^{\dagger}_{\bmk} = \mfp_{-\bmk}$. Keeping track of the $\bmk$ and $-\bmk$ dependence is tedious, so switching to the real variables defined in \S\ref{App:GaussianProperty} gives
\bea \label{app:abcd}
\partial_{\eta} \varrho(\eta) & = & - \big( A + i a \big) \Bigl[\tilde\zeta ,\ \tilde\zeta \varrho \Bigr] - \big( A - i a \big) \Bigl[ \varrho \tilde\zeta, \tilde\zeta \Bigr]   - ( B + i b ) \Bigl[ \tilde\zeta,\ \tilde\mfp \varrho \Bigr] - ( B - i b ) \Bigl[ \varrho\tilde\mfp , \tilde\zeta \Bigr] \  \\
&&\qquad - (C+ i c) \Bigl[\tilde\mfp ,\ \tilde\zeta \varrho\Bigr]  - (C - i c) \Bigl[ \varrho\tilde \zeta ,\tilde\mfp \Bigr] 
 - ( D + i d ) \Bigl[\tilde \mfp ,\ \tilde\mfp \varrho\Bigr] - ( D - i d ) \Bigl[ \varrho\tilde\mfp, \tilde\mfp \Bigr] \,, \nn
\eea
where the common subscript $\bmk$ on each quantity in these equations is suppressed.

Using the simple identities
\be
\ [ X,X\varrho] + [ \varrho X, X ]  =- 2 \Big( X \rho X - \tfrac{1}{2} \{ X^2 , \rho \} \Big) \qquad \hbox{and} \qquad
  \ [ X , X \varrho ] - [ \varrho X, X ]  =  [ X^2 , \rho ]
\ee
implies
\begin{equation}
\begin{aligned}[c]
- [ \tilde\zeta, \tilde\mfp \varrho ]
- [ \varrho \tilde\mfp, \tilde\zeta ]
- [ \tilde\mfp, \tilde\zeta \varrho ]
- [ \varrho \tilde\zeta, \tilde\mfp ]
&=
2(
\tilde\zeta \varrho \tilde\mfp
-\tfrac12\{\tilde\mfp\tilde\zeta,\varrho\}
)
+
2(
\tilde\mfp \varrho \tilde\zeta
-\tfrac12\{\tilde\zeta\tilde\mfp,\varrho\}
)
\\[1ex]
- [ \tilde\zeta, \tilde\mfp \varrho ]
+ [ \varrho \tilde\mfp, \tilde\zeta ]
+ [ \tilde\mfp, \tilde\zeta \varrho ]
- [ \varrho \tilde\zeta, \tilde\mfp ]
&=
-2(
\tilde\zeta \varrho \tilde\mfp
-\tfrac12\{\tilde\mfp\tilde\zeta,\varrho\}
)
+
2(
\tilde\mfp \varrho \tilde\zeta
-\tfrac12\{\tilde\zeta\tilde\mfp,\varrho\})
\end{aligned}
\end{equation}
and
\begin{equation}
\begin{aligned}[c]
- [\tilde\zeta,\tilde\mfp\varrho]
- [\varrho\tilde\mfp,\tilde\zeta]
+ [\tilde\mfp,\tilde\zeta\varrho]
+ [\varrho\tilde\zeta,\tilde\mfp]
&=
-\bigl[[\tilde\zeta,\tilde\mfp],\varrho\bigr],
\\[1ex]
- [\tilde\zeta,\tilde\mfp\varrho]
+ [\varrho\tilde\mfp,\tilde\zeta]
- [\tilde\mfp,\tilde\zeta\varrho]
+ [\varrho\tilde\zeta,\tilde\mfp]
&=
-\bigl[\{\tilde\zeta,\tilde\mfp\},\varrho\bigr].
\end{aligned}
\end{equation}
and so equation (\ref{app:abcd}) becomes
\bea
\partial_{\eta} \varrho (\eta) & = & - i [ a \tilde \zeta^2 , \varrho  ] + 2 A \Big( \tilde \zeta \rho \tilde \zeta - \tfrac{1}{2} \{ \tilde \zeta^2 , \rho \} \Big) \\
& \ & + ( B - i b + C + i c ) \Big( \tilde \zeta \varrho \tilde \mfp - \tfrac{1}{2} \{ \tilde \mfp \tilde \zeta, \varrho \}  \Big) + ( B + i b + C - i c ) \Big( \tilde \mfp \varrho \tilde \zeta - \tfrac{1}{2} \{ \tilde \zeta \tilde \mfp, \varrho \} \Big) \nn \\
& \ & + \tfrac12({B-C}) \Big[ [ \tilde \zeta,\tilde \mfp ],\varrho\Big] + \tfrac{i}{2}{ (b+c)} \Big[ \{ \tilde \zeta, \tilde \mfp\} ,\tilde \mfp ],\varrho\Big]  
  - i [ d \tilde \mfp^2 , \varrho_{\bmk} ] + 2 D \Big( \tilde \mfp \rho \tilde \mfp - \tfrac{1}{2} \{ \tilde \mfp^2 , \rho \} \Big) \nn\,.
\eea
This further simplifies once $[ \tilde \zeta,\tilde \mfp ] = i $ is used, leaving
\bea
\partial_{\eta} \varrho (\eta) & = & - i \left[ a \tilde \zeta^2 -\tfrac12 (b+c) \{\tilde \zeta,\tilde \mfp \}  + d \tilde \mfp^2 , \varrho  \right] + 2 A \Big( \tilde \zeta \varrho \tilde \zeta - \tfrac{1}{2} \{ \tilde \zeta^2 , \varrho \} \Big) \\
& \ & + ( B - i b + C + i c ) \Big( \tilde \zeta \varrho \tilde \mfp - \tfrac{1}{2} \{ \tilde \mfp \tilde \zeta, \varrho \}  \Big) + ( B + i b + C - i c ) \Big( \tilde \mfp \varrho \tilde \zeta - \tfrac{1}{2} \{ \tilde \zeta \tilde \mfp, \varrho \} \Big) \nn \\
& \ & + 2 D \Big( \tilde \mfp \varrho \tilde \mfp - \tfrac{1}{2} \{ \tilde \mfp^2 , \varrho \} \Big) \notag
\eea

In terms of the original notation this becomes
\bea
\partial_{\eta} \varrho_{\bmk }(\eta) & = & - i \Bigl[ \, \mathrm{Im}[ \mathcal{J}^{\zeta \zeta}_{k0}] \tilde \zeta_\bmk^2 - \mathrm{Im}[\mathcal{J}^{\zeta \mfp}_{k0} + \mathcal{J}^{ \mfp \zeta}_{k0}]  \tfrac12 \{\tilde \zeta_\bmk,\tilde \mfp_\bmk \}  + \mathrm{Im}[\mathcal{J}^{ \mfp  \mfp}_{k0}] \tilde \mfp_\bmk^2 , \,  \varrho_{\bmk} \, \Bigr] \\
&&\qquad + 2 \mathrm{Re}[ \mathcal{J}^{\zeta \zeta}_{k0} ] \Big( \tilde \zeta_\bmk \varrho_{\bmk} \tilde \zeta_{\bmk} - \tfrac{1}{2} \{ \tilde \zeta_{\bmk}^2 , \varrho_{\bmk} \} \Big)   + ( \mathcal{J}^{\zeta  \mfp \ast}_{k0} + \mathcal{J}^{\mfp \zeta}_{k 0} ) \Big( \tilde \zeta_{\bmk} \varrho_{\bmk} \tilde \mfp_{\bmk} - \tfrac{1}{2} \{ \tilde \mfp_{\bmk} \tilde \zeta_{\bmk}, \varrho_{\bmk} \}  \Big) \nn\\
&&\qquad\qquad + ( \mathcal{J}^{ \zeta  \mfp}_{k0} + \mathcal{J}^{ \mfp \zeta \ast}_{k 0} )  \Big( \tilde \mfp_{\bmk} \varrho_{\bmk} \tilde \zeta_{\bmk} - \tfrac{1}{2} \{ \tilde \zeta_{\bmk} \tilde \mfp_{\bmk}, \varrho_{\bmk} \} \Big)  + 2 \mathrm{Re}[\mathcal{J}^{ \mfp \mfp}_{k 0}] \Big( \tilde \mfp_{\bmk} \varrho_{\bmk} \tilde \mfp_{\bmk} - \tfrac{1}{2} \{ \tilde \mfp_{\bmk}^2 , \varrho_{\bmk} \} \Big) \,.\nn
\eea
This expression has the Lindblad-like form
\bea
\partial_{\eta} \varrho_{\bmk } & = & - i [ \scrH_{\rm eff} , \varrho_{\bmk} ] + \sum_{r,s = 1,2} \mathrm{h}_{rs} \Big( \mathrm{O}_{r} \varrho_{\bmk} \mathrm{O}_{s} - \tfrac{1}{2} \{ \mathrm{O}_{s} \mathrm{O}_{r}, \varrho_{\bmk}  \} \Big)
\eea
if we define\footnote{Notice the different font $\mathrm{O}$ used here, since these definitions differ from those of the main text by not including factors of $Z$.} $\mathrm{O}_{1} = \zeta$ and $\mathrm{O}_{2} = \mfp$, as well as
\begin{eqnarray} \label{app:HeffCorr}
\scrH_{\rm eff} &=&  a \tilde\zeta_\bmk^2 - (b+c) \tfrac12 \{\tilde \zeta_\bmk, \tilde \mfp_\bmk \}  + d \tilde \mfp_\bmk^2 \\
&=&  \mathrm{Im}[ \mathcal{J}^{\zeta \zeta}_{k0}] \tilde \zeta_\bmk^2 - \mathrm{Im}[\mathcal{J}^{\zeta \mfp}_{k0} + \mathcal{J}^{\mfp \zeta}_{k0}]  \tfrac12 \{\tilde\zeta_\bmk, \tilde \mfp_\bmk \}  + \mathrm{Im}[\mathcal{J}^{\mfp \mfp}_{k0}] \tilde \mfp_\bmk^2 \nn
\end{eqnarray}
and
\begin{equation} \label{app:Kosmatv1}
\left[ \begin{matrix} \mathrm{h}^{11} & \mathrm{h}^{12} \\ \mathrm{h}^{21} & \mathrm{h}^{22} \end{matrix} \right] = \left[ \begin{matrix} 2A & B - i b + C + i c \\ B + i b + C - i c & 2D \end{matrix} \right] = \left[ \begin{matrix} 2\mathrm{Re}[\mathcal{J}^{\zeta \zeta}_{k0} ] & \mathcal{J}^{\zeta \mfp \ast}_{k0} + \mathcal{J}^{\mfp \zeta}_{k 0}  \\ \mathcal{J}^{\zeta \mfp}_{k0} + \mathcal{J}^{\mfp \zeta \ast}_{k 0}  & 2 \mathrm{Re}[\mathcal{J}^{\mfp \mfp}_{k0} ] \end{matrix} \right] \,.
\end{equation}
Notice that \pref{app:HeffCorr} contains just the contributions to $\scrH_{\rm eff}$ coming from the cubic interactions, but (unlike for applications to decoherence) quartic interactions can also contribute to $\scrH_{\rm eff}$ at the same order in $1/\Mp$. We notice in passing that the determinant of the coupling matrix is
\bea
\det \left[ \begin{matrix} \mathrm{h}^{11} & \mathrm{h}^{12} \\ \mathrm{h}^{21} & \mathrm{h}^{22} \end{matrix} \right] & = & 4 \mathrm{Re}[\mathcal{J}^{\zeta \zeta}_{k0} ] \mathrm{Re}[\mathcal{J}^{\mfp \mfp}_{k0} ] - \left| \mathcal{J}^{\zeta \mfp \ast}_{k0} + \mathcal{J}^{\mfp \zeta}_{k 0} \right|^2 \\
& = & 4 \mathrm{Re}[\mathcal{J}^{\zeta \zeta}_{k0} ] \mathrm{Re}[\mathcal{J}^{\mfp \mfp}_{k0} ] - \mathrm{Re}[ \mathcal{J}^{\zeta \mfp \ast}_{k0} + \mathcal{J}^{\mfp \zeta}_{k 0}]^2 - \mathrm{Im}[ \mathcal{J}^{\zeta \mfp \ast}_{k0} - \mathcal{J}^{\mfp \zeta}_{k 0}]^2 
\eea

In the main text, we instead work with the rescaled operators
\begin{equation}
\mathcal{O}_{1} = Z^{-1} \zeta \quad \mathrm{and}  \quad \mathcal{O}_{2} = Z \mfp \qquad \mathrm{with} \ Z =\frac{1}{\sqrt{2 \epsilon_1 k^3}} \frac{H}{M_p}  \,,
\end{equation}
and in terms of these, our Lindblad-like equation is 
\begin{equation}\label{app:LindbladGaussian}
\partial_{\eta} \varrho_{\bmk} = -i\Bigl[ \scrH_{\rm eff}(\eta), \varrho_{\bmk}(\eta)\Bigr] +\sum_{r,s=1}^2 h^{s r}_\bmk\left[\mathcal{O}_{\bmk,s}\varrho_{\bmk}\mathcal{O}^{\dag}_{\bmk,r}-\frac{1}{2}\left\{\mathcal{O}^{\dag}_{\bmk,r}\mathcal{O}_{\bmk,s},\varrho_{\bmk}\right\}\right] \,.
\end{equation}
where  
\bea
\left[ \begin{matrix} h_{\bmk}^{11} & h_{\bmk}^{12} \\ h_{\bmk}^{21} & h_{\bmk}^{22} \end{matrix} \right]  &=& \left[ \begin{matrix} Z^2 \mathrm{h}_{11} & \mathrm{h}_{12} \\ \mathrm{h}_{21} & Z^{-2}\mathrm{h}_{22} \end{matrix} \right] = \left[ \begin{matrix} 2Z^{2}\mathrm{Re}[\mathcal{J}^{\zeta \zeta}_{k0} ] & \mathcal{J}^{\zeta \mfp \ast}_{k0} + \mathcal{J}^{\mfp \zeta}_{k 0}  \\ \mathcal{J}^{\zeta \mfp}_{k0} + \mathcal{J}^{\mfp \zeta \ast}_{k 0}  & 2 Z^{-2} \mathrm{Re}[\mathcal{J}^{\mfp \mfp}_{k0} ] \end{matrix} \right]  \nn\\
 &=& k\left(\frac{H}{\Mp}\right)^2  \left[ \begin{matrix} \frac{1}{ \epsilon_1 k^4} \mathrm{Re}[\mathcal{J}^{\zeta \zeta}_{k0} ] & \frac{M_p^2}{ k H^2} ( \mathcal{J}^{\zeta \mfp \ast}_{k0} + \mathcal{J}^{\mfp \zeta}_{k 0} )   \\ \frac{ M_p^2}{ k H^2} ( \mathcal{J}^{\zeta \mfp}_{k0} + \mathcal{J}^{\mfp \zeta \ast}_{k 0} ) & \frac{4 \slrl k^2 M_p^4}{H^4} \mathrm{Re}[\mathcal{J}^{\mfp \mfp}_{k0} ] \end{matrix} \right] \,.
\eea
In terms of the dimensionless integrals defined in \pref{app:curlyJdimless} this becomes
\begin{equation} \label{app:Kossmat}
\left[ \begin{matrix} h_{\bmk}^{11} & h_{\bmk}^{12} \\ h_{\bmk}^{21} & h_{\bmk}^{22} \end{matrix} \right] = k\left(\frac{H}{\Mp}\right)^2  \left[ \begin{matrix}  \mathrm{Re}[\widetilde{\mathcal{J}}^{\zeta \zeta}_{k0} ] & \widetilde{\mathcal{J}}^{\zeta \mfp \ast}_{k0}+\slrl \widetilde{\mathcal{J}}^{\mfp \zeta }_{k0}   \\  \widetilde{\mathcal{J}}^{\zeta \mfp}_{k0}+ \slrl \widetilde{\mathcal{J}}^{\mfp \zeta \ast}_{k0}  &  4 \slrl \mathrm{Re}[\widetilde{\mathcal{J}}^{\mfp \mfp}_{k0} ] \end{matrix} \right] \,.
\end{equation}
Eqs.~\pref{app:LindbladGaussian} and \pref{app:Kossmat} agree with eqs.~\pref{app:eq:lindbladlikeform} and \pref{app:eq:decohmatrixh} of the earlier Appendix and appear as (\ref{eq:lindbladlikeform}) and (\ref{eq:decohmatrixh}) in the main text.

\subsection{Transport Equations}
\label{Appssec:Transport}

We now derive the transport equations, which are equations of motion for the equal-time 2-point correlation functions $\boldsymbol{\Sigma}$ with components
\begin{equation}
\Sigma_{ij} = \mathrm{Tr}\big[ \tfrac{1}{2} \{ O_{i}, O_{j} \} \varrho \big] \,.
\end{equation}
More explicitily
\begin{equation}
\boldsymbol{\Sigma} = \left[ \begin{matrix} \Sigma_{11} & \Sigma_{12} \\  \Sigma_{21} & \Sigma_{22} \end{matrix} \right] =  \left[ \begin{matrix} \mathrm{Tr}\big[ \tilde\zeta \tilde\zeta \varrho \big] & \tfrac12 \mathrm{Tr}\big[\{ \tilde\zeta,\tilde \mfp \}  \varrho \big] \\  \tfrac12 \mathrm{Tr}\big[ \{ \tilde \zeta, \tilde\mfp \}  \varrho \big]  & \mathrm{Tr}\big[\tilde \mfp \tilde \mfp \varrho \big] \end{matrix} \right]
\end{equation}
and the anti-commutator ensures the correlation matrix is symmetric: $ \Sigma_{12}=  \Sigma_{21}$. 

The evolution equations for $\Sigma_{ij}$ is given by direct differentiation of the definition,
\be
\partial_{\eta} \Sigma_{ij} = \tfrac12 \mathrm{Tr}[\{ \partial_\eta O_{i} , O_j \} \varrho ] + \tfrac12 \mathrm{Tr}[\{ O_{i} , \partial_\eta O_j\} \varrho ] + \tfrac12 \mathrm{Tr}[\{ O_{i} , O_j \} \partial_\eta \varrho ] \,,
\ee
where the time evolution of the interaction-picture operators is computed using the free Hamiltonian $\mathscr{H}_0$ given in (\ref{freeHclassical})), which leads to
\be
\partial_\eta \zeta_{\bmk} = \frac{\mfp_{\bmk}}{2 \epsilon_1 M_p^2 a^2} = \frac{\mfp_{\bmk}}{\mfz_s^2}  \qquad \mathrm{and} \qquad \partial_\eta \mfp_\bmk =   - 2 \epsilon_1 M_p^2 a^{2} k^2 \zeta_{\bmk} =  - \mfz_s^2 k^2 \zeta_{\bmk}   \,,
\ee
where  $\mfz_s = a\Mp \sqrt{2\slrl }$ is defined in (\ref{eq:defzeta}). These both are equivalent to the usual Klein-Gordon equation  $\zeta''_{\bmk} + \frac{2a'}{a} \zeta'_{\bmk} + k^2 \zeta_{\bmk} = 0$, leading to
\begin{eqnarray} \label{app:matTrans1}
\frac{\partial}{ \partial \eta } \left[ \begin{matrix} \Sigma_{11} \\ \Sigma_{12} \\ \Sigma_{22} \end{matrix}  \right] =\left[ \begin{matrix} 0 & 2 \mfz_s^{-2} & 0 \\
 - \mfz_s^2 k^2 & 0 & \mfz_s^{-2} \\
 0 & -2 \mfz_s^{2} k^2 & 0 \end{matrix}  \right] \left[ \begin{matrix} \Sigma_{11} \\ \Sigma_{12} \\ \Sigma_{22} \end{matrix}  \right] + \left[ \begin{matrix} \mathrm{Tr}[ \tilde\zeta_\bmk \tilde\zeta_\bmk \partial_\eta \varrho_\bmk ]  \\  \mathrm{Tr}[ \tfrac12 \{\tilde \zeta_\bmk,\tilde \mfp_\bmk\}  \partial_\eta \varrho_\bmk ] \\ \mathrm{Tr}[\tilde \mfp \tilde \mfp \partial_\eta \varrho ]  \end{matrix}  \right] \,.
\end{eqnarray}

It remains to determine $\mathrm{Tr}\big( S \partial_\eta \varrho \big)$ for the three choices $S = \tilde\zeta_\bmk \tilde\zeta_\bmk$, $\tfrac12 \{ \tilde\zeta_\bmk, \tilde\mfp_\bmk \}, \tilde\mfp_\bmk \tilde\mfp_\bmk$, which in the Gaussian case can be computed explicitly by evaluating $\partial_\eta \varrho_\bmk$ using  (\ref{app:abcd}) or \pref{app:LindbladGaussian}.  Using (\ref{app:abcd}) leads traces of the form $\mathrm{Tr}\big( S [ X, Y\varrho] \big) = \mathrm{Tr}\big( [S,X] Y \varrho \big)$, where:
\begin{eqnarray}
\mathrm{Tr}( S \partial_\eta \rho ) & = & - \big( A + i a \big) \mathrm{Tr}\Big( [ S, \tilde\zeta_\bmk ] \tilde \zeta_\bmk \varrho_{\bmk} \Big)  - (B+ i b)  \mathrm{Tr}\Big( [ S, \tilde \zeta_\bmk ]\tilde \mfp_\bmk \varrho_{\bmk} \Big)  \\
&& \qquad  - (C+ i c)  \mathrm{Tr}\Big( [ S, \tilde\mfp_\bmk ]\tilde \zeta_\bmk \varrho_{\bmk} \Big) - \big( D + i d \big) \mathrm{Tr}\Big( [ S,\tilde \mfp_\bmk ]\tilde \mfp_\bmk \varrho_{\bmk} \Big) + \mathrm{h.c.} \nn
\end{eqnarray}
The interaction picture equal time commutators are easily evaluated using the canonical commutation relation $[\tilde\zeta, \tilde\mfp] = i$, giving
\bea 
      && [ \tilde \zeta, \tilde \zeta \tilde \zeta ]  = [ \tilde \mfp , \tilde \mfp  \tilde \mfp ] = 0 \,, \quad
  [ \tilde \zeta, \tfrac12\{\tilde \zeta,\tilde \mfp \} ]   =  + i \tilde \zeta \,, \qquad
  [ \tilde \zeta, \tilde \mfp \tilde \mfp  ]   =  + 2 i \tilde \mfp  \,, \nn\\
  && \qquad\quad\qquad [ \tilde \mfp , \tilde \zeta \tilde \zeta ]  =  - 2 i \tilde \zeta \,, \qquad
 [ \tilde \mfp , \tfrac12 \{\tilde \zeta,\tilde \mfp \} ]  =  -  i \tilde \mfp  \,.
\eea 

We then find that with $S = \tilde \zeta \tilde \zeta$ we have
\begin{small}
\begin{eqnarray}
\mathrm{Tr}( \tilde \zeta \tilde \zeta \partial_\eta \rho ) & = & - \big( A + i a \big) \mathrm{Tr}\Big( [ \tilde \zeta \tilde \zeta, \tilde \zeta ]  \tilde \zeta \varrho_{\bmk} \Big)  - (B+ i b)  \mathrm{Tr}\Big( [ \tilde \zeta \tilde \zeta, \tilde \zeta ] \tilde \mfp  \varrho_{\bmk} \Big)  \\
&&\qquad\qquad\qquad  - (C+ i c)  \mathrm{Tr}\Big( [ \tilde \zeta \tilde \zeta, \tilde \mfp  ] \tilde \zeta \varrho_{\bmk} \Big) - \big( D + i d \big) \mathrm{Tr}\Big( [ \tilde \zeta \tilde \zeta, \tilde \mfp  ] \tilde \mfp  \varrho_{\bmk} \Big) + \mathrm{h.c.} \nn \\
& = & - 2 i  (C+ i c)  \mathrm{Tr}\Big(  \tilde \zeta \tilde \zeta \varrho_{\bmk} \Big) + 2 i  (C - i c)  \mathrm{Tr}\Big(  \tilde \zeta \tilde \zeta \varrho_{\bmk} \Big)  - 2 i \big( D + i d \big) \mathrm{Tr}\Big( \tilde \zeta \tilde \mfp  \varrho_{\bmk} \Big) + 2 i \big( D - i d \big) \mathrm{Tr}\Big( \tilde \mfp  \tilde \zeta  \varrho_{\bmk} \Big) \nn \\
& = & 2 (c - i C) \Sigma_{11} + 2 ( c + i C) \Sigma_{11} + 2 \big(d - i D\big) \mathrm{Tr}\Big( \tilde \zeta \tilde \mfp  \varrho_{\bmk} \Big) + 2 \big( d + i D  \big) \mathrm{Tr}\Big( \tilde \mfp  \tilde \zeta  \varrho_{\bmk} \Big) \nn \\
& = & 4 c  \Sigma_{11} + 4 d \Sigma_{12} + 2 D \,,\nn
\end{eqnarray}
\end{small}\ignorespaces
where the term involving $D$ in the final line is proportional to $[\tilde \zeta, \tilde \mfp]$ and so becomes field-independent once the canonical commutation relations are used. Following similar steps gives the result for the other choices for $S$:
\be 
 \tfrac12\mathrm{Tr}( \{ \tilde \zeta,\tilde \mfp  \} \partial_\eta \rho )   =   - 2 a \Sigma_{11} + 2( -  b + c ) \Sigma_{12} + 2 d \Sigma_{22} - B - C  
\ee
and
\be 
 \mathrm{Tr}( \tilde \mfp  \tilde \mfp  \partial_\eta \rho )   =   - 4  a \Sigma_{12}  - 4 b \Sigma_{22} +2 A \,.
\ee

Inserting these into eq.~(\ref{app:matTrans1}) leads to our main transport equation:
\begin{eqnarray} \label{app:matTrans2}
\frac{\partial}{ \partial \eta } \left[ \begin{matrix} \Sigma_{11} \\ \Sigma_{12} \\ \Sigma_{22} \end{matrix}  \right] =\left[ \begin{matrix} 4 c & 2 \mfz_s^{-2} + 4 d & 0 \\
 - \mfz_s^2 k^2 - 2 a & 2(-b + c) & \mfz_s^{-2} + 2 d \\
 0 & -2 \mfz_s^{2} k^2 - 4 a & - 4 b \end{matrix}  \right] \left[ \begin{matrix} \Sigma_{11} \\ \Sigma_{12} \\ \Sigma_{22} \end{matrix}  \right] + \left[ \begin{matrix} 2 D  \\ - B - C \\ + 2 A  \end{matrix}  \right] 
\end{eqnarray}
Rewritten in terms of the notation of the main text this becomes
\begin{small}
\be \label{app:matTrans3}
\frac{\partial}{ \partial \eta } \left[ \begin{matrix} \Sigma_{11} \\ \Sigma_{12} \\ \Sigma_{22} \end{matrix}  \right] =\left[ \begin{matrix} 4 \mathrm{Im}[\mathcal{J}^{\mfp  \zeta}_{k0} ]  & 2 \mfz_s^{-2} + 4 \mathrm{Im}[\mathcal{J}^{\mfp \mfp}_{k0} ] & 0 \\
 - \mfz_s^2 k^2 - 2 \mathrm{Im}[\mathcal{J}^{\zeta \zeta}_{k0} ] & 2\mathrm{Im}[-\mathcal{J}^{\zeta \mfp}_{k0}  + \mathcal{J}^{\mfp \zeta}_{k0} ] & \mfz_s^{-2} + 2 \mathrm{Im}[\mathcal{J}^{\mfp \mfp}_{k0} ] \\
 0 & -2 \mfz_s^{2} k^2 - 4 \mathrm{Im}[\mathcal{J}^{\zeta \zeta}_{k0} ] & - 4 \mathrm{Im}[\mathcal{J}^{\zeta \mfp}_{k0} ] \end{matrix}  \right] \left[ \begin{matrix} \Sigma_{11} \\ \Sigma_{12} \\ \Sigma_{22} \end{matrix}  \right] + \left[ \begin{matrix} 2 \mathrm{Re}[\mathcal{J}^{\mfp \mfp}_{k0} ]  \\ - \mathrm{Re}[\mathcal{J}^{\zeta \mfp}_{k0}  + \mathcal{J}^{\mfp \zeta}_{k0} ] \\ + 2 \mathrm{Re}[\mathcal{J}^{\zeta \zeta}_{k0}]   \end{matrix}  \right] 
\ee
\end{small}\ignorespaces

\subsection{Gaussian Evolution of the Purity}
\label{ssecApp:GaussianEvo}

The purity for a Gaussian theory is given in terms of $\boldsymbol{\Sigma}$ by
\begin{eqnarray}
\gamma_\bmk =\frac{1}{\sqrt{ 4 \det \boldsymbol{\Sigma_\bmk} }} \,,
\end{eqnarray}
where we re-emphasize the $\bmk$-dependence of $\boldsymbol{\Sigma}$. For applications to decoherence it is useful to use these expressions to compute the evolution of the determinant of the covariance matrix 
\begin{eqnarray}
\frac{\partial \det \boldsymbol{\Sigma}}{\partial \eta} & = & \frac{\partial \Sigma_{11}}{\partial \eta} \Sigma_{22} + \Sigma_{11} \frac{\partial \Sigma_{22}}{\partial \eta} - 2 \frac{\partial \Sigma_{12}}{\partial \eta} \Sigma_{12} \nn\\
& = & 4 (c-b) \det \boldsymbol{\Sigma} + 2\Bigl[ A \Sigma_{11} + (B+C)\Sigma_{12} + D \Sigma_{22} \Bigr] \,.
\end{eqnarray}
Notice in particular that the inhomogeneous source terms in \pref{app:matTrans2} contribute to this last equation terms linear in the correlators. In terms of the notation from the main body this is
\begin{equation}
\frac{\partial \det \boldsymbol{\Sigma}}{\partial \eta} = - 4 \mathrm{Im}[\mathcal{J}^{\zeta \mfp}_{k0} - \mathcal{J}^{\mfp \zeta}_{k0}] \det \boldsymbol{\Sigma} + 2\Big( \mathrm{Re}[\mathcal{J}^{\zeta \zeta}_{k0} ] \Sigma_{11} + \mathrm{Re}[\mathcal{J}^{\zeta \mfp}_{k0} + \mathcal{J}^{\mfp \zeta}_{k0}] \Sigma_{12} + \mathrm{Re}[\mathcal{J}^{\mfp \mfp}_{k0} ] \Sigma_{22} \Big)
\end{equation}

In perturbation theory it suffices to use the leading order behaviour of the equal-time correlators, $\Sigma_{ij} \simeq \sigma_{ij}$, on the right-hand side of this last result. Recall that the mode functions $\zeta$ and $\mfp$ are respectively
\begin{eqnarray}
\widetilde{u}_{k}(\eta) =  \frac{ u_{k}(\eta) }{\mfz(\eta)} = \frac{ H e^{- i k\eta} ( i - k \eta) }{2 M_p \sqrt{\epsilon_1}k^{3/2}  } \qquad \mathrm{and} \qquad \widetilde{p}_{k}(\eta) = \mfz^2(\eta) u'_{k}(\eta) = \frac{i \sqrt{k\epsilon_1} e^{- i k\eta} M_p}{H \eta}
\end{eqnarray}
we see that
\begin{eqnarray} \label{FreeSigma}
\sigma_{11}(\eta) & \simeq & \widetilde{u}_{k}(\eta) \widetilde{u}^{\ast}_{k}(\eta)  = \frac{H^2 \left(\eta ^2 k^2+1\right)}{4 k^3 M_p^2 \epsilon _1}   \\
\sigma_{12}(\eta) & \simeq & \mathrm{Re}[ \widetilde{u}_{k}(\eta) \widetilde{p}^{\ast}_{k}(\eta) ]  = \frac{1}{2 k \eta}   \\
\sigma_{22}(\eta) & \simeq & \widetilde{p}_{k}(\eta) \widetilde{p}^{\ast}_{k}(\eta)  = \frac{k \epsilon _1 M_p^2}{\eta ^2 H^2}  \,.
\end{eqnarray}
These imply the leading result $\det \boldsymbol{\sigma}  = \frac{1}{4}$ as required for a pure state.

\bibliographystyle{JHEP}
\bibliography{dSDecoherenceGWDom.bib}

\end{document}